\documentclass[]{imag-ms-template}
\usepackage{comment}

\usepackage{tikz}
\usepackage{url}

\title{Converting T1-weighted MRI from 3T to 7T quality using deep learning}

\author{Malo Gicquel$^{1,2}$, Ruoyi Zhao$^{1}$, Anika Wuestefeld$^{3}$, 
Nicola Spotorno$^{3}$,\\  Olof Strandberg$^{3}$, Kalle Åström$^{4}$, Yu Xiao$^{1}$, Laura EM Wisse$^{5}$, Danielle van Westen$^{5}$, \\ Rik Ossenkoppele$^{3,6,7}$ Niklas Mattsson-Carlgren$^{3,8}$, David Berron$^{3,9,10}$,   
  Oskar Hansson$^{3}$,\\ Gabrielle Flood$^{4,11}$\#, Jacob Vogel$^{1}\ast\#$ \\
{\small $^{1}$ Department of Clinical Sciences Malmö, SciLifeLab, Lund University, Lund, Sweden}\\
{\small $^{2}$ Univ Rennes, CNRS, Inria, Inserm, IRISA UMR 6074, Empenn ERL U 1228, Rennes, France}\\
{\small $^{3}$ Clinical Memory Research Unit, Department of Clinical Sciences Malmö, Lund University, Lund, Sweden}\\
{\small $^{4}$ Centre for Mathematical Sciences, Lund University, Lund, Sweden}\\
{\small $^{5}$ Diagnostic Radiology Unit, Department of Clinical Sciences Lund, Lund University, Lund, Sweden}\\
{\small $^{6}$ Alzheimer Center Amsterdam, Vrije Universiteit Amsterdam, Amsterdam UMC, Amsterdam, The Netherlands}\\
{\small $^{7}$  Amsterdam Neuroscience, Neurodegeneration, Amsterdam, The Netherlands}\\
{\small $^{8}$ Memory Clinic, Skåne University Hospital, Malmö, Sweden}\\
{\small $^{9}$ German Center for Neurodegenerative Diseases, Magdeburg, Germany}\\
{\small $^{10}$ Center for Behavioral Brain Sciences, Otto-von-Guericke University Magdeburg, Magdeburg, Germany}\\
{\small $^{11}$ Visual Recognition Group, Faculty of Electrical Engineering, Czech Technical University in Prague,} \vspace{-2mm}\\{\small Prague, Czech Republic}\\
{\small $^\ast$ Correspondence:  jacob.vogel@med.lu.se
} \\
{\small $\#$Denotes equal contribution}
}

\begin{document} 
\date{}
\maketitle

\newpage
\begin{abstract}
   
Ultra-high resolution 7 tesla (7T) magnetic resonance imaging (MRI) provides detailed anatomical views, offering better signal-to-noise ratio, resolution and tissue contrast than 3T MRI, though at the cost of accessibility. We present an advanced deep learning model for synthesizing 7T brain MRI from 3T brain MRI. Paired 7T and 3T T1-weighted images were acquired from 172 participants (124 cognitively unimpaired, 48 impaired) from the Swedish BioFINDER-2 study. To synthesize 7T MRI from 3T images, we trained two models: a specialized U-Net, and a U-Net integrated with a generative adversarial network (GAN U-Net). Our models outperformed two previous state-of-the-art 3T-to-7T models in image-based evaluation metrics. Four blinded MRI professionals judged our synthetic 7T images as comparable in detail to real 7T images, and superior in subjective visual quality to 7T images, due to the reduction of artifacts. Using both SynthSeg and NextBrain, automated segmentations of the synthetic 7T images were more similar to real 7T segmentations than automated segmentations from the 3T images that were used to synthesize the 7T images. Finally, synthetic 7T images showed similar performance to real 3T images in downstream prediction of cognitive status using MRI derivatives (n=3,168). In all, we show that synthetic T1-weighted brain images approaching 7T quality can be generated from 3T images, which may improve image quality and segmentation, without compromising performance in downstream tasks. Future directions, possible clinical use cases, and limitations are discussed.
\end{abstract}


\newpage
\section{Introduction}
Magnetic Resonance Imaging (MRI) is an in vivo, non-invasive medical imaging technique used to visualize detailed internal structures of the body using magnetic fields. Brain MRI is critical for diagnosis and treatment planning of a wide range of neurological disorders, including normal pressure hydrocephalus \citep{Hashimoto2010}, glioma \citep{Weller2021}, epilepsy \citep{Wang2020}, neurovascular disease \citep{Debette2019}, and dementia \citep{Barkhof2011}, among many others \citep{Morrisb3016, Kuoy2022}. T1-weighted MRI scans are used to visualize and/or quantify anatomical structures, providing an excellent contrast between different tissue types. Besides its use in clinical care, structural MRI provides insights into regional brain structure and morphometry, allowing researchers to track alterations to gray and white matter over the course of normal brain development \citep{Bethlehem2022}, brain aging and disease \citep{Yang2024}.

The quality of MR images is determined by the strength of the magnet, measured in teslas (T). Common field strengths within clinical care are 1.5T and 3T, whereas 7T is used primarily in research settings. Compared to 3T MRI, 7T offers a superior signal-to-noise ratio, spatial resolution and contrast, helping to visualize more fine-grained brain structures with greater fidelity. This visual improvement can aid clinicians in detecting many types of brain pathology \citep{Duzel2019,Opheim2021}. For instance, in multiple sclerosis (MS), 7T MRI enables the detection of small and subtle cortical lesions, and has a higher iron and myelin susceptibility that enables a better characterization of those lesions \citep{MS7T}. In epilepsy, where successful surgical treatment relies on correct localization of epilectic lesions, 7T scans offer more confidence in identification, especially in cases with more subtle pathology \citep{Zampeli2022,Sharma2021}. 7T can also benefit Alzheimer's disease (AD) research by providing in vivo information about changes to structures that are difficult to image with 3T MRI, such as the locus coeruleus \citep{Priovoulos2018} and substructures of the medial temporal lobe \citep{Kenkhuis2019, Berron2017, perera20237TALZ}, while also allowing better visualization of microinfarcts and microbleeds \citep{vanVeluw2012,vanVeluw2015}. Despite these benefits, 7T scanners are rare, expensive, and technically challenging to employ, partly due to their powerful magnetic fields. At the time of writing, estimates suggest that there are less than 150 7T MRI scanners worldwide\footnote{\url{https://google.com/maps/d/viewer?ll=1.9418261240470145\%2C0\&z=2\&mid=1dXG84OZIAOxjsqh3x2tGzWL1bNU}}, used primarily for research. In contrast, the regular use of 3T MRI for routine clinical care has led to datasets of tens or even hundreds of thousands of 3T MRI images becoming available to researchers \citep{Bethlehem2022, wang2025triadvisionfoundationmodel}.

To make high-resolution clinical imaging more accessible and clinically viable, many ``super resolution'' techniques have been developed (\cite{SRclinical}), among which the most efficient rely on deep learning. Most existing super resolution models seek to convert images from 1.5T quality to 3T quality (\cite{wang2023inversesr}, \cite{liao2022comparativestudy15t3tmri}) or from portable very low field 0.064T to higher quality images from higher field strength acquisitions \citep{Islam2023, Iglesias2023, Lucas2023}. While less common, other researchers have also investigated synthesis of 7T quality T1-weighted brain MR images from 3T images (\cite{Bahrami2017}, \cite{Qu2020}, \cite{cui20247tmrisynthesization3t}, \cite{SR3T7T}). However, most of these 3T to 7T models were trained on small datasets, often with less than 20 participants. Here, we present a deep super resolution model to convert T1-weighted MRI images from 3T to 7T quality, trained on 172 participants with matched 3T and 7T data, and built atop previous architectures (\cite{Bahrami2017},\cite{cui20247tmrisynthesization3t}). Aside from classical image evaluation metrics, we also investigate three methods for practical evaluation of synthetic images. This includes visual quality evaluation from MRI professionals, a segmentation task, and downstream performance on machine learning based automatic dementia diagnosis. 

\section{Methods}
\subsection{Study population and datasets}
\label{sec: dataset}

\textbf{Main dataset.} Our dataset is composed of pairs of 3T and 7T images from 172 participants  of the Swedish BioFINDER-2 study (NCT03174938) \citep{Palmqvist2020}. The study was approved by the ethical review board in Lund, Sweden, and all study participants provided written informed consent.
Participant age (average: $61.93\pm 11.8$), gender ($48\%$ females) and diagnosis were recorded. Participant diagnosis was either cognitively normal (CN; 74 participants), subjective cognitive decline (SCD; 50 patients), mild cognitive impairment (MCI; 46 patients) or dementia (2 patients). SCD indicates a subjective experience of impaired cognition that could not be corroborated with objective cognitive testing. SCD, MCI and dementia patients were recruited from a memory clinic and their diagnoses were attributed according to the DSM-5 criteria \citep{arvidsson2024MCI}. The unimpaired participants were recruited from the population in and around the city of Malmö, Sweden. We separated the diagnoses into two groups: cognitively unimpaired (CN and SCD together) and cognitively impaired (MCI and dementia together). Age distribution according to diagnosis and gender can be found in Appendix Fig.\ref{fig: BF2 charac}.

\begin{figure}[!ht]
    \centering
    \includegraphics[width=0.75\textwidth]{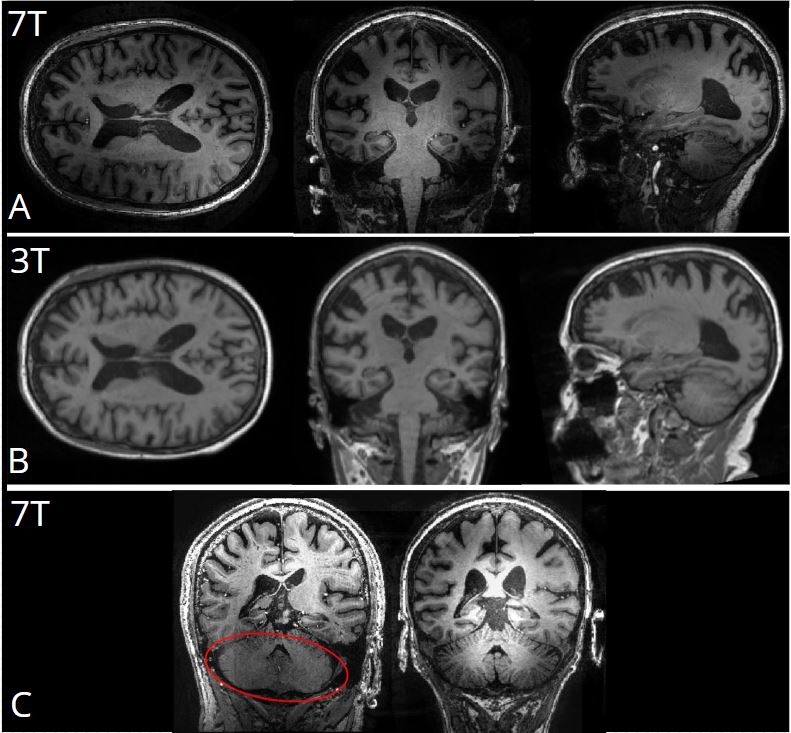}
\caption{ A and B: Three raw T1-weighted MRI slices, one along each dimension, of the same patient's scan in 7T (A) and 3T (B) quality. The images were aligned for this figure to make the comparison easier. C: Example of a slice of a 7T scan with (on the left) and without (on the right) artifacts in the cerebellum, which we identified as systematic issues that could affect model training.}
    \label{fig: ex3T 7T}
\end{figure}

Each patient had only one 7T scan, but most had multiple 3T scans. Examples of the data can be seen in Figure \ref{fig: ex3T 7T}A and B. We selected the 3T scan acquired closest in time of 7T acquisition (mean difference between 7T and 3T acquisition dates : 3.6 $\pm$ 9.4 months), in order to limit age- and disease-related effects on brain morphometry. We performed visual quality control (QC) of all images, to exclude patients with excess motion or distortion and overall scan quality. From the full dataset, we removed seven participants with 7T acquisitions. We further quarantined 29 pairs as our testing dataset, and did not use these participants in model training. This train/test split was stratified by the age, gender and diagnosis (``impaired'' or ``unimpaired''). As a result, we built our model on 136 3T-7T pairs and tested it on 29 pairs. Notably, several of the 7T scans showed artifacts that selectively but dramatically affected the cerebellum (Figure~\ref{fig: ex3T 7T}C), but otherwise passed visual QC. These artifacts are likely caused by reduced sensitivity of the headcoil around the cerebellum, and variability across participants may relate to placement of the participant head in the scanner. Out of the 172 participants, 44 7T scans showed at least some evidence of this scanning artifact in the cerebellum. We discuss our mitigation strategy for these artifacts below.

\textbf{Diagnostic prediction dataset}. An additional dataset was available, composed of 3,574 3T T1-weighted scans from the same Swedish BioFINDER-2 study, obtained from the same scanner as the training images. After filtering out scans with missing age and those from participants already used to build the model, 3,168 scans were retained. Neither the images nor the participants were previously seen by the model during training, ensuring independent evaluation. We did not remove any of the 29 participants used to test the models, as the output is not influenced by a prior exposure to a participants' scan. These scans were used to assess the generalizability and utility of the models. In the appendix, we include Table \ref{Participants} which contains the characteristics of the participants in this dataset.

\subsection{MRI acquisition}
3T T1-weighted images were acquired on a Siemens Prisma scanner (Siemens Medical Solutions) with a 64-channel head coil using an MPRAGE sequence (in-plane resolution = $1\times1$ mm$^2$, slice thickness = $1$ mm, repetition time = $1,900$ ms, echo time = $2.54$ ms, flip-angle = $9$\textdegree) (\cite{ref3t}). 7T images were acquired on a 7T MRI scanner (Philips Acheiva, Best, the Netherlands) at Skåne University Hospital in Lund. The scanner was equipped with a gradient head coil with $32$ receive channels and two transmit channels using single channel transmit (Nova Medical, Wilmington, MA). Data were acquired using standard procedure implemented at the National 7T facility of Lund University Hospital. To obtain T1-weighted images, a 3D magnetization prepared-rapid gradient echo (MPRAGE) sequence (resolution = $0.7 \times 0.7 \times 0.7 \,\text{mm}^3$, TR = $8$ ms, shot duration = $2,200$ ms, echo time (TE) = $2.7$ ms, flip angle = $7$\textdegree) was used. Examples of 3T and 7T T1-weighted MRI scans are shown in Figure~\ref{fig: ex3T 7T}.

\subsection{Image processing}
\label{processing}
\begin{figure}[!ht]
    \centering
    \includegraphics[width=0.65\textwidth]{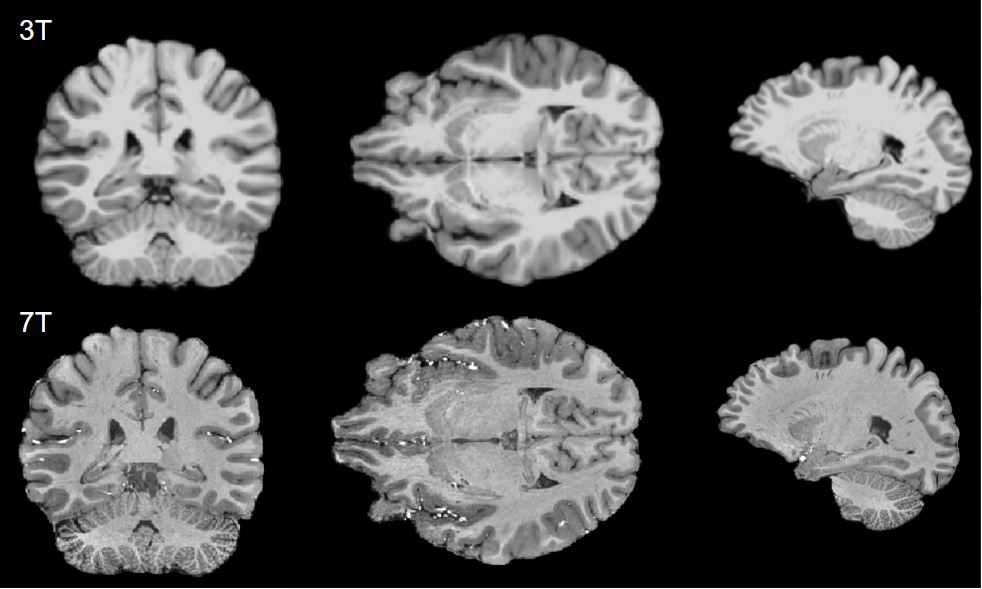}
    \caption{Aligned and preprocessed T1 weighted MRI slices of a patient scan in 3T and 7T MRI. We show one slice along each dimension for each image type. Notice how we can see hyperintense voxels (probably blood vessels) in the 7T, which are not visible in the 3T image.}
    \label{fig: ex3T 7T proc}
\end{figure}

Our 3T images were bias field corrected and skull stripped using FreeSurfer v6.0\footnote{\url{https://surfer.nmr.mgh.harvard.edu/}} using recon -all. For 7T scans, we generated preliminary brain masks using MRI SynthStrip \citep{SynthStriphoopes2022}. Then, we applied a masked bias field correction using the n4 bias field correction algorithm \citep{BIASCORRECTION} with the following parameters: Convergence Threshold = 1e-7; Maximum Number of Iterations = [150,150,150,150]; Bias Field Full Width At Half Maximum = 0.18; Wiener Filter Noise = 0.2. We then skull stripped the images once again after n4 bias field correction for an improved brain mask. Next, each 3T image was registered to its 7T counterpart using the affine followed by a “SyN” nonlinear registration from ANTs \citep{ANTSSYN}, using a linear interpolator for the initial upsampling and a mutual information metric. Finally, we normalized all the images using a clipped min-max normalization. We could not use a regular min-max normalization, as the maximum intensity is unstable due to certain hyperintense voxels (e.g.\ from blood vessels). To solve this issue, we took inspiration from \cite{ScannerAug} and we neglected the background and the intensities above the $99^{\text{th}}$ percentile when doing a min-max normalization. We show an example of a resulting processed pair of scans in Figure~\ref{fig: ex3T 7T proc}.

The models were trained on 3T images registered to their 7T counterpart with an affine transform and a small non-linear deformation map. However, during inference, there is no 7T image to which the 3T image can be registered. Yet, a deep learning model performs better with images similar to its training dataset. We thus created a simple template-based registration framework for inference. Using ANTs, we created a study-specific 3T template based on 30 3T images from our training set. We then created a study-specific 7T template with the 30 paired 7T images. We performed an affine registration of the 3T template to the 7T template, and stored the transform. During inference, 3T images were first rigidly coregistered to the 3T template, and the 3T$\rightarrow$7T affine transform was subsequently applied. This process ensured that any 3T T1w image could be moved to the general space expected by our models. The 3T template and 3T$\rightarrow$7T transform were made available on the project github page\footnote{\url{https://github.com/DeMONLab-BioFINDER/SuperResolutionMalo}}.

\subsection{Model}

We use two different models: a U-Net and a U-Net generative adversarial network (GAN). The input and output of the models are images slice-by-slice along the axial dimension. The GAN U-Net combines the U-Net architecture as the generator with the generative adversarial network framework, enhancing the quality of the synthesized images. The U-Net is built on code by Lopez Pinaya et al., who created a U-Net to use as a denoiser in a diffusion model generating synthetic brains \citep{pinaya2022brainGen}. We made three changes to Lopez Pinaya et al.'s code:

\begin{itemize}
    \item We turned the 3D model into a 2D model (moving along axial slices). Compared to models using full 3D images as input, a 2D slice approach reduces GPU memory requirements, while providing more samples for training and ensuring those samples still contain coherent macro-scale anatomical information. We anticipated that this choice might result in slicing artifacts along sagittal and coronal planes, and attempted to mitigate this with a 3-channel sandwich approach as described below.
    \item We changed the residual blocks to include AdaDM layers \citep{liu2021adadm} at each residual block, to reduce the blurriness caused by the normalization layers.
    \item We removed the last normalization layer, as it can make the results more blurry.
\end{itemize}

The U-Net uses residual blocks and attention mechanisms. The attention mechanisms allow conditioning on external variables. We initially decided to condition on the age, gender and diagnosis (i.e.\ cognitively impaired or cognitively unimpaired) of the participants. This was partly motivated by \cite{wang2023inversesr}'s choice to condition on age, gender, ventricular volume and brain volume, but we also note that age and sex can influence T1 contrast \citep{lewis2019cortical,corpataux2025effect}. However, we determined that conditioning on the diagnosis can limit utility in practice (i.e., then a diagnosis is needed in all cases where one wishes to perform inference), and so we opted not to condition on diagnosis in the end. We also conditioned on the approximate slice location, calculated as
\begin{equation*}
    \frac{2s-(top+bot)}{top-bot},
\end{equation*}
where we let $s$ denote the slice index and the indexes of the highest and lowest brain 
axial slice (i.e., the indexes above and under which there is only background) are denoted $top$ and $bot$, respectively.
We chose this simple formula so that the middle slice of index $s=\frac{top+bot}{2}$ has a slice location of 0, $s=top$ of 1 and $s=bot$ of -1.

The discriminator we used in our GAN U-Net is the discriminator from the patch GAN in the Python package MONAI generative \citep{pinaya2023generativeaimedicalimaging,cardoso2022monai}. We also used a WGAN-GP \citep{gulrajani2017WGANGP} that adds a gradient penalty to the GAN loss, to prevent a mode collapse. The gradient penalty of the WGAN-GP was implemented using existing code from \cite{wgangp_github}. Detailed model architectures are available in Appendix \ref{Drawings} (Figures~\ref{fig: self block} - \ref{fig: GAN discr}).

\subsection{Losses}

With regard to model loss function, we let $I$ denote a ground truth 2D image (slice), $I'$ the corresponding generated image, and $\mathcal{V}$ the set of all pixel indexes in these images. We use the L1 loss as it makes the resulting image sharper than the L2 loss, which can make the results blurry and smooth. Furthermore, there is an issue with voxel hyperintensity endemic to the 7T images, and the L1 loss reduces the impact of the hyperintensities compared to the L2 loss. The L1 loss is given by

\begin{equation*}
    \mathcal{L}_1(I,I') =  \frac{1}{|\mathcal{V}|}\sum_{v\in\mathcal{V}}|I(v)-I'(v)|,
\end{equation*}
with $|\mathcal{V}|$ denoting the cardinality of $\mathcal{V}$.

The L1 and L2 losses are based on pixel-to-pixel differences. To look at image differences on broader scales, many computer vision tasks use a perceptual loss \citep{zhang2018Perc}. The idea is to compare the features generated by a CNN trained on another task, such as segmentation or classification, by computing the mean squared difference of the features generated from the two images. Let $L$ be the set of all the layers of the perceptual model, $\mathcal{V}_i$ be the set of all the feature indices of the $i^{\text{th}}$ layer and let $\phi_{l}(I)$ be the features of $I$ generated by the layer $l$ of the pretrained CNN. Then, the perceptual loss is given by

\begin{equation*}
    \mathcal{L}_{perc}(I,I') = \frac{1}{|L|}\sum_{l\in L}\frac{1}{| \mathcal{V}_i|}\sum_{v\in\mathcal{V}_i}(\phi_{l}(I)(v)-\phi_{l}(I')(v))^2.
\end{equation*}

Many papers (\cite{pinaya2022brainGen} and \cite{wang2023inversesr}) use a perceptual loss from the Python package LPIPS \citep{zhang2018Perc}. It contains \enquote{radimagenet\_resnet50} (\enquote{radimagnet} for short). We decided to use it as it has been trained on 2D slices of different parts of the body and scanner modalities, including brain MRI \citep{radimagenet}. 

The final loss $\mathcal{L}_{U\text{-}Net}$ for the U-Net is constructed as a weighted sum of the perceptual loss and the L1 loss, using the perceptual weight $\lambda_{perc}$
\begin{equation*}
    \mathcal{L}_{U\text{-}Net} = \mathcal{L}_1 + \lambda_{perc}\mathcal{L}_{perc}.
\end{equation*}

We also define a GAN loss $\mathcal{L}_{GAN}$ with a weighted gradient penalty (from the WGAN-GP \citep{gulrajani2017WGANGP}), where $I,I'$ are, respectively, the real and fake 7T slices, $D$ is the discriminator (that we want to train such that $D(I)=1$ and $D(I')=0$) and $\lambda_{GP}$ is the weight of the loss
\begin{equation*}
    \mathcal{L}_{GAN}(I,I') = \log(D(I))+\log(1-D(I'))+\lambda_{GP}(\lVert \nabla_x D(x)\rVert_2-1)^2,
\end{equation*}
where $x = \alpha I + (1-\alpha) I'$, $\alpha\sim\mathcal{U}(0,1)$, and with $\mathcal{U}$ being the uniform distribution.

\noindent The loss used to train the generator is, with $\lambda_{GAN}$ the weight of the GAN loss,

\begin{equation*}
    \mathcal{L}(I,I') = \mathcal{L}_{U\text{-}Net}(I,I')+\lambda_{GAN}\mathcal{L}_{GAN}(I,I').
\end{equation*}

We show a detailed description of the model and loss hyperparameters $\lambda_{GP}, \lambda_{GAN}, \lambda_{perc}$ in Appendix~\ref{Hyperpara}, along with the values we chose and the reasoning behind our choices. 

\subsection{Baseline models}

We compared our models to the state-of-the-art models WATNet (\cite{Qu2020}) and V-Net (\cite{cui20247tmrisynthesization3t}). The training and inference code for both models provided by Cui et al., was applied directly on our preprocessed dataset without any data augmentations. We only used the default V-Net and not the GAN or SynthSeg loss V-Nets, as it had the best results in the original paper \citep{cui20247tmrisynthesization3t}. To take into account the fact that our dataset is very large, we decided to add a learning rate decay of 0.8 per epoch and used an initial learning rate of $10^{-3}$ for the V-Net and $10^{-4}$ for the WATNet, during 20~epochs.

\subsection{Training details}

We chose to slice our input images along the axial dimension for two reasons. First, it allowed for more slices per subject compared to slicing along the sagittal plane, and therefore allowed more overall samples in model training. Second, the axial slices containing the cerebellar artifacts had less brain in them (mostly brainstem and pons) than the coronal slices containing the cerebellar artifact (parietal and occipital cortex, posterior ventricles). Importantly, we explicitly did not train on slices containing the artifacts, which were determined separately for each subject using visual assessment. So, using training slices along the axial dimension allowed us to minimize information loss when censoring slices containing the artifacts. To optimize memory usage, we cropped most of the image background. This resulted in slices of size (288,224). These numbers are both divisible by $2^5$, allowing us to perform five downsamplings by a factor of two, which is important as each stage of the U-Net uses a downsampling layer (see Figure~\ref{fig: attention Unet}). To reduce potential slicing artifacts along non-axial dimensions, we increased the number of input channels by also including the two neighboring 2D slices. This resulted in inputs of size (3,288,224) and  outputs of size (1,288,224). To train our model, we used PyTorch on eight A100 SXM4 80GB NVIDIA GPUs.

\subsection{Assessment metrics}

Once we have trained a model, its performance needs to be assessed using different methods, by comparing ground truth 7T images to their synthetic counterparts. All comparisons and segmentations are performed on the entire 3D images and not on the individual 2D slices. Below, we present our different assessment metrics.

\subsubsection{Mathematical comparisons} 
\label{sec : math compa}

We used two mathematical comparison tools: PSNR and SSIM. However, we compute these slightly differently from the classical PSNR and SSIM, as we used a clipped min-max normalization instead of the usual min-max normalization. We denote $I$, $I'$ two normalized 3D images to be compared, $\mathcal{V}$ the set of all the voxel indexes and $|\cdot|$ the function that gives the cardinality of a set. The PSNR is then given by

\begin{equation*}
    \text{PSNR}(I,I') = -10 \log_{10}\left(\text{MSE}(I,I')\right),
\end{equation*}
where
\begin{equation*}
     \text{MSE}(I,I') = \frac{1}{|\mathcal{V}|} \sum_{v\in \mathcal{V}}(I(v)-I'(v))^2.
\end{equation*}

We also define
\begin{equation*}
        \text{SSIM}(I,I') = \frac{(2\mathbb{E}(I)\mathbb{E}(I')+c_1)(2\mathbb{C}(I,I')+c_2)}{(\mathbb{E}(I)^2+\mathbb{E}(I')^2+c_1)(\mathbb{V}(I)+\mathbb{V}(I')+c_2)},
\end{equation*}
where $c_1 = 10^{-4}, c_2 = 9.10^{-4}$ and 
$\mathbb{E}$, $\mathbb{C}$ and $\mathbb{V}$ are, respectively, the intensity expectancy, covariance and variance.

The PSNR evaluates voxel-to-voxel squared differences, while the SSIM compares the contrast, luminance and structure of the images. To account for the aforementioned cerebellum corruptions (see Section \ref{sec: dataset}) and the amount of background that can vary from one dataset to another, we calculated these metrics on the images without background and artifacts. To remove the artifacts, we excluded the axial slices in which we saw any artifact in the 7T images, just like during training.

\subsubsection{Visual assessment}
\label{sec : visual survey}

A classic way to assess the performances of natural image super resolution tools is to rely on human qualitative assessment, as the ultimate goal is to make the images look better to humans, which is difficult to investigate and characterize mathematically. When it comes to medical image super resolution, this is not the only goal, but one application could be to help professionals assess the scans by making them look better and be more detailed. We thus decided to conduct a small survey, by asking four neuroradiologists and MRI scientists to qualitatively rate sets of real and synthetic images. Our survey included 28 queries -- one for each of the images in the test set, but excluding an image in the test set that showed low outlying values across all metrics and all models. Each query included a randomly chosen 2D slice in any dimension for one participant, displayed in six variants: real 3T, real 7T and synthetic 7T using four different models (our U-Net, our GAN U-Net, a V-Net and a WATNet). We then asked the professionals to rank the six images from best to worst according to a criterion. We had two criteria (left intentionally open, given the subjective nature of the query):

\begin{itemize}
    \item Rank based on how good the image looks.
    \item Rank based on how detailed the image is.
\end{itemize}

Our hypothesis was that it would be easy to spot the 7T, mostly because of the noise, blood vessels and artifacts. We also anticipated that the 7T would be perceived as the most detailed but not necessarily the best looking, especially as the 7T scans can often contain artifacts. To assess whether rankings were statistically different across images, we performed repeated-measure ANOVAs with rank as the dependent variable, image type as the within-subject effect, and rater as the grouping variable. Image to image differences were quantified using Tukey's posthoc tests. 

\subsubsection{Automatic segmentation}
\label{sec:automatic_segmentation}

As 7T scans offer a better tissue contrast and a better spatial resolution than 3T, they have the potential to improve anatomical segmentation \citep{Bahrami20167TBetterSeg}. To know if our models can improve upon 3T segmentations, we assessed segmentations using the Dice score across two assessments softwares. 
Denoting two segmentated 3D regions to be compared $V$ and $V'$, respectively, the Dice score is defined as
\begin{equation*}
\mathrm{Dice}(V, V') =
\frac{2|V \cap V'|}{|V| + |V'|},
\label{eq:dice}
\end{equation*}
where $|\cdot|$ again represents the cardinality of, or in this case the number of voxels in, the set. We used the average Dice score over all regions for a certain participant as an overall measure of segmentation quality.

Inspired by \cite{cui20247tmrisynthesization3t}, we first used SynthSeg \citep{synthseg}, a deep learning segmentation tool, to automatically segment the ground truth 7T scans and compare the results given by the same tool applied to the associated 3T registered scan and the synthetic 7T scans generated by the different models. The comparison is done using the Dice metric. We used SynthSeg V1, as it excludes the cerebrospinal fluid (CSF) surrounding the brain. SynthSeg unfortunately works on images of sample size $1\times 1\times 1 \text{ mm}^3$, so our 7T and synthetic 7T images (sample size: $0.7\times 0.7\times 0.7$) were downsampled, which reduces the advantages of using 7T MRI. Yet, the better tissue contrast could theoretically still improve the results. To account for the image artifacts in the cerebellum, we simply removed the segments related to the cerebellum. 

It should be noted that SynthSeg is a deep learning model, trained on various MRI sequences (including T1-weighted) and on various field strengths including 3T, but not 7T. Since these tools rely on data augmentation, they do work on our 7T images. Yet, it is possible that this reduces the benefits of 7T MRI for automatic segmentation. Therefore, we repeated this same task using NextBrain \citep{casamitjana2025probabilistic}, a newer segmentation tool designed for segmentation of high-resolution MRI and histology images. We used the version of NextBrain packaged with Freesurfer v.8.1.0. Similar to our analysis with SynthSeg, we obtained segmentations from the 3T, 7T and all synthetic images in the test set with NextBrain, and used the Dice metrics to compare labels from each image to the labels of its respective 7T image. Importantly, NextBrain resamples input images to $0.4\times 0.4\times 0.4 \text{ mm}^3$ for segmentation, and the Dice comparisons were done there in its native resolution to avoid biasing specifically to 7T or 3T resolution. Initially, we ran Dice metrics across all labels (216 L+R pairs), which includes many small nuclei and sub-sections of larger structures. For post hoc analysis investigating efficacy of segmentation of larger structures, we additionally aggregated these smaller regions into ten larger regions (bilateral amygdala, hippocampus, putamen, pallidum, caudate and accumbens, thalamus, claustrum, brainstem, white matter, and gray matter), mapping small labels to larger labels based on Freesurfer recommendations\begin{samepage}\footnote{\url{https://github.com/freesurfer/freesurfer/blob/dev/mri_histo_util/ERC_bayesian_segmentation/ERC_bayesian_segmentation/relabeling.py}}\end{samepage}.

\subsubsection{Downstream diagnostic predictions}

7T scans offer a better tissue contrast and allow visualization of smaller brain features in greater detail. These properties could possibly facilitate more accurate diagnoses. To test what effect our model would have on diagnostic capabilities, we evaluated its potential in automatic downstream predictions, using the Diagnostic Prediction previously described in Section \ref{sec: dataset}. The principle behind this evaluation is that regional brain volumes, particularly cortical and subcortical segmentations, are valuable biomarkers for dementia prediction. The idea is to use the volume calculation feature of SynthSeg \citep{synthseg}, for different brain regions in various types of images: 3T; enhanced images by our U-Net; and enhanced images by our GAN U-Net. These volumes will be used as input to a simple machine learning model that predicts the diagnosis of the patient. We used both cortical (i.e.\ ``aparc'') and subcortical (i.e.\ ``aseg'') parcellations from SynthSeg V1. A random forest classifier was utilized to predict patient diagnosis from the SynthSeg extracted input features. In this case, patient diagnosis was given as Cognitively Normal Control (CN or SCD), Mild Cognitive Impairment (MCI) or Alzheimer's disease dementia (AD). Note that all AD cases had confirmed AD pathology using CSF or PET biomarkers. The random forest machine learning model was chosen because of its robustness, ability to handle nonlinear relationships, and interpretability. Our goal here was not to derive a very good diagnostic model, but rather to use a familiar machine learning objective to benchmark performance across real and synthetic images.

To assess segmentation performance, we measured the prediction accuracy of the segmentation results derived from the synthesized images and compared them against the ground truth 3T images. The model performance was evaluated using 10-fold cross-validated accuracy and balanced accuracy, the latter being more reliable than the former given the unequal representation among diagnosis categories. To establish confidence bands around performance, we retrained the model 1000 times for each image type (3T, synthetic 7T UNet, synthetic 7T UNet GAN). Additionally, we also aimed to examine how the feature importance varies across real and synthetic images. Since we try to predict the patients' diagnoses, we made sure to use the U-Net and GAN U-Net models that were not conditioned on diagnosis.

\subsection{Ablation experiments}
We wished to assess whether some of our particular modeling decisions strongly influenced the performance of our models. Here, we focused on the GAN U-Net. We trained two additional models. First we trained a model that was the same, except that it was no longer conditioned on age and sex (though we retained conditioning on slice location). In the second model, we removed the AdaDM layers from the residual blocks, and added the final normalization layer back. We compared performance of these models to our final GAN U-Net, using both traditional metrics (PSNR, SSIM) and segmentation performance (Dice scores for both SynthSeg and NextBrain). 

\section{Results}

\subsection{Quantitative evaluation of super-resolution models}

\begin{figure}[!ht]
    \centering
    \includegraphics[width=1\textwidth]{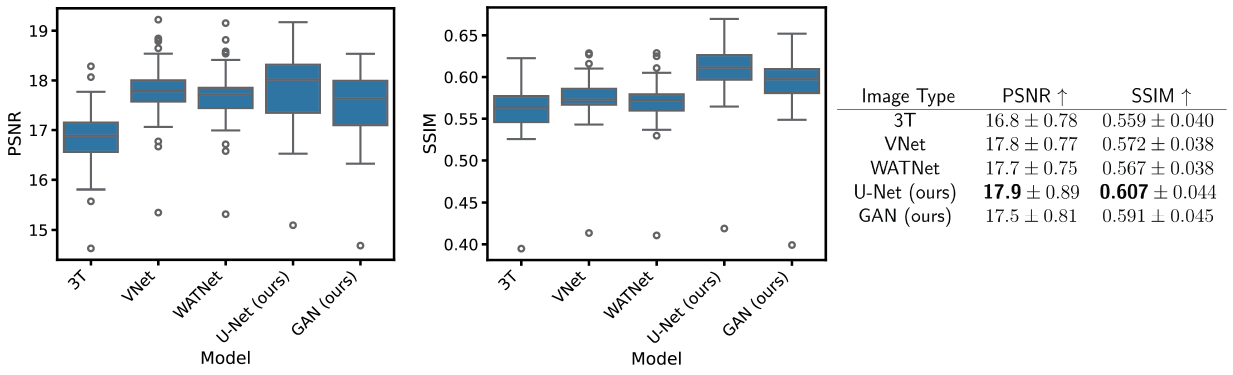}
    \caption{Model performance comparing synthetic 7T to real 7T images. Box plots represent the PSNR (left) and SSIM (middle) of the 29 test participants. A comparison between the 3T image and the real 7T is included as a baseline. Computations are done on the 3D images (excluding background and slices containing cerebellum artifacts). The table shows the mean and standard deviation of PSNR and SSIM scores, with bold indicating the best performing model.}
    \label{fig: math test}
\end{figure}

The quantitative results (PSNR, SSIM) for all the models -- i.e.\ WATNet, V-Net, U-Net, GAN U-Net -- are displayed in Figure~\ref{fig: math test}. We also display the results by comparing the normalized 3T directly to the normalized 7T (PSNR 16.8, SSIM 0.559), to indicate whether the models improve the images beyond an objective baseline. The best model in terms of both SSIM and PSNR was our U-Net (PSNR 17.9, SSIM 0.607). Our GAN showed the second best SSIM, but did not show a PSNR advantage over the other models (PSNR 17.5, SSIM 0.591). The V-Net and WATNet models improve the PSNR, but only slightly improve the SSIM over baseline.

\subsection{Synthetic 7T images judged as visually comparable or improved compared to true 7T images}
\label{visu eval}

We display a qualitative comparison of 3T and 7T MRI and their synthetic versions generated by four models, namely the V-Net, the WATNet, our U-Net and our GAN U-Net in Figure~\ref{fig: axial, hippo ex}. In Figure~\ref{fig: axial, hippo ex}A, we show an example of a 2D slice along the axial dimension from the validation dataset. As a reminder, the model was trained along the axial dimension, and we therefore expected the results to be best along this dimension. Images generated using the V-Net and the WATNet are qualitatively similar to the 3T image from which they were generated. In contrast, our U-Net and GAN U-Net appeared to produce visually sharper results, especially the GAN U-Net. Notably, motion-based and cerebellar artifacts in the 7T image were not carried over to the synthetic images. In Figure~\ref{fig: axial, hippo ex}B, we show another example along the coronal dimension together with a hippocampus segmentation performed using SynthSeg (on the 3D image). Results along this dimension appear blurrier for the GAN U-Net in particular, perhaps since the models were trained on 2D slices along the axial dimension and not along the coronal dimension (except for the V-Net, which was trained on 3D blocks of size $64\times 64\times 64$), though this was also the case for the U-Net, which showed less slicing bias. Figure~\ref{fig:supp_qualcomp} shows further side-by-side examples of real and synthetic images. 

\begin{figure}[!ht]
    \centering
    \includegraphics[width=\textwidth]{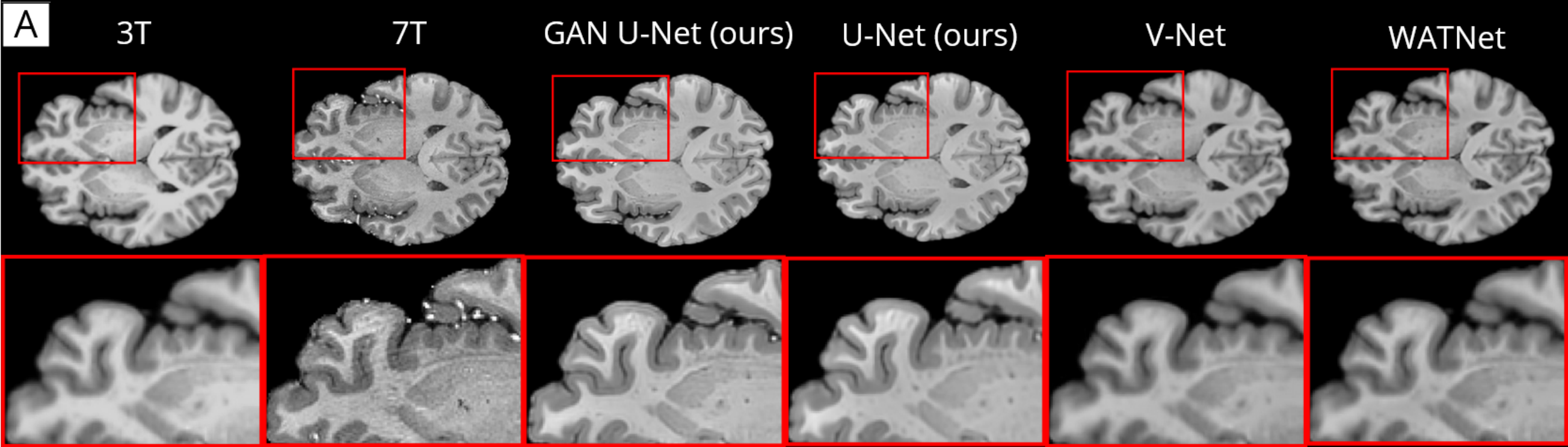}
    \includegraphics[width=\textwidth]{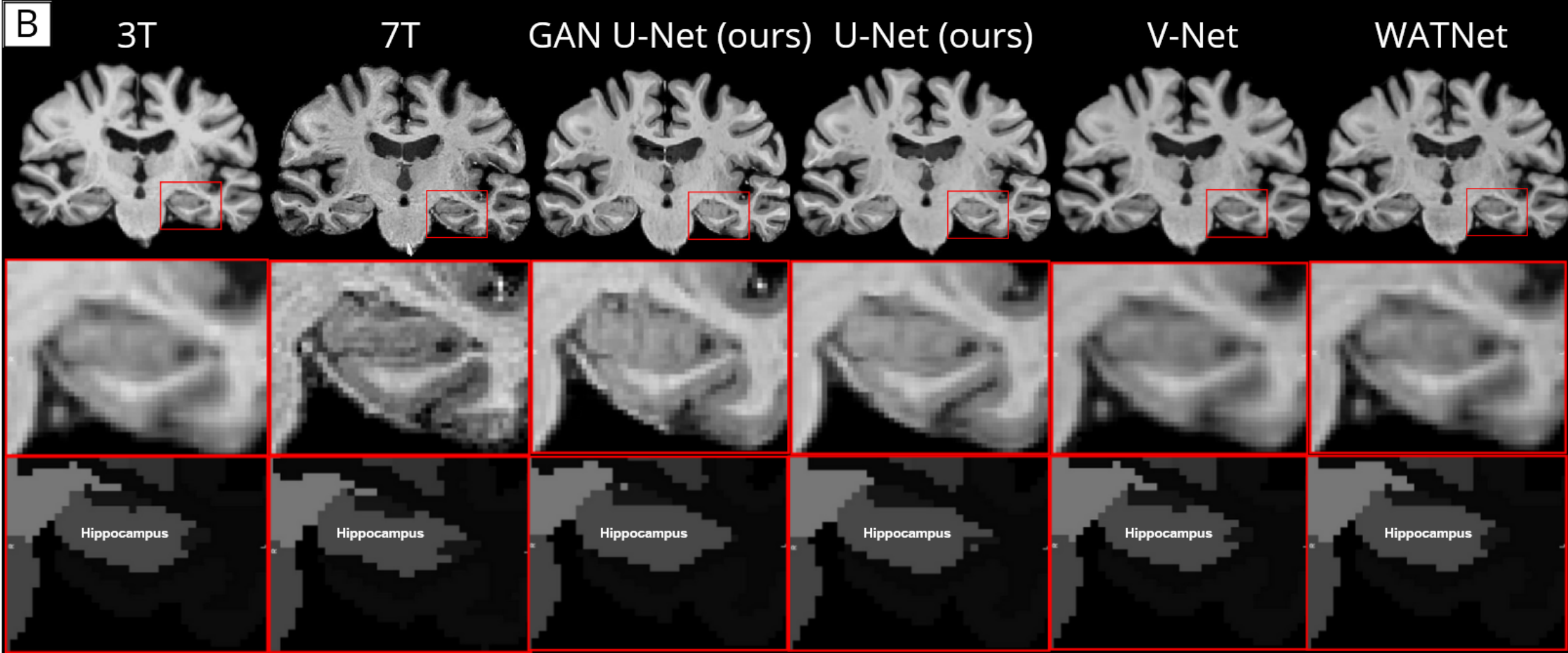}
    \begin{minipage}{0.45\textwidth}
        \begin{tikzpicture}
            \node[anchor=north west, inner sep=0] (img)
                {\includegraphics[width=0.9\textwidth]{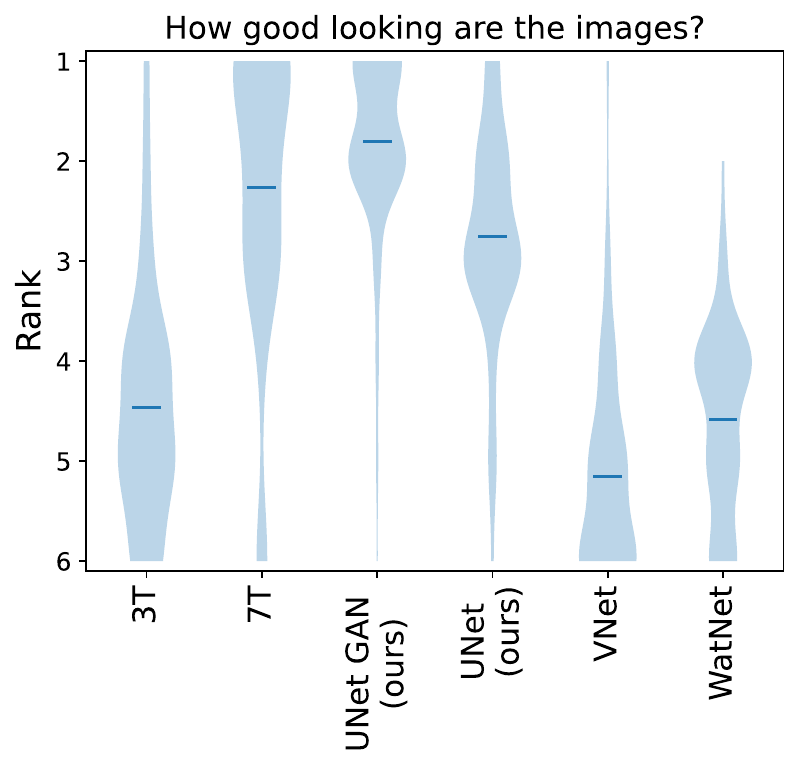}};
            \node[anchor=north east, font=\large]
            at ([xshift=0em,yshift=0em]img.north west) {C};
        \end{tikzpicture}
    \end{minipage}
    \begin{minipage}[!ht]{0.45\textwidth}
        \includegraphics[width=0.9\textwidth]{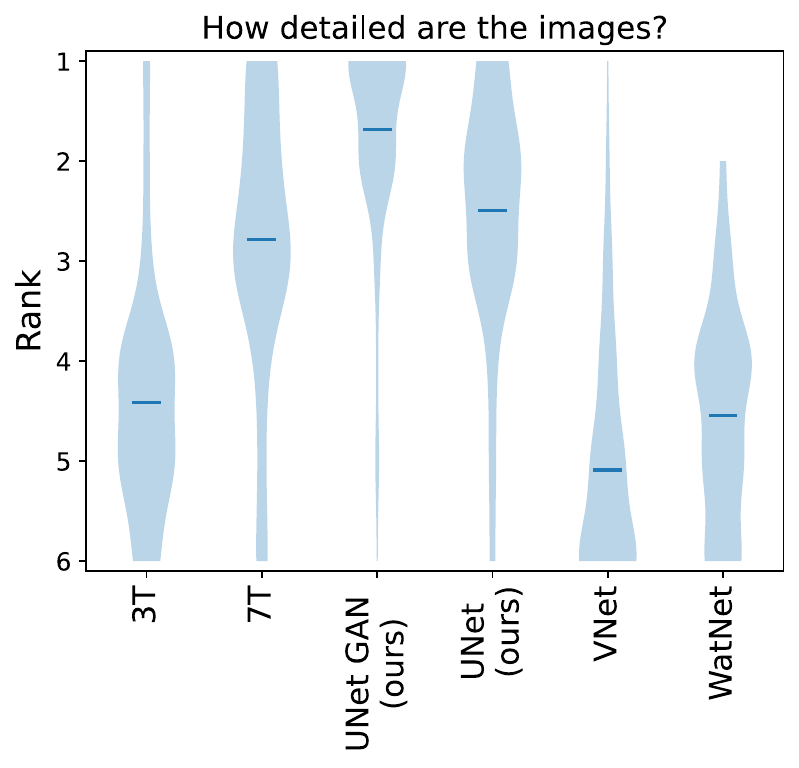}
    \end{minipage}
    \caption{{Qualitative evaluation of synthetic 7T images. A: a 2D axial slice along with a close-up of the 3T, real 7T and synthetic 7T generated by the four models. B: a 2D coronal slice along with a close-up and a hippocampus segmentation performed using SynthSeg \citep{synthseg} of the 3T, real 7T and synthetic 7T versions from the four models. C:~Violin graphs of all the ranks given to an image type (across 28 images) and a bar indicating the mean, according to the indicated criteria.}}
    \label{fig: axial, hippo ex}
\end{figure}

Given the qualitative nature of the above evaluations, we asked a group of neuroradiologists and MRI scientists to rank the 3T, 7T and synthetic images based on quality across two criteria -- how ``good looking'' the images were, and how "detailed" they were (see Section \ref{sec : visual survey}). We display a violin graph of the ranks given to each image type in Figure~\ref{fig: axial, hippo ex}C (a more detailed view of the results is available in the Appendix (Figures~\ref{fig: survey 1} - \ref{fig: survey 4}). The survey demonstrated that the GAN U-Net synthesized images were rated the best looking (average rank $1.7\pm1.0$), followed by U-Net synthesized (average rank $2.5\pm1.2$) and the real 7T (average rank $2.8\pm1.4$), then the 3T (average rank $4.4\pm1.2$), WATNet (average rank $4.5\pm1.1$) and V-Net (average rank $5.1\pm1.1$). A repeated-measures ANOVA showed a difference in mean rank across image types (F=31.02, p$<$0.0001). Posthoc tests (with Bonferroni correction) showed the GAN U-Net and U-Net to be ranked significantly higher than the 3T, V-Net and WATNet, while the 3T was ranked higher than the V-Net (all p[adj.]$<$0.05). Regarding assessment of how detailed the image types are, the survey demonstrated that the GAN U-Net (average rank $1.8\pm0.92$) was rated better than the U-Net (average rank $2.8\pm1.1$), while it was rated similarly to the real 7T images (average rank $2.3\pm1.5$). Posthoc comparisons once again showed the same significant relationships as the first condition — the GAN U-Net and U-Net were ranked higher than 3T, V-Net and WATNet, while the 3T was ranked higher than the V-Net. These results did not necessarily by themselves provide a clear indication of which model is best. However, we can conclude that the real 7T images were judged as being quite detailed, but not necessarily ``good looking'', perhaps due to the number of artifacts (motion, susceptibility, etc.). The 3T (average rank: $4.5\pm1.2$), V-Net (average rank $5.2\pm1.1$) and WATNet (average rank $4.6\pm1.0$) also performed the worst for this criterion, and the V-Net and WATNet were not rated as any better than the 3T.

\subsection{Segmentations of synthetic 7T better resemble 7T segmentations compared to 3T}
\label{sec:results_quantitative_metrics}
\begin{figure}[!ht]
    \centering
    \includegraphics[width=1\textwidth]{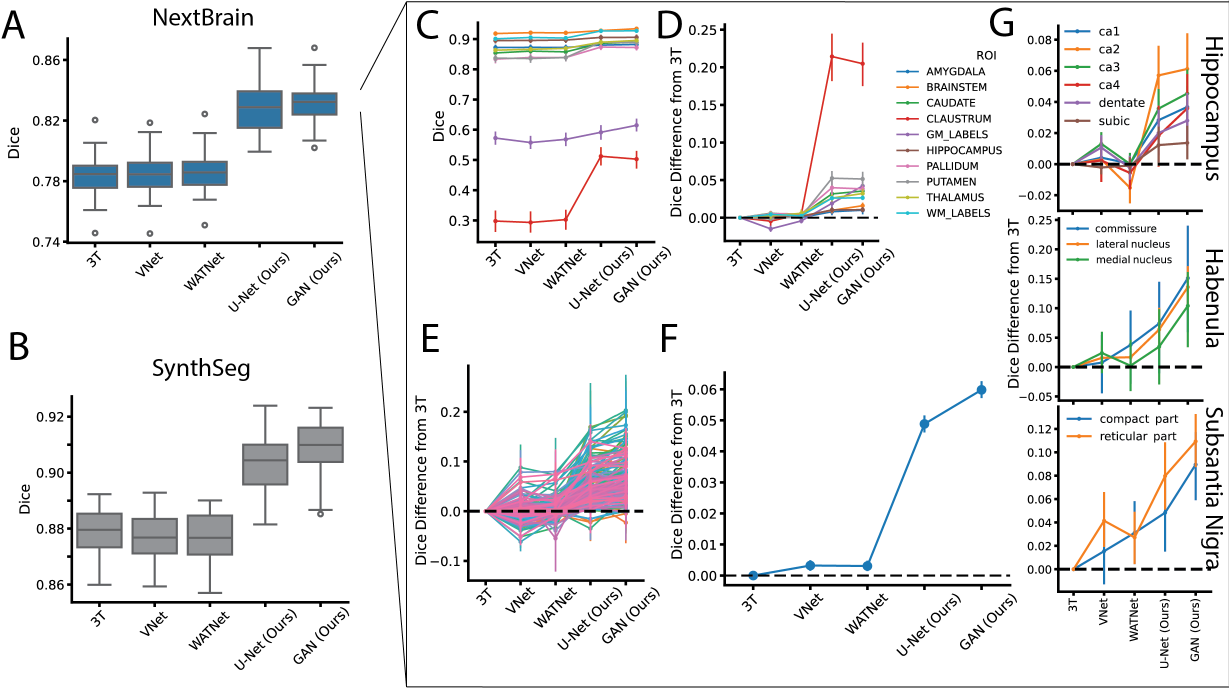}
    \caption{Comparison of automated segmentation on synthetic images to automated segmentation on 7T images. A: Patient average NextBrain Dice scores across models. Dice scores compare each subject's segmentation to the 7T segmentation in the same subject. B: The same thing, but using SynthSeg for segmentation instead of NextBrain. C: Dice scores across models for individual aggregated brain regions from the NextBrain segmentation. D: Dice score difference from 3T for each model, across the same aggregated brain regions. E: Dice score difference from 3T for each model, across all NextBrain brain regions. F: The same information but collapsed over all brain regions. G. Dice score difference from 3T across models, focusing on NextBrain subregions in the hippocampus, habenula and substantia nigra. }
    \label{fig: Dice}
\end{figure}

Besides visual reads, another common use of brain MRI is segmentations, with 7T theoretically offering advantages in segmenting smaller regions and subregions. We performed automated segmentations on the 3T, 7T and all synthetic 7T images in the test set using NextBrain, a software for precise sub-region segmentation on high-resolution brain images \citep{casamitjana2025probabilistic}. The objective was to test whether segmentations on the synthetic 7T were more similar to true 7T segmentations than segmentations on the original 3T images. The average patient Dice score (defined in Section \ref{sec:automatic_segmentation}) was substantially higher than the 3T for our U-Net and GAN U-Net models, whereas the V-Net and WATNet did not show appreciable Dice score differences from the 3T (Figure~\ref{fig: Dice}A, Table~\ref{tab:supp_synthseg_nextbrain_dice_main}). To ensure that this effect was not software dependent, we repeated the analysis using SynthSeg \citep{synthseg}, a very fast automated segmentation program that performs well on a variety of sequences and conditions. The results were very similar to the NextBrain segmentation, with our U-Net and GAN U-Net models performing well and the other models performing similarly to the 3T (Figure~\ref{fig: Dice}B). Next, we further investigated performance of the NextBrain segmentation. When collapsing the many NextBrain regions into a set of larger ROIs, we found that Dice scores were generally high (Figure~\ref{fig: Dice}C), but improvement over 3T was mainly seen in our U-Net and GAN U-Net models (Figure~\ref{fig: Dice}D). A two-way within-subjects ANOVA showed a significant main effect of model and ROI, as well as a model x ROI interaction (all p$<$0.0001). Post-hoc tests showed that our GAN U-Net showed better Dice scores than our U-Net in the gray matter, thalamus, brainstem, and caudate/accumbens regions (p[FDR]$<$0.05). Similarly, when evaluating every NextBrain region, our GAN U-Net and U-Net models showed segmentations more similar to 7T on average across nearly all regions (Figure~\ref{fig: Dice}E), and in aggregate (Figure~\ref{fig: Dice}F). This effect was consistent within regions that are difficult to segment with 3T, including hippocampal subfields, habenula subregions and the substantia nigra (Figure~\ref{fig: Dice}G).

\subsection{Ablation experiments}
\label{sec:results_ablation}
After seeing promising performance of our U-Net and GAN U-Net models, we next wished to understand whether specific choices we made in model construction influenced performance. Here, we focused specifically on the GAN model, and we evaluated quantitative metrics (PSNR, SSIM, Dice score). Our changes to the normalization layers did not appear to have a strong influence on the results (Figure~\ref{fig:supp_ablation}, Table~\ref{tab:supp_psnr_ssim_ablation}). Conditioning on age and sex did seem to improve model performance, though it could not fully explain the improvement of our model over other 3T to 7T models.

\subsection{Derivatives generated from synthetic images achieve performance similar to 3T in downstream prediction tasks}

\begin{figure}[!ht]
    \centering
    \includegraphics[width=1\textwidth]{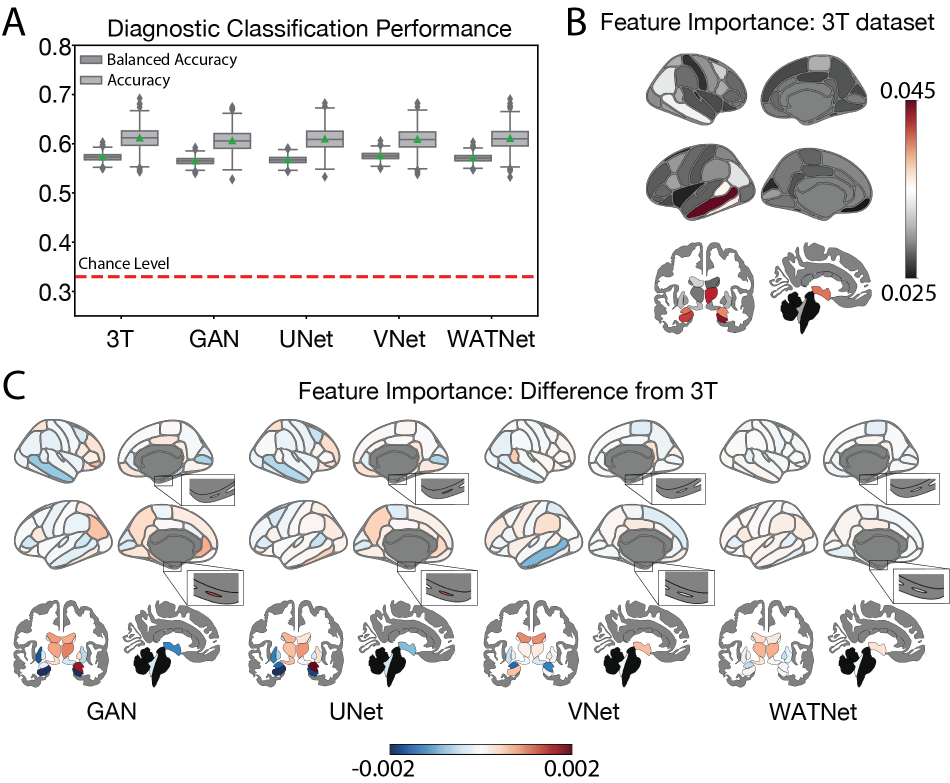}
    \caption{Performance and feature importances from diagnostic prediction task. The U-Net and GAN U-Net models were used to synthesize 7T images from 3,168 3T images, each of which had a clinical diagnosis of cognitively normal (CN), mild cognitive impairment (MCI) or Alzheimer’s disease dementia (AD). Each dataset was automatically parcellated with SynthSeg and the cortical thickness derivatives and subcortical volumes were entered with age and gender into a multiclass random forest classifier predicting clinical diagnosis. A: Boxplots show accuracy and balanced accuracy scores (“Score” on y-axis) across 1000 train-test splits. B: Brains showing the regional mean feature importance across the 1000 models trained on the 3T dataset. Deeper red regions were more important for making predictions, whereas darker gray regions were less important. C: Plots show deviation in feature importance between models trained on synthetic data vs.\ models trained on the 3T data.}
    \label{fig: ML}
\end{figure}

We conducted a straightforward and standard downstream analysis procedure to assess the utility of the segmentation results from the 3T and synthetic 7T images.
A random forest classification model was developed to predict the participants' clinical diagnosis (CN, MCI or AD) based on automated regional segmentation results from the T1-weighted image, along with demographic features such as age and gender.
The accuracy of the predictive models across 1000 bootstrap samples ranged from 0.55 to 0.69 across all the image types, while balanced accuracy ranged from 0.55 to 0.60 (Figure~\ref{fig: ML}A). Performance did not differ significantly across image types.

The random forest model's feature importance analysis provided insights into the key brain regions contributing to dementia prediction. Models tended to use temporal, medial temporal and ventricular regions for prediction (Figure~\ref{fig: ML}B), consistent with known areas of brain atrophy in AD \citep{Pini2016,Schwarz2016,Ossenkoppele2019}, and similar to other MRI-based AD prediction models \citep{Cuingnet2011,Davatzikos2009,Vogel2018,Tam2019}. Interestingly, the feature importance was similar but not identical across the image types (Figure~\ref{fig: ML}C). Classification models trained on derivatives from our U-Net and GAN U-Net synthesizers placed more importance on entorhinal cortex, precuneus, thalamus, lateral ventricles and right amygdala, and placed less importance on the hippocampus, right lateral temporal cortex and left putamen.

\section{Discussion}

7T MRI holds promise for both clinical enhancement and research into brain structure and function in both health and disease 
\citep{Opheim2021,Duzel2019}. However, these efforts are limited by the low number of 7T scanners worldwide, which reduces the speed of research and development using this technology. We present an advanced U-Net model that can synthesize high-resolution 7T T1w MR images from 3T T1w acquisitions. Compared to previous studies describing 7T data synthesis \citep{cui20247tmrisynthesization3t,wang2025triadvisionfoundationmodel,Bahrami2017,Zhang2018Cascade,Qu2020}, we train our model on a much larger set of paired 3T-7T data, and our models achieve performance on our dataset that exceeds performance of previously described models. Anecdotally, we noted that expanding the overall size of our architecture was the most influential modeling decision, which was undoubtedly supported by a larger sample size. Synthetic 7T images from our study improve the contrast and sharpness of the 3T images, enhancing their subjective appearance to MRI scientists, and preserving anatomical details while avoiding motion and magnetic field imhomogeneity artifacts common to real 7T scans. Most importantly, we show evidence that the synthetic increase in resolution results in more 7T-like brain region segmentation, showing that the value of increasing resolution might go beyond improved visual appearance.

Our work adds to the growing literature describing AI-based super resolution models synthesizing 3T brain imaging data to 7T quality \citep{cui20247tmrisynthesization3t,wang2025triadvisionfoundationmodel,Bahrami2017,Zhang2018Cascade,Qu2020}. We uniquely employ a U-Net and GAN U-Net, and each showed strengths and limitations. The GAN U-Net clearly outperformed all other models when evaluated for visual appearance, and its segmentations bore the closest resemblance to 7T segmentations. However, interestingly, the U-Net showed better performance on traditional evaluation metrics, PSNR and SSIM. One interpretation of this is that these traditional metrics may only represent one dimension of performance — one that might not generalize to qualities most relevant to the task at hand. This emphasizes the need for integrating data visualization, human assessment, and real-life downstream tasks and utilities into the evaluation process. However, metrics like PSNR and SSIM do still provide important information relevant to the evaluation of synthetic images. The higher performance of the U-Net on these metrics may indicate slightly better preservation of anatomical information in images synthesized with this approach. For example, our models were trained on axial slices, and our investigation of coronal slices of the medial temporal lobe revealed what appeared to be mild slicing artifacts in the hippocampus of GAN U-Net images that were less prevalent in the U-Net images. While this potential improvement in the U-Net generated images did not lead to enhanced downstream tasks, the findings do help to inform our next steps, which might be to investigate methods for training on 3D images, or on multi-view 2D training \citep{Zuo2021}. This work altogether points to evaluating super resolution models along multiple dimensions.

One of the most significant findings of this study was that automated segmentations taken from synthetic 7T images more closely matched 7T segmentations than automated segmentations from the source 3T images. Most classic MRI-based automated brain segmentation tools need a great deal of user input and quality control when applied to 7T data in order to achieve intended performance \citep{Svanera2021,Chu2024}, which is why manual segmentation remains popular for this modality \citep{Berron2017}. However, with cleaner segmentations, there is precedent for 7T data outperforming 3T data in recognizing age-related changes to the brain \citep{Chu2024}. While our results point to some optimism in synthetic 7T images helping to improve segmentation of 3T data without any manual adjustment, we cannot conclude from the present work whether the segmentations themselves were truly more precise. This would require ground truth manual segmentations performed using the NextBrain or SynthSeg protocol to be available. It is noteworthy that our synthetic 7T afforded no advantages over the original 3T in using brain structure derivatives from automated segmentations to predict cognitive diagnosis. However, this does not necessarily provide any information on segmentation quality as incorrect segmentations can actually lead to better prediction. For example, old versions of Freesurfer produce age bias on medial temporal lobe segmentation that over-inflate aging effects (i.e.\ over-segmentation of younger brains and/or under-segmentation of older brains) \citep{Srinivasan2020,Wenger2014}, and AI approaches are also well-known to capitalize on bias to make predictions \citep{Mihan2024}. Other brain imaging work has shown that age and different pipelines introduce bias on brain morphology that nonetheless does not influence performance of downstream prediction \citep{Debiasi2023}. Therefore, we cannot conclude that similar downstream prediction performance is reflective of segmentation quality. However, we are encouraged that synthetic 7T images produced by our model do not suffer any decrement in prediction performance that might be suggestive of hallucination. Further investigation is needed to confirm the finding of synthetic 7T improving brain region segmentation over 3T images used to generate them, as this would carry great potential toward the utility of our model to the general research community.

It is important to consider how a 3T-to-7T super resolution model can be used to aid research or clinical management. These models require the existence of a 3T image in the first place, which differentiates it from the growing literature on brain image synthesis from other medical data \citep{Tudosiu2024,Wang2024,Khader2023}. It also provides a different use-case to the literature describing synthesis of 3T images from low-field MRI \citep{Islam2023,Iglesias2023,Lucas2023}. While these tools have potential to enhance the reach of MR imaging to underserved communities, the requirement of a 3T image in our model restricts its use to clinical centers with greater access to resources. However, compared to these other use-cases, a 3T-to-7T model is far less prone to hallucination, making it easier to rely on for actual clinical use. For instance, 7T MRI is used to confirm lesion locations in circumstances when the location is not visible or ambiguous from lower-field MRI \citep{Sharma2021,Zampeli2022,Hangel2023,Klodowski2025}, which can help inform surgical decisions \citep{zampeli2025does}. Prospective clinical studies will be needed to verify whether synthetic 7T can serve this same purpose. However, we also found that humans seem to, in many cases, prefer the appearance of the synthetic 7T over both the original 3T and original 7T. In scenarios requiring a neurologist or radiologist to simply read a patient’s structural image, the sharpness and enhanced contrast of the synthetic 7T may be more favorable. This, too, would require further study to validate. Finally, an unexpected outcome of our synthetic 7T images was that they were devoid of the many magnetic and motion-related artifacts that are so common for 7T MRI. Where acquisition of both 3T and 7T data from one patient is possible, our model might be useful in ``cleaning'' the 7T image.

Many of the aforementioned clinical use cases would benefit from an even better performing model, and will likely only be possible with a model that is generalizable beyond our dataset, which was acquired on only one pair of 3T-7T scanners. While our model can be adapted to other sites with paired 3T and 7T data, we would like to build a model that can synthesize 7T images from 3T images from any source. We are continuing to train our model on additional datasets to help make that possible. Many clinical and research tasks would also benefit from synthesis of both T1-weighted and T2-weighted data \citep{Li2024}, and we are currently working toward this goal. Other future directions include enhancing our models through augmentation, and through transfer learning from MRI-based foundation models \citep{Sun2024,Tak2024,Cox2024,wang2025triadvisionfoundationmodel,Su2025}. Our study also comes with a number of limitations that can be improved upon in future efforts. As mentioned above, our model may or may not have been limited by its 2D-slice approach, and would also benefit from models that mitigate or eliminate this slicing bias. Our segmentation task did not use manual segmentation as a ground truth, preventing us from drawing conclusions as to whether segmentations from synthetic 7Ts are actually more precise than the 3T segmentations. The true 7T segmentations may not always be more precise, and other factors like co-registration imprecision could influence Dice score similarities. We also note that the high time-of-flight contrast in the 7T makes blood vessels visible on the 7T images where they are not visible on the 3T images. We did not investigate whether our synthetic 7T images could learn to accurately visualize blood vessels, though this may be an interesting pursuit for future studies. There was also a possibility of bias in the survey answers, since the 7T images were easy to identify by the MRI professionals. Additionally, despite being good, there is still room for improvements of 7T acquisitions, and future models could be made even better by training on 7T data with even greater contrast, sharpness or resolution, and with less artifacts. Another major limitation affecting this work and its future application is the difficulty in systematically screening for hallucinations. Hallucinations are often subtle enough to evade detection using traditional performance metrics \citep{buugday2026triggering,kim2026hallugen}, and yet can have legitimate clinical consequences. For example, a 2D model like ours could depict the most inferior point of a gyrus as a white matter lesion, as both would appear as dark spots surrounded by white matter on the 3T image. Anecdotally, we did not observe hallucinations in our data, but we did not do a systematic screening. We are encouraged by recent progress in developing frameworks for detection of hallucinations in medical images \citep{kim2026hallugen,kim2025tackling,tivnan2024hallucination}. Applications of these frameworks will be absolutely necessary for translation of this and similar work to clinical practice. Finally, the road to regulatory approval of a new clinical tool is long and especially arduous for AI-based approaches. While we strive to build a tool with clinical utility, there is still much work to be done to validate the usefulness of this approach and approaches like it.
	
In conclusion, we present a 3T-to-7T synthesis model for T1-weighted brain MRI that shows improved quantitative similarity to real 7T over state-of-the-art models, improves subjective quality of images, and shows some evidence for enhancing segmentation of medial temporal lobe structures. Future work will continue to develop this model toward generalizability and will seek to test its value in real-life clinical use cases.

\section*{Author Contributions}

\textbf{Conceptualization:}
Jacob Vogel,
Malo Gicquel,
Gabrielle Flood

\textbf{Data analysis:}
Malo Gicquel,
Ruoyi Zhou,
Jacob Vogel,
Gabrielle Flood
 
\textbf{Design and Interpretation:}
Malo Gicquel,
Jacob Vogel,
Gabrielle Flood,
Kalle Åström,
Xiao Yu,
Nicola Spotorno,
David Berron,
Laura Wisse,
Danielle van Westen

\textbf{Data generation:}
Anika Wuestefeld,
David Berron,
Olof Strandberg,
Nicola Spotorno,
Danielle van Westen, Niklas Mattsson-Carlgren, Rik Ossenkoppele,
Oskar Hansson

\textbf{Drafting the manuscript:}
Malo Gicquel,
Jacob Vogel,
Gabrielle Flood

\section*{Declaration of Competing Interests}

OH is an employee of Lund University and Eli Lilly. R.O. has received research funding/support from Avid Radiopharmaceuticals, Janssen Research \& Development, Roche, Quanterix and Optina Diagnostics, has given lectures in symposia sponsored by GE Healthcare, received speaker fees from Springer, and is an advisory board/steering committee member for Asceneuron, Biogen and Bristol Myers Squibb. All the aforementioned has been paid to his institutions. NMC has received consultancy/speaker fees from Biogen, Eli Lilly, Owkin and Merck. DB is co-founder and shareholder of neotiv GmbH. All other authors have nothing to declare.

\section*{Data Availability Statement} 
The BioFINDER data are not publicly available, but access requests of anonymized data can be made to the study’s steering group bf$\_$executive@med.lu.se. Access to the data will be granted in compliance with European Union legislation on the General Data Protection Regulation (GDPR) and decisions by the Ethical Review Board of Sweden and Region Skåne. Data transfer will be regulated under a material transfer agreement.

\section*{Code Availability}

The code used to conduct the analyses is available at:\\ https://github.com/DeMONLab-BioFINDER/SuperResolutionMalo .

\section*{Acknowledgments}
This work was supported by the SciLifeLab \& Wallenberg Data Driven Life Science Program (grant: KAW 2020.0239), the Crafoord Foundation (20230790) and the Swedish Alzheimer Foundation (AF-994626). The National 7T facility at Lund University Bioimaging Center is gratefully acknowledged for providing experimental resources. We would also like to acknowledge Emil Ljungberg for manuscript review and insights relating to 7T cerebellar artifacts. The computations were enabled in project Berzelius-2024-156 by the Berzelius resource provided by the Knut and Alice Wallenberg Foundation at the National Supercomputer Centre. The data handling was enabled by resources in project sens2023026 provided by the National Academic Infrastructure for Supercomputing in Sweden (NAISS) at UPPMAX, funded by the Swedish Research Council through grant agreement no.~2022-06725. LW was supported by MultiPark, a Strategic Research Area at Lund University, the Swedish Research Council (2022-00900) and the Crafoord Foundation (20210690). D.B.\ was supported by funding from the European Union’s Horizon 2020 research and innovation programme under the Marie Sk\l{}odowska-Curie grant agreement No 843074 and the donors of Alzheimer's Disease Research, a program of the BrightFocus Foundation. The BioFINDER-2 study was funded by the National Institute of Aging (R01AG083740), European Research Council (ADG-101096455), Alzheimer’s Association (ZEN24-1069572, SG-23-1061717), GHR Foundation, Swedish Research Council (2021-02219, 2022-00775), ERA PerMed (ERAPERMED2021-184), Knut and Alice Wallenberg foundation (2022-0231), Strategic Research Area MultiPark (Multidisciplinary Research in Parkinson’s disease) at Lund University, Swedish Alzheimer Foundation (AF-980907,  AF-994229, AF-1011799), Swedish Brain Foundation (FO2021-0293, FO2023-0163), WASP and DDLS Joint call for research projects (WASP/DDLS22-066), Parkinson foundation of Sweden (1412/22), Cure Alzheimer’s fund, Rönström Family Foundation, Konung Gustaf V:s och Drottning Victorias Frimurarestiftelse, Michael J Fox Foundation (MJFF-025507), Lilly Research Award Program, Skåne University Hospital Foundation (2020-O000028), Regionalt Forskningsstöd (2022-1259) and Swedish federal government under the ALF agreement (2022-Projekt0080, 2022-Projekt0107). The precursor of 18F-flutemetamol was sponsored by GE Healthcare. The precursor of 18F-RO948 was provided by Roche.

\printbibliography

@misc{wang2023inversesr,
      title={InverseSR: 3D Brain MRI Super-Resolution Using a Latent Diffusion Model}, 
      author={Jueqi Wang and Jacob Levman and Walter Hugo Lopez Pinaya and Petru-Daniel Tudosiu and M. Jorge Cardoso and Razvan Marinescu},
      year={2023},
      eprint={2308.12465},
      archivePrefix={arXiv},
      primaryClass={eess.IV}
}

@misc{pinaya2023generativeaimedicalimaging,
      title={Generative AI for Medical Imaging: extending the MONAI Framework}, 
      author={Walter H. L. Pinaya and Mark S. Graham and Eric Kerfoot and Petru-Daniel Tudosiu and Jessica Dafflon and Virginia Fernandez and Pedro Sanchez and Julia Wolleb and Pedro F. da Costa and Ashay Patel and Hyungjin Chung and Can Zhao and Wei Peng and Zelong Liu and Xueyan Mei and Oeslle Lucena and Jong Chul Ye and Sotirios A. Tsaftaris and Prerna Dogra and Andrew Feng and Marc Modat and Parashkev Nachev and Sebastien Ourselin and M. Jorge Cardoso},
      year={2023},
      eprint={2307.15208},
      archivePrefix={arXiv},
      primaryClass={eess.IV},
      url={https://arxiv.org/abs/2307.15208}, 
}

@ARTICLE{ScannerAug,

AUTHOR={Meyer, Maria Ines  and de la Rosa, Ezequiel  and Pedrosa de Barros, Nuno  and Paolella, Roberto  and Van Leemput, Koen  and Sima, Diana M. },

TITLE={A Contrast Augmentation Approach to Improve Multi-Scanner Generalization in MRI},

JOURNAL={Frontiers in Neuroscience},

VOLUME={15},

YEAR={2021},

URL={https://www.frontiersin.org/journals/neuroscience/articles/10.3389/fnins.2021.708196},

DOI={10.3389/fnins.2021.708196},

ISSN={1662-453X}}

@misc{SR3T7T,
      title={High-resolution 3T to 7T MRI Synthesis with a Hybrid CNN-Transformer Model}, 
      author={Zach Eidex and Jing Wang and Mojtaba Safari and Eric Elder and Jacob Wynne and Tonghe Wang and Hui-Kuo Shu and Hui Mao and Xiaofeng Yang},
      year={2023},
      eprint={2311.15044},
      archivePrefix={arXiv},
      primaryClass={physics.med-ph}
}

@misc{liu2021adadm,
      title={AdaDM: Enabling Normalization for Image Super-Resolution}, 
      author={Jie Liu and Jie Tang and Gangshan Wu},
      year={2021},
      eprint={2111.13905},
      archivePrefix={arXiv},
      primaryClass={eess.IV}
}

@misc{vaswani2017AttentionIsAllYouNeed,
      title={Attention Is All You Need}, 
      author={Ashish Vaswani and Noam Shazeer and Niki Parmar and Jakob Uszkoreit and Llion Jones and Aidan N. Gomez and Lukasz Kaiser and Illia Polosukhin},
      year={2017},
      eprint={1706.03762},
      archivePrefix={arXiv},
      primaryClass={cs.CL}
}

@misc{zhang2018Perc,
      title={The Unreasonable Effectiveness of Deep Features as a Perceptual Metric}, 
      author={Richard Zhang and Phillip Isola and Alexei A. Efros and Eli Shechtman and Oliver Wang},
      year={2018},
      eprint={1801.03924},
      archivePrefix={arXiv},
      primaryClass={cs.CV}
}

@ARTICLE{BIASCORRECTION,
  author={Tustison, Nicholas J. and Avants, Brian B. and Cook, Philip A. and Zheng, Yuanjie and Egan, Alexander and Yushkevich, Paul A. and Gee, James C.},
  journal={IEEE Transactions on Medical Imaging}, 
  title={N4ITK: Improved N3 Bias Correction}, 
  year={2010},
  volume={29},
  number={6},
  pages={1310-1320},
  doi={10.1109/TMI.2010.2046908}}

@article{SynthStriphoopes2022,
    title = {SynthStrip: skull-stripping for any brain image},
    journal = {NeuroImage},
    volume = {260},
    pages = {119474},
    year = {2022},
    issn = {1053-8119},
    doi = {https://doi.org/10.1016/j.neuroimage.2022.119474},
    url = {https://www.sciencedirect.com/science/article/pii/S1053811922005900},
    author = {Andrew Hoopes and Jocelyn S. Mora and Adrian V. Dalca and Bruce Fischl and Malte Hoffmann},
}

@misc{pinaya2022brainGen,
      title={Brain Imaging Generation with Latent Diffusion Models}, 
      author={Walter H. L. Pinaya and Petru-Daniel Tudosiu and Jessica Dafflon and Pedro F da Costa and Virginia Fernandez and Parashkev Nachev and Sebastien Ourselin and M. Jorge Cardoso},
      year={2022},
      eprint={2209.07162},
      archivePrefix={arXiv},
      primaryClass={eess.IV}
}

@misc{ho2020denoising,
      title={Denoising Diffusion Probabilistic Models}, 
      author={Jonathan Ho and Ajay Jain and Pieter Abbeel},
      year={2020},
      eprint={2006.11239},
      archivePrefix={arXiv},
      primaryClass={cs.LG}
}

@article{radimagenet,
author = {Mei, Xueyan and Liu, Zelong and Robson, Philip M. and Marinelli, Brett and Huang, Mingqian and Doshi, Amish and Jacobi, Adam and Cao, Chendi and Link, Katherine E. and Yang, Thomas and Wang, Ying and Greenspan, Hayit and Deyer, Timothy and Fayad, Zahi A. and Yang, Yang},
title = {RadImageNet: An Open Radiologic Deep Learning Research Dataset for Effective Transfer Learning},
journal = {Radiology: Artificial Intelligence},
volume = {0},
number = {ja},
pages = {e210315},
year = {2022},
doi = {10.1148/ryai.210315},

URL = { 
        https://doi.org/10.1148/ryai.210315
    
},
eprint = { 
        https://doi.org/10.1148/ryai.210315
}
}

@article{ANTSSYN,
title = {Symmetric diffeomorphic image registration with cross-correlation: Evaluating automated labeling of elderly and neurodegenerative brain},
journal = {Medical Image Analysis},
volume = {12},
number = {1},
pages = {26-41},
year = {2008},
note = {Special Issue on The Third International Workshop on Biomedical Image Registration – WBIR 2006},
issn = {1361-8415},
doi = {https://doi.org/10.1016/j.media.2007.06.004},
url = {https://www.sciencedirect.com/science/article/pii/S1361841507000606},
author = {B.B. Avants and C.L. Epstein and M. Grossman and J.C. Gee}
}

@article{ref3t,
    author = {Berron, David and Vogel, Jacob W and Insel, Philip S and Pereira, Joana B and Xie, Long and Wisse, Laura E M and Yushkevich, Paul A and Palmqvist, Sebastian and Mattsson-Carlgren, Niklas and Stomrud, Erik and Smith, Ruben and Strandberg, Olof and Hansson, Oskar},
    title = "{Early stages of tau pathology and its associations with functional connectivity, atrophy and memory}",
    journal = {Brain},
    volume = {144},
    number = {9},
    pages = {2771-2783},
    year = {2021},
    month = {03},
    issn = {0006-8950},
    doi = {10.1093/brain/awab114},
    url = {https://doi.org/10.1093/brain/awab114},
    eprint = {https://academic.oup.com/brain/article-pdf/144/9/2771/40880323/awab114.pdf},
}

@misc{cardoso2022monai,
      title={MONAI: An open-source framework for deep learning in healthcare}, 
      author={M. Jorge Cardoso and Wenqi Li and Richard Brown and Nic Ma and Eric Kerfoot and Yiheng Wang and Benjamin Murrey and Andriy Myronenko and Can Zhao and Dong Yang and Vishwesh Nath and Yufan He and Ziyue Xu and Ali Hatamizadeh and Andriy Myronenko and Wentao Zhu and Yun Liu and Mingxin Zheng and Yucheng Tang and Isaac Yang and Michael Zephyr and Behrooz Hashemian and Sachidanand Alle and Mohammad Zalbagi Darestani and Charlie Budd and Marc Modat and Tom Vercauteren and Guotai Wang and Yiwen Li and Yipeng Hu and Yunguan Fu and Benjamin Gorman and Hans Johnson and Brad Genereaux and Barbaros S. Erdal and Vikash Gupta and Andres Diaz-Pinto and Andre Dourson and Lena Maier-Hein and Paul F. Jaeger and Michael Baumgartner and Jayashree Kalpathy-Cramer and Mona Flores and Justin Kirby and Lee A. D. Cooper and Holger R. Roth and Daguang Xu and David Bericat and Ralf Floca and S. Kevin Zhou and Haris Shuaib and Keyvan Farahani and Klaus H. Maier-Hein and Stephen Aylward and Prerna Dogra and Sebastien Ourselin and Andrew Feng},
      year={2022},
      eprint={2211.02701},
      archivePrefix={arXiv},
      primaryClass={cs.LG}
}

@misc{kong2022dropout,
      title={Reflash Dropout in Image Super-Resolution}, 
      author={Xiangtao Kong and Xina Liu and Jinjin Gu and Yu Qiao and Chao Dong},
      year={2022},
      eprint={2112.12089},
      archivePrefix={arXiv},
      primaryClass={cs.CV}
}

@misc{gulrajani2017WGANGP,
      title={Improved Training of Wasserstein GANs}, 
      author={Ishaan Gulrajani and Faruk Ahmed and Martin Arjovsky and Vincent Dumoulin and Aaron Courville},
      year={2017},
      eprint={1704.00028},
      archivePrefix={arXiv},
      primaryClass={cs.LG}
}

@misc{basu2024Collapse,
      title={Fortifying Fully Convolutional Generative Adversarial Networks for Image Super-Resolution Using Divergence Measures}, 
      author={Arkaprabha Basu and Kushal Bose and Sankha Subhra Mullick and Anish Chakrabarty and Swagatam Das},
      year={2024},
      eprint={2404.06294},
      archivePrefix={arXiv},
      primaryClass={id='eess.IV' full_name='Image and Video Processing' is_active=True alt_name=None in_archive='eess' is_general=False description='Theory, algorithms, and architectures for the formation, capture, processing, communication, analysis, and display of images, video, and multidimensional signals in a wide variety of applications. Topics of interest include: mathematical, statistical, and perceptual image and video modeling and representation; linear and nonlinear filtering, de-blurring, enhancement, restoration, and reconstruction from degraded, low-resolution or tomographic data; lossless and lossy compression and coding; segmentation, alignment, and recognition; image rendering, visualization, and printing; computational imaging, including ultrasound, tomographic and magnetic resonance imaging; and image and video analysis, synthesis, storage, search and retrieval.'}
}

@misc{cui20247tmrisynthesization3t,
      title={7T MRI Synthesization from 3T Acquisitions}, 
      author={Qiming Cui and Duygu Tosun and Pratik Mukherjee and Reza Abbasi-Asl},
      year={2024},
      eprint={2403.08979},
      archivePrefix={arXiv},
      primaryClass={eess.IV},
      url={https://arxiv.org/abs/2403.08979}, 
}

@article{synthseg,
title = {SynthSeg: Segmentation of brain MRI scans of any contrast and resolution without retraining},
journal = {Medical Image Analysis},
volume = {86},
pages = {102789},
year = {2023},
issn = {1361-8415},
doi = {https://doi.org/10.1016/j.media.2023.102789},
url = {https://www.sciencedirect.com/science/article/pii/S1361841523000506},
author = {Benjamin Billot and Douglas N. Greve and Oula Puonti and Axel Thielscher and Koen {Van Leemput} and Bruce Fischl and Adrian V. Dalca and Juan Eugenio Iglesias}
}

@misc{liao2022comparativestudy15t3tmri,
      title={A Comparative Study on 1.5T-3T MRI Conversion through Deep Neural Network Models}, 
      author={Binhua Liao and Yani Chen and Zhewei Wang and Charles D. Smith and Jundong Liu},
      year={2022},
      eprint={2210.06362},
      archivePrefix={arXiv},
      primaryClass={eess.IV},
      url={https://arxiv.org/abs/2210.06362}, 
}

@misc{kingma2017adammethodstochasticoptimization,
      title={Adam: A Method for Stochastic Optimization}, 
      author={Diederik P. Kingma and Jimmy Ba},
      year={2017},
      eprint={1412.6980},
      archivePrefix={arXiv},
      primaryClass={cs.LG},
      url={https://arxiv.org/abs/1412.6980}, 
}

@article{SRclinical,
title = {Medical image super-resolution for smart healthcare applications: A comprehensive survey},
journal = {Information Fusion},
volume = {103},
pages = {102075},
year = {2024},
issn = {1566-2535},
doi = {https://doi.org/10.1016/j.inffus.2023.102075},
url = {https://www.sciencedirect.com/science/article/pii/S1566253523003913},
author = {Sabina Umirzakova and Shabir Ahmad and Latif U. Khan and Taegkeun Whangbo}
}

@article{arvidsson2024MCI,
  title={Comparing a pre-defined versus deep learning approach for extracting brain atrophy patterns to predict cognitive decline due to Alzheimer’s disease in patients with mild cognitive symptoms},
  author={Arvidsson, I. and Strandberg, O. and Palmqvist, S. and others},
  journal={Alzheimer's Research \& Therapy},
  volume={16},
  number={61},
  year={2024},
  doi={10.1186/s13195-024-01428-5},
  url={https://doi.org/10.1186/s13195-024-01428-5}
}

@article{Bahrami2017,
  author    = {Bahrami, K. and Shi, F. and Rekik, I. and Gao, Y. and Shen, D.},
  title     = {7T-guided super-resolution of 3T MRI},
  journal   = {Medical Physics},
  volume    = {44},
  number    = {5},
  pages     = {1661--1677},
  year      = {2017},
  month     = {5},
  doi       = {10.1002/mp.12132},
  pmid      = {28177548},
  pmcid     = {PMC5686784},
  note      = {Epub 2017 Apr 22}
}

@inproceedings{Bahrami20167TBetterSeg,
  author    = {Bahrami, K. and Rekik, I. and Shi, F. and Gao, Y. and Shen, D.},
  title     = {7T-Guided Learning Framework for Improving the Segmentation of 3T MR Images},
  booktitle = {Medical Image Computing and Computer-Assisted Intervention -- MICCAI 2016},
  year      = {2016},
  volume    = {9901},
  pages     = {572--580},
  doi       = {10.1007/978-3-319-46723-8_66},
  pmid      = {28149968},
  pmcid     = {PMC5278835},
  note      = {Epub 2016 Oct 2}
}

@article{MS7T,
    author = {Harrison, Daniel M and Sati, Pascal and Klawiter, Eric C and Narayanan, Sridar and Bagnato, Francesca and Beck, Erin S and Barker, Peter and Calvi, Alberto and Cagol, Alessandro and Donadieu, Maxime and Duyn, Jeff and Granziera, Cristina and Henry, Roland G and Huang, Susie Y and Hoff, Michael N and Mainero, Caterina and Ontaneda, Daniel and Reich, Daniel S and Rudko, David A and Smith, Seth A and Trattnig, Siegfried and Zurawski, Jonathan and Bakshi, Rohit and Gauthier, Susan and Laule, Cornelia, on behalf of the NAIMS Cooperative},
    title = {The use of 7T MRI in multiple sclerosis: review and consensus statement from the North American Imaging in Multiple Sclerosis Cooperative},
    journal = {Brain Communications},
    volume = {6},
    number = {5},
    pages = {fcae359},
    year = {2024},
    month = {10},
    issn = {2632-1297},
    doi = {10.1093/braincomms/fcae359},
    url = {https://doi.org/10.1093/braincomms/fcae359},
    eprint = {https://academic.oup.com/braincomms/article-pdf/6/5/fcae359/59983144/fcae359.pdf},
}

@article{perera20237TALZ,
  title = {Seven Tesla MRI in Alzheimer’s disease research: State of the art and future directions: A narrative review},
  volume = {10},
  ISSN = {2373-7972},
  url = {http://dx.doi.org/10.3934/Neuroscience.2023030},
  DOI = {10.3934/neuroscience.2023030},
  number = {4},
  journal = {AIMS Neuroscience},
  publisher = {American Institute of Mathematical Sciences (AIMS)},
  author = {Perera Molligoda Arachchige,  Arosh S. and Garner,  Anton Kristoffer},
  year = {2023},
  pages = {401–422}
}

@article{Palmqvist2020,
    author = {Palmqvist, Sebastian and Janelidze, Shorena and Quiroz, Yakeel T. and Zetterberg, Henrik and Lopera, Francisco and Stomrud, Erik and Su, Yi and Chen, Yinghua and Serrano, Geidy E. and Leuzy, Antoine and Mattsson-Carlgren, Niklas and Strandberg, Olof and Smith, Ruben and Villegas, Andres and Sepulveda-Falla, Diego and Chai, Xiyun and Proctor, Nicholas K. and Beach, Thomas G. and Blennow, Kaj and Dage, Jeffrey L. and Reiman, Eric M. and Hansson, Oskar},
    title = {Discriminative Accuracy of Plasma Phospho-tau217 for Alzheimer Disease vs Other Neurodegenerative Disorders},
    journal = {JAMA},
    volume = {324},
    number = {8},
    pages = {772-781},
    year = {2020},
    month = {8},
    issn = {0098-7484},
    doi = {10.1001/jama.2020.12134},
    url = {https://doi.org/10.1001/jama.2020.12134},
    eprint = {https://jamanetwork.com/journals/jama/articlepdf/2768841/jama_palmqvist_2020_oi_200077_1597950344.77871.pdf}
}

@article{Hashimoto2010,
  author = {Hashimoto, M and Ishikawa, M and Mori, E and Kuwana, N and Study of INPH on Neurological Improvement (SINPHONI)},
  title = {Diagnosis of idiopathic normal pressure hydrocephalus is supported by MRI-based scheme: a prospective cohort study},
  journal = {Cerebrospinal Fluid Research},
  year = {2010},
  volume = {7},
  pages = {18},
  doi = {10.1186/1743-8454-7-18},
  pmid = {21040519},
  pmcid = {PMC2987762},
  month = {10}
}

@article{Weller2021,
  author = {Weller, M. and van den Bent, M. and Preusser, M. and others},
  title  = {EANO guidelines on the diagnosis and treatment of diffuse gliomas of adulthood},
  journal = {Nature Reviews Clinical Oncology},
  year   = {2021},
  volume = {18},
  pages  = {170--186},
  doi    = {10.1038/s41571-020-00447-z},
  month  = {02}
}

@article{Wang2020,
  title = {MRI essentials in epileptology: a review from the ILAE Imaging Taskforce},
  volume = {22},
  ISSN = {1950-6945},
  url = {http://dx.doi.org/10.1684/epd.2020.1174},
  DOI = {10.1684/epd.2020.1174},
  number = {4},
  journal = {Epileptic Disorders},
  publisher = {Wiley},
  author = {Wang,  Irene and Bernasconi,  Andrea and Bernhardt,  Boris and Blumenfeld,  Hal and Cendes,  Fernando and Chinvarun,  Yotin and Jackson,  Graeme and Morgan,  Victoria and Rampp,  Stefan and Vaudano,  Anna Elisabetta and Federico,  Paolo},
  year = {2020},
  month = aug,
  pages = {421–437}
}

@article{Debette2019,
  author  = {Debette, S and Schilling, S and Duperron, M. G. and Larsson, S. C. and Markus, H. S.},
  title   = {Clinical Significance of Magnetic Resonance Imaging Markers of Vascular Brain Injury: A Systematic Review and Meta-analysis},
  journal = {JAMA Neurology},
  year    = {2019},
  volume  = {76},
  number  = {1},
  pages   = {81--94},
  doi     = {10.1001/jamaneurol.2018.3122},
  month   = {01}
}

@book{Barkhof2011,
author = {Barkhof, Frederik and Fox, Nick and Bastos-Leite, António and Scheltens, Ph},
year = {2011},
month = {01},
pages = {},
title = {Neuroimaging in Dementia},
isbn = {978-3-642-00818-4},
journal = {Neuroimaging in Dementia},
doi = {10.1007/978-3-642-00818-4}
}

@article{Kuoy2022,
  author  = {Kuoy, E and Glavis-Bloom, J and Hovis, G and Yep, B and Biswas, A and Masudathaya, L. A and Norrick, L. A and Limfueco, J and Soun, J. E and Chang, P. D and Chu, E and Akbari, Y and Yaghmai, V and Fox, J. C and Yu, W and Chow, D. S},
  title   = {Point-of-Care Brain MRI: Preliminary Results from a Single-Center Retrospective Study},
  journal = {Radiology},
  year    = {2022},
  volume  = {305},
  number  = {3},
  pages   = {666--671},
  doi     = {10.1148/radiol.211721},
  pmid    = {35916678},
  pmcid   = {PMC9713449},
  month   = {12},
  note    = {Epub 2022 Aug 2}
}

@article {Morrisb3016,
	author = {Morris, Zoe and Whiteley, William N and Longstreth, W T and Weber, Frank and Lee, Yi-Chung and Tsushima, Yoshito and Alphs, Hannah and Ladd, Susanne C and Warlow, Charles and Wardlaw, Joanna M and Al-Shahi Salman, Rustam},
	title = {Incidental findings on brain magnetic resonance imaging: systematic review and meta-analysis},
	volume = {339},
	elocation-id = {b3016},
	year = {2009},
	doi = {10.1136/bmj.b3016},
	publisher = {BMJ Publishing Group Ltd},
	issn = {0959-8138},
	URL = {https://www.bmj.com/content/339/bmj.b3016},
	eprint = {https://www.bmj.com/content/339/bmj.b3016.full.pdf},
	journal = {BMJ}
}

@article{Yang2024,
  author  = {Yang, Z and Wen, J and Erus, G and Govindarajan, S. T. and Melhem, R and Mamourian, E and Cui, Y and Srinivasan, D and Abdulkadir, A and Parmpi, P and Wittfeld, K and Grabe, H. J and B{\"u}low, R and Frenzel, S and Tosun, D and Bilgel, M and An, Y and Yi, D and Marcus, D. S and LaMontagne, P and Benzinger, T. L. S and Heckbert, S. R and Austin, T. R and Waldstein, S. R and Evans, M. K and Zonderman, A. B and Launer, L. J and Sotiras, A and Espeland, M. A and Masters, C. L and Maruff, P and Fripp, J and Toga, A. W and O'Bryant, S and Chakravarty, M. M and Villeneuve, S and Johnson, S. C and Morris, J. C and Albert, M. S and Yaffe, K and V{\"o}lzke, H and Ferrucci, L and Bryan, R. N and Shinohara, R. T and Fan, Y and Habes, M and Lalousis, P. A and Koutsouleris, N and Wolk, D. A and Resnick, S. M and Shou, H and Nasrallah, I. M and Davatzikos, C},
  title   = {Brain aging patterns in a large and diverse cohort of 49,482 individuals},
  journal = {Nature Medicine},
  year    = {2024},
  volume  = {30},
  number  = {10},
  pages   = {3015--3026},
  doi     = {10.1038/s41591-024-03144-x},
  pmid    = {39147830},
  pmcid   = {PMC11483219},
  month   = {10},
  note    = {Epub 2024 Aug 15}
}

@article{Zampeli2022,
  title = {Structural association between heterotopia and cortical lesions visualised with 7 T MRI in patients with focal epilepsy},
  volume = {101},
  ISSN = {1059-1311},
  url = {http://dx.doi.org/10.1016/j.seizure.2022.08.008},
  DOI = {10.1016/j.seizure.2022.08.008},
  journal = {Seizure: European Journal of Epilepsy},
  publisher = {Elsevier BV},
  author = {Zampeli,  Ariadne and Hansson,  Boel and Bloch,  Karin Markenroth and Englund,  Elisabet and K\"{a}llén,  Kristina and Strandberg,  Maria Compagno and Bj\"{o}rkman-Burtscher,  Isabella M.},
  year = {2022},
  month = oct,
  pages = {177–183}
}

@article{Sharma2021,
  title = {Utility of 7 tesla MRI brain in 16 “MRI Negative” epilepsy patients and their surgical outcomes},
  volume = {15},
  ISSN = {2589-9864},
  url = {http://dx.doi.org/10.1016/j.ebr.2020.100424},
  DOI = {10.1016/j.ebr.2020.100424},
  journal = {Epilepsy \& Behavior Reports},
  publisher = {Elsevier BV},
  author = {Sharma,  Himanshu K. and Feldman,  Rebecca and Delman,  Bradley and Rutland,  John and Marcuse,  Lara V. and Fields,  Madeline C. and Ghatan,  Saadi and Panov,  Fedor and Singh,  Anuradha and Balchandani,  Priti},
  year = {2021},
  pages = {100424}
}

@article{Priovoulos2018,
  title = {High-resolution in vivo imaging of human locus coeruleus by magnetization transfer MRI at 3T and 7T},
  volume = {168},
  ISSN = {1053-8119},
  url = {http://dx.doi.org/10.1016/j.neuroimage.2017.07.045},
  DOI = {10.1016/j.neuroimage.2017.07.045},
  journal = {NeuroImage},
  publisher = {Elsevier BV},
  author = {Priovoulos,  Nikos and Jacobs,  Heidi I.L. and Ivanov,  Dimo and Uludağ,  K\^amil and Verhey,  Frans R.J. and Poser,  Benedikt A.},
  year = {2018},
  month = mar,
  pages = {427–436}
}

@article{Kenkhuis2019,
  title = {7T MRI allows detection of disturbed cortical lamination of the medial temporal lobe in patients with Alzheimer’s disease},
  volume = {21},
  ISSN = {2213-1582},
  url = {http://dx.doi.org/10.1016/j.nicl.2019.101665},
  DOI = {10.1016/j.nicl.2019.101665},
  journal = {NeuroImage: Clinical},
  publisher = {Elsevier BV},
  author = {Kenkhuis,  Boyd and Jonkman,  Laura E. and Bulk,  Marjolein and Buijs,  Mathijs and Boon,  Baayla D.C. and Bouwman,  Femke H. and Geurts,  Jeroen J.G. and van de Berg,  Wilma D.J. and van der Weerd,  Louise},
  year = {2019},
  pages = {101665}
}

@article{Berron2017,
  title = {A protocol for manual segmentation of medial temporal lobe subregions in 7 Tesla MRI},
  volume = {15},
  ISSN = {2213-1582},
  url = {http://dx.doi.org/10.1016/j.nicl.2017.05.022},
  DOI = {10.1016/j.nicl.2017.05.022},
  journal = {NeuroImage: Clinical},
  publisher = {Elsevier BV},
  author = {Berron,  D. and Vieweg,  P. and Hochkeppler,  A. and Pluta,  J.B. and Ding,  S.-L. and Maass,  A. and Luther,  A. and Xie,  L. and Das,  S.R. and Wolk,  D.A. and Wolbers,  T. and Yushkevich,  P.A. and D\"{u}zel,  E. and Wisse,  L.E.M.},
  year = {2017},
  pages = {466–482}
}

@article{Bethlehem2022,
  title = {Brain charts for the human lifespan},
  volume = {604},
  ISSN = {1476-4687},
  url = {http://dx.doi.org/10.1038/s41586-022-04554-y},
  DOI = {10.1038/s41586-022-04554-y},
  number = {7906},
  journal = {Nature},
  publisher = {Springer Science and Business Media LLC},
  author = {Bethlehem,  R. A. I. and Seidlitz,  J. and White,  S. R. and Vogel,  J. W. and Anderson,  K. M. and Adamson,  C. and Adler,  S. and Alexopoulos,  G. S. and Anagnostou,  E. and Areces-Gonzalez,  A. and Astle,  D. E. and Auyeung,  B. and Ayub,  M. and Bae,  J. and Ball,  G. and Baron-Cohen,  S. and Beare,  R. and Bedford,  S. A. and Benegal,  V. and Beyer,  F. and Blangero,  J. and Blesa Cábez,  M. and Boardman,  J. P. and Borzage,  M. and Bosch-Bayard,  J. F. and Bourke,  N. and Calhoun,  V. D. and Chakravarty,  M. M. and Chen,  C. and Chertavian,  C. and Chetelat,  G. and Chong,  Y. S. and Cole,  J. H. and Corvin,  A. and Costantino,  M. and Courchesne,  E. and Crivello,  F. and Cropley,  V. L. and Crosbie,  J. and Crossley,  N. and Delarue,  M. and Delorme,  R. and Desrivieres,  S. and Devenyi,  G. A. and Di Biase,  M. A. and Dolan,  R. and Donald,  K. A. and Donohoe,  G. and Dunlop,  K. and Edwards,  A. D. and Elison,  J. T. and Ellis,  C. T. and Elman,  J. A. and Eyler,  L. and Fair,  D. A. and Feczko,  E. and Fletcher,  P. C. and Fonagy,  P. and Franz,  C. E. and Galan-Garcia,  L. and Gholipour,  A. and Giedd,  J. and Gilmore,  J. H. and Glahn,  D. C. and Goodyer,  I. M. and Grant,  P. E. and Groenewold,  N. A. and Gunning,  F. M. and Gur,  R. E. and Gur,  R. C. and Hammill,  C. F. and Hansson,  O. and Hedden,  T. and Heinz,  A. and Henson,  R. N. and Heuer,  K. and Hoare,  J. and Holla,  B. and Holmes,  A. J. and Holt,  R. and Huang,  H. and Im,  K. and Ipser,  J. and Jack,  C. R. and Jackowski,  A. P. and Jia,  T. and Johnson,  K. A. and Jones,  P. B. and Jones,  D. T. and Kahn,  R. S. and Karlsson,  H. and Karlsson,  L. and Kawashima,  R. and Kelley,  E. A. and Kern,  S. and Kim,  K. W. and Kitzbichler,  M. G. and Kremen,  W. S. and Lalonde,  F. and Landeau,  B. and Lee,  S. and Lerch,  J. and Lewis,  J. D. and Li,  J. and Liao,  W. and Liston,  C. and Lombardo,  M. V. and Lv,  J. and Lynch,  C. and Mallard,  T. T. and Marcelis,  M. and Markello,  R. D. and Mathias,  S. R. and Mazoyer,  B. and McGuire,  P. and Meaney,  M. J. and Mechelli,  A. and Medic,  N. and Misic,  B. and Morgan,  S. E. and Mothersill,  D. and Nigg,  J. and Ong,  M. Q. W. and Ortinau,  C. and Ossenkoppele,  R. and Ouyang,  M. and Palaniyappan,  L. and Paly,  L. and Pan,  P. M. and Pantelis,  C. and Park,  M. M. and Paus,  T. and Pausova,  Z. and Paz-Linares,  D. and Pichet Binette,  A. and Pierce,  K. and Qian,  X. and Qiu,  J. and Qiu,  A. and Raznahan,  A. and Rittman,  T. and Rodrigue,  A. and Rollins,  C. K. and Romero-Garcia,  R. and Ronan,  L. and Rosenberg,  M. D. and Rowitch,  D. H. and Salum,  G. A. and Satterthwaite,  T. D. and Schaare,  H. L. and Schachar,  R. J. and Schultz,  A. P. and Schumann,  G. and Sch\"{o}ll,  M. and Sharp,  D. and Shinohara,  R. T. and Skoog,  I. and Smyser,  C. D. and Sperling,  R. A. and Stein,  D. J. and Stolicyn,  A. and Suckling,  J. and Sullivan,  G. and Taki,  Y. and Thyreau,  B. and Toro,  R. and Traut,  N. and Tsvetanov,  K. A. and Turk-Browne,  N. B. and Tuulari,  J. J. and Tzourio,  C. and Vachon-Presseau,  É. and Valdes-Sosa,  M. J. and Valdes-Sosa,  P. A. and Valk,  S. L. and van Amelsvoort,  T. and Vandekar,  S. N. and Vasung,  L. and Victoria,  L. W. and Villeneuve,  S. and Villringer,  A. and Vértes,  P. E. and Wagstyl,  K. and Wang,  Y. S. and Warfield,  S. K. and Warrier,  V. and Westman,  E. and Westwater,  M. L. and Whalley,  H. C. and Witte,  A. V. and Yang,  N. and Yeo,  B. and Yun,  H. and Zalesky,  A. and Zar,  H. J. and Zettergren,  A. and Zhou,  J. H. and Ziauddeen,  H. and Zugman,  A. and Zuo,  X. N. and Rowe,  C. and Frisoni,  G. B. and Binette,  A. Pichet and Bullmore,  E. T. and Alexander-Bloch,  A. F.},
  year = {2022},
  month = {04},
  pages = {525–533}
}

@misc{wang2025triadvisionfoundationmodel,
      title={Triad: Vision Foundation Model for 3D Magnetic Resonance Imaging}, 
      author={Shansong Wang and Mojtaba Safari and Qiang Li and Chih-Wei Chang and Richard LJ Qiu and Justin Roper and David S. Yu and Xiaofeng Yang},
      year={2025},
      eprint={2502.14064},
      archivePrefix={arXiv},
      primaryClass={cs.CV},
      url={https://arxiv.org/abs/2502.14064}, 
}

@article{Islam2023,
  title = {Improving portable low-field MRI image quality through image-to-image translation using paired low- and high-field images},
  volume = {13},
  ISSN = {2045-2322},
  url = {http://dx.doi.org/10.1038/s41598-023-48438-1},
  DOI = {10.1038/s41598-023-48438-1},
  number = {1},
  journal = {Scientific Reports},
  publisher = {Springer Science and Business Media LLC},
  author = {Islam,  Kh Tohidul and Zhong,  Shenjun and Zakavi,  Parisa and Chen,  Zhifeng and Kavnoudias,  Helen and Farquharson,  Shawna and Durbridge,  Gail and Barth,  Markus and McMahon,  Katie L. and Parizel,  Paul M. and Dwyer,  Andrew and Egan,  Gary F. and Law,  Meng and Chen,  Zhaolin},
  year = {2023},
  month = dec 
}

@article{Iglesias2023,
  title = {Quantitative Brain Morphometry of Portable Low-Field-Strength MRI Using Super-Resolution Machine Learning},
  volume = {306},
  ISSN = {1527-1315},
  url = {http://dx.doi.org/10.1148/radiol.220522},
  DOI = {10.1148/radiol.220522},
  number = {3},
  journal = {Radiology},
  publisher = {Radiological Society of North America (RSNA)},
  author = {Iglesias,  Juan Eugenio and Schleicher,  Riana and Laguna,  Sonia and Billot,  Benjamin and Schaefer,  Pamela and McKaig,  Brenna and Goldstein,  Joshua N. and Sheth,  Kevin N. and Rosen,  Matthew S. and Kimberly,  W. Taylor},
  year = {2023},
  month = mar 
}

@article{Lucas2023,
  title = {Multi-contrast high-field quality image synthesis for portable low-field MRI using generative adversarial networks and paired data},
  url = {http://dx.doi.org/10.1101/2023.12.28.23300409},
  DOI = {10.1101/2023.12.28.23300409},
  publisher = {Cold Spring Harbor Laboratory},
  author = {Lucas,  Alfredo and Arnold,  T. Campbell and Okar,  Serhat V. and Vadali,  Chetan and Kawatra,  Karan D. and Ren,  Zheng and Cao,  Quy and Shinohara,  Russell T. and Schindler,  Matthew K. and Davis,  Kathryn A. and Litt,  Brian and Reich,  Daniel S. and Stein,  Joel M.},
  year = {2023},
  month = dec 
}

@article{Opheim2021,
  title = {7T Epilepsy Task Force Consensus Recommendations on the Use of 7T MRI in Clinical Practice},
  volume = {96},
  ISSN = {1526-632X},
  url = {http://dx.doi.org/10.1212/WNL.0000000000011413},
  DOI = {10.1212/wnl.0000000000011413},
  number = {7},
  journal = {Neurology},
  publisher = {Ovid Technologies (Wolters Kluwer Health)},
  author = {Opheim,  Giske and van der Kolk,  Anja and Bloch,  Karin Markenroth and Colon,  Albert J. and Davis,  Kathryn A. and Henry,  Thomas R. and Jansen,  Jacobus F.A. and Jones,  Stephen E. and Pan,  Jullie W. and R\"{o}ssler,  Karl and Stein,  Joel M. and Strandberg,  Maria C. and Trattnig,  Siegfried and Van de Moortele,  Pierre-Francois and Vargas,  Maria Isabel and Wang,  Irene and Bartolomei,  Fabrice and Bernasconi,  Neda and Bernasconi,  Andrea and Bernhardt,  Boris and Bj\"{o}rkman-Burtscher,  Isabella and Cosottini,  Mirco and Das,  Sandhitsu R. and Hertz-Pannier,  Lucie and Inati,  Sara and Jurkiewicz,  Michael T. and Khan,  Ali R. and Liang,  Shuli and Ma,  Ruoyun Emily and Mukundan,  Srinivasan and Pardoe,  Heath and Pinborg,  Lars H. and Polimeni,  Jonathan R. and Ranjeva,  Jean-Philippe and Steijvers,  Esther and Stufflebeam,  Steven and Veersema,  Tim J. and Vignaud,  Alexandre and Voets,  Natalie and Vulliemoz,  Serge and Wiggins,  Christopher J. and Xue,  Rong and Guerrini,  Renzo and Guye,  Maxime},
  year = {2021},
  month = feb,
  pages = {327–341}
}

@article{Duzel2019,
  title = {European Ultrahigh‐Field Imaging Network for Neurodegenerative Diseases (EUFIND)},
  volume = {11},
  ISSN = {2352-8729},
  url = {http://dx.doi.org/10.1016/j.dadm.2019.04.010},
  DOI = {10.1016/j.dadm.2019.04.010},
  number = {1},
  journal = {Alzheimer’s \& Dementia: Diagnosis,  Assessment \& Disease Monitoring},
  publisher = {Wiley},
  author = {D\"{u}zel,  Emrah and Acosta‐Cabronero,  Julio and Berron,  David and Biessels,  Geert Jan and Bj\"{o}rkman‐Burtscher,  Isabella and Bottlaender,  Michel and Bowtell,  Richard and Buchem,  Mark v and Cardenas‐Blanco,  Arturo and Boumezbeur,  Fawzi and Chan,  Dennis and Clare,  Stuart and Costagli,  Mauro and de Rochefort,  Ludovic and Fillmer,  Ariane and Gowland,  Penny and Hansson,  Oskar and Hendrikse,  Jeroen and Kraff,  Oliver and Ladd,  Mark E. and Ronen,  Itamar and Petersen,  Esben and Rowe,  James B. and Siebner,  Hartwig and Stoecker,  Tony and Straub,  Sina and Tosetti,  Michela and Uludag,  Kamil and Vignaud,  Alexandre and Zwanenburg,  Jaco and Speck,  Oliver},
  editor = {Jovicich,  Jorge and Frisoni,  Giovanni B.},
  year = {2019},
  month = {07},
  pages = {538–549}
}

@inbook{Zhang2018Cascade,
  title = {Dual-Domain Cascaded Regression for Synthesizing 7T from 3T MRI},
  ISBN = {9783030009281},
  ISSN = {1611-3349},
  url = {http://dx.doi.org/10.1007/978-3-030-00928-1_47},
  DOI = {10.1007/978-3-030-00928-1_47},
  booktitle = {Medical Image Computing and Computer Assisted Intervention – MICCAI 2018},
  publisher = {Springer International Publishing},
  author = {Zhang,  Yongqin and Cheng,  Jie-Zhi and Xiang,  Lei and Yap,  Pew-Thian and Shen,  Dinggang},
  year = {2018},
  pages = {410–417}
}

@article{Qu2020,
  title = {Synthesized 7T MRI from 3T MRI via deep learning in spatial and wavelet domains},
  volume = {62},
  ISSN = {1361-8415},
  url = {http://dx.doi.org/10.1016/j.media.2020.101663},
  DOI = {10.1016/j.media.2020.101663},
  journal = {Medical Image Analysis},
  publisher = {Elsevier BV},
  author = {Qu,  Liangqiong and Zhang,  Yongqin and Wang,  Shuai and Yap,  Pew-Thian and Shen,  Dinggang},
  year = {2020},
  month = may,
  pages = {101663}
}

@article{Zuo2021,
  title = {Unsupervised MR harmonization by learning disentangled representations using information bottleneck theory},
  volume = {243},
  ISSN = {1053-8119},
  url = {http://dx.doi.org/10.1016/j.neuroimage.2021.118569},
  DOI = {10.1016/j.neuroimage.2021.118569},
  journal = {NeuroImage},
  publisher = {Elsevier BV},
  author = {Zuo,  Lianrui and Dewey,  Blake E. and Liu,  Yihao and He,  Yufan and Newsome,  Scott D. and Mowry,  Ellen M. and Resnick,  Susan M. and Prince,  Jerry L. and Carass,  Aaron},
  year = {2021},
  month = nov,
  pages = {118569}
}

@article{Svanera2021,
  title = {CEREBRUM‐7T: Fast and Fully Volumetric Brain Segmentation of 7 Tesla MR Volumes},
  volume = {42},
  ISSN = {1097-0193},
  url = {http://dx.doi.org/10.1002/hbm.25636},
  DOI = {10.1002/hbm.25636},
  number = {17},
  journal = {Human Brain Mapping},
  publisher = {Wiley},
  author = {Svanera,  Michele and Benini,  Sergio and Bontempi,  Dennis and Muckli,  Lars},
  year = {2021},
  month = oct,
  pages = {5563–5580}
}

@article{Chu2024,
  title = {Brain morphometrics correlations with age among 352 participants imaged with both 3T and 7T MRI: 7T improves statistical power and reduces required sample size},
  url = {http://dx.doi.org/10.1101/2024.10.28.24316292},
  DOI = {10.1101/2024.10.28.24316292},
  publisher = {Cold Spring Harbor Laboratory},
  author = {Chu,  Cong and Santini,  Tales and Liou,  Jr-Jiun and Cohen,  Ann D. and Maki,  Pauline M. and Marsland,  Anna L. and Thurston,  Rebecca C. and Gianaros,  Peter J. and Ibrahim,  Tamer S.},
  year = {2024},
  month = nov 
}

@article{Srinivasan2020,
  title = {A comparison of Freesurfer and multi-atlas MUSE for brain anatomy segmentation: Findings about size and age bias,  and inter-scanner stability in multi-site aging studies},
  volume = {223},
  ISSN = {1053-8119},
  url = {http://dx.doi.org/10.1016/j.neuroimage.2020.117248},
  DOI = {10.1016/j.neuroimage.2020.117248},
  journal = {NeuroImage},
  publisher = {Elsevier BV},
  author = {Srinivasan,  Dhivya and Erus,  Guray and Doshi,  Jimit and Wolk,  David A. and Shou,  Haochang and Habes,  Mohamad and Davatzikos,  Christos},
  year = {2020},
  month = dec,
  pages = {117248}
}

@article{Wenger2014,
  title = {Comparing manual and automatic segmentation of hippocampal volumes: Reliability and validity issues in younger and older brains},
  volume = {35},
  ISSN = {1097-0193},
  url = {http://dx.doi.org/10.1002/hbm.22473},
  DOI = {10.1002/hbm.22473},
  number = {8},
  journal = {Human Brain Mapping},
  publisher = {Wiley},
  author = {Wenger,  Elisabeth and Mårtensson,  Johan and Noack,  Hannes and Bodammer,  Nils Christian and K\"{u}hn,  Simone and Schaefer,  Sabine and Heinze,  Hans‐Jochen and D\"{u}zel,  Emrah and B\"{a}ckman,  Lars and Lindenberger,  Ulman and L\"{o}vdén,  Martin},
  year = {2014},
  month = feb,
  pages = {4236–4248}
}

@article{Mihan2024,
  title = {Artificial intelligence bias in the prediction and detection of cardiovascular disease},
  volume = {1},
  ISSN = {2948-2836},
  url = {http://dx.doi.org/10.1038/s44325-024-00031-9},
  DOI = {10.1038/s44325-024-00031-9},
  number = {1},
  journal = {npj Cardiovascular Health},
  publisher = {Springer Science and Business Media LLC},
  author = {Mihan,  Ariana and Pandey,  Ambarish and Van Spall,  Harriette G. C.},
  year = {2024},
  month = nov 
}

@article{Debiasi2023,
  title = {The effect of processing pipelines,  input images and age on automatic cortical morphology estimates},
  volume = {242},
  ISSN = {0169-2607},
  url = {http://dx.doi.org/10.1016/j.cmpb.2023.107825},
  DOI = {10.1016/j.cmpb.2023.107825},
  journal = {Computer Methods and Programs in Biomedicine},
  publisher = {Elsevier BV},
  author = {Debiasi,  Giulia and Mazzonetto,  Ilaria and Bertoldo,  Alessandra},
  year = {2023},
  month = dec,
  pages = {107825}
}

@article{Tudosiu2024,
  title = {Realistic morphology-preserving generative modelling of the brain},
  volume = {6},
  ISSN = {2522-5839},
  url = {http://dx.doi.org/10.1038/s42256-024-00864-0},
  DOI = {10.1038/s42256-024-00864-0},
  number = {7},
  journal = {Nature Machine Intelligence},
  publisher = {Springer Science and Business Media LLC},
  author = {Tudosiu,  Petru-Daniel and Pinaya,  Walter H. L. and Ferreira Da Costa,  Pedro and Dafflon,  Jessica and Patel,  Ashay and Borges,  Pedro and Fernandez,  Virginia and Graham,  Mark S. and Gray,  Robert J. and Nachev,  Parashkev and Ourselin,  Sebastien and Cardoso,  M. Jorge},
  year = {2024},
  month = jul,
  pages = {811–819}
}

@article{Wang2024,
  title = {Self-improving generative foundation model for synthetic medical image generation and clinical applications},
  volume = {31},
  ISSN = {1546-170X},
  url = {http://dx.doi.org/10.1038/s41591-024-03359-y},
  DOI = {10.1038/s41591-024-03359-y},
  number = {2},
  journal = {Nature Medicine},
  publisher = {Springer Science and Business Media LLC},
  author = {Wang,  Jinzhuo and Wang,  Kai and Yu,  Yunfang and Lu,  Yuxing and Xiao,  Wenchao and Sun,  Zhuo and Liu,  Fei and Zou,  Zixing and Gao,  Yuanxu and Yang,  Lei and Zhou,  Hong-Yu and Miao,  Hanpei and Zhao,  Wenting and Huang,  Lisha and Zeng,  Lingchao and Guo,  Rui and Chong,  Ieng and Deng,  Boyu and Cheng,  Linling and Chen,  Xiaoniao and Luo,  Jing and Zhu,  Meng-Hua and Baptista-Hon,  Daniel and Monteiro,  Olivia and Li,  Ming and Ke,  Yu and Li,  Jiahui and Zeng,  Simiao and Guan,  Taihua and Zeng,  Jin and Xue,  Kanmin and Oermann,  Eric and Luo,  Huiyan and Yin,  Yun and Zhang,  Kang and Qu,  Jia},
  year = {2024},
  month = dec,
  pages = {609–617}
}

@article{Khader2023,
  title = {Denoising diffusion probabilistic models for 3D medical image generation},
  volume = {13},
  ISSN = {2045-2322},
  url = {http://dx.doi.org/10.1038/s41598-023-34341-2},
  DOI = {10.1038/s41598-023-34341-2},
  number = {1},
  journal = {Scientific Reports},
  publisher = {Springer Science and Business Media LLC},
  author = {Khader,  Firas and M\"{u}ller-Franzes,  Gustav and Tayebi Arasteh,  Soroosh and Han,  Tianyu and Haarburger,  Christoph and Schulze-Hagen,  Maximilian and Schad,  Philipp and Engelhardt,  Sandy and Baeßler,  Bettina and Foersch,  Sebastian and Stegmaier,  Johannes and Kuhl,  Christiane and Nebelung,  Sven and Kather,  Jakob Nikolas and Truhn,  Daniel},
  year = {2023},
  month = may 
}

@article{Hangel2023,
  title = {Implementation of a 7T Epilepsy Task Force consensus imaging protocol for routine presurgical epilepsy work-up: effect on diagnostic yield and lesion delineation},
  volume = {271},
  ISSN = {1432-1459},
  url = {http://dx.doi.org/10.1007/s00415-023-11988-5},
  DOI = {10.1007/s00415-023-11988-5},
  number = {2},
  journal = {Journal of Neurology},
  publisher = {Springer Science and Business Media LLC},
  author = {Hangel,  Gilbert and Kasprian,  Gregor and Chambers,  Stefanie and Haider,  Lukas and Lazen,  Philipp and Koren,  Johannes and Diehm,  Robert and Moser,  Katharina and Tomschik,  Matthias and Wais,  Jonathan and Winter,  Fabian and Zeiser,  Vitalij and Gruber,  Stephan and Aull-Watschinger,  Susanne and Traub-Weidinger,  Tatjana and Baumgartner,  Christoph and Feucht,  Martha and Dorfer,  Christian and Bogner,  Wolfgang and Trattnig,  Siegfried and Pataraia,  Ekaterina and Roessler,  Karl},
  year = {2023},
  month = oct,
  pages = {804–818}
}

@article{Klodowski2025,
  title = {Parallel transmit 7<scp>T MRI</scp> for adult epilepsy pre‐surgical evaluation},
  ISSN = {1528-1167},
  url = {http://dx.doi.org/10.1111/epi.18353},
  DOI = {10.1111/epi.18353},
  journal = {Epilepsia},
  publisher = {Wiley},
  author = {Klodowski,  Krzysztof and Zhang,  Minghao and Jen,  Jian P. and Scoffings,  Daniel J. and Morris,  Robert and Lupson,  Victoria and Mauconduit,  Franck and Massire,  Aurélien and Gras,  Vincent and Boulant,  Nicolas and Rodgers,  Christopher T. and Cope,  Thomas E.},
  year = {2025},
  month = mar 
}

@article{Sun2024,
  title = {A foundation model for enhancing magnetic resonance images and downstream segmentation,  registration and diagnostic tasks},
  volume = {9},
  ISSN = {2157-846X},
  url = {http://dx.doi.org/10.1038/s41551-024-01283-7},
  DOI = {10.1038/s41551-024-01283-7},
  number = {4},
  journal = {Nature Biomedical Engineering},
  publisher = {Springer Science and Business Media LLC},
  author = {Sun,  Yue and Wang,  Limei and Li,  Gang and Lin,  Weili and Wang,  Li},
  year = {2024},
  month = dec,
  pages = {521–538}
}

@article{Tak2024,
  title = {A foundation model for generalized brain MRI analysis},
  url = {http://dx.doi.org/10.1101/2024.12.02.24317992},
  DOI = {10.1101/2024.12.02.24317992},
  publisher = {Cold Spring Harbor Laboratory},
  author = {Tak,  Divyanshu and Garomsa,  Biniam A. and Chaunzwa,  Tafadzwa L. and Zapaishchykova,  Anna and Climent Pardo,  Juan Carlos and Ye,  Zezhong and Zielke,  John and Ravipati,  Yashwanth and Vajapeyam,  Sri and Mahootiha,  Maryam and Smith,  Ceilidh and Familiar,  Ariana M. and Liu,  Kevin X. and Prabhu,  Sanjay and Bandopadhayay,  Pratiti and Nabavizadeh,  Ali and Mueller,  Sabine and Aerts,  Hugo JWL and Huang,  Raymond Y. and Poussaint,  Tina Y. and Kann,  Benjamin H.},
  year = {2024},
  month = dec 
}

@article{Cox2024,
  title = {BrainSegFounder: Towards 3D foundation models for neuroimage segmentation},
  volume = {97},
  ISSN = {1361-8415},
  url = {http://dx.doi.org/10.1016/j.media.2024.103301},
  DOI = {10.1016/j.media.2024.103301},
  journal = {Medical Image Analysis},
  publisher = {Elsevier BV},
  author = {Cox,  Joseph and Liu,  Peng and Stolte,  Skylar E. and Yang,  Yunchao and Liu,  Kang and See,  Kyle B. and Ju,  Huiwen and Fang,  Ruogu},
  year = {2024},
  month = oct,
  pages = {103301}
}

@article{Su2025,
  title = {From Slices to Volumes: A Scalable Pipeline for Developing General-Purpose Brain MRI Foundation Models},
  url = {http://dx.doi.org/10.1101/2025.04.12.25325728},
  DOI = {10.1101/2025.04.12.25325728},
  publisher = {Cold Spring Harbor Laboratory},
  author = {Su,  Feng and Yi,  Xiaoping and Cheng,  Ye and Ma,  Yongjie and Zu,  Wenqiang and Zhao,  Qing and Huang,  Gengdi and Ma,  Lei},
  year = {2025},
  month = apr 
}

@article{Pini2016,
  title = {Brain atrophy in Alzheimer’s Disease and aging},
  volume = {30},
  ISSN = {1568-1637},
  url = {http://dx.doi.org/10.1016/j.arr.2016.01.002},
  DOI = {10.1016/j.arr.2016.01.002},
  journal = {Ageing Research Reviews},
  publisher = {Elsevier BV},
  author = {Pini,  Lorenzo and Pievani,  Michela and Bocchetta,  Martina and Altomare,  Daniele and Bosco,  Paolo and Cavedo,  Enrica and Galluzzi,  Samantha and Marizzoni,  Moira and Frisoni,  Giovanni B.},
  year = {2016},
  month = sep,
  pages = {25–48}
}

@article{Schwarz2016,
  title = {A large-scale comparison of cortical thickness and volume methods for measuring Alzheimer’s disease severity},
  volume = {11},
  ISSN = {2213-1582},
  url = {http://dx.doi.org/10.1016/j.nicl.2016.05.017},
  DOI = {10.1016/j.nicl.2016.05.017},
  journal = {NeuroImage: Clinical},
  publisher = {Elsevier BV},
  author = {Schwarz,  Christopher G. and Gunter,  Jeffrey L. and Wiste,  Heather J. and Przybelski,  Scott A. and Weigand,  Stephen D. and Ward,  Chadwick P. and Senjem,  Matthew L. and Vemuri,  Prashanthi and Murray,  Melissa E. and Dickson,  Dennis W. and Parisi,  Joseph E. and Kantarci,  Kejal and Weiner,  Michael W. and Petersen,  Ronald C. and Jack,  Clifford R.},
  year = {2016},
  pages = {802–812}
}

@article{Ossenkoppele2019,
  title = {Associations between tau,  A$\beta$,  and cortical thickness with cognition in Alzheimer disease},
  volume = {92},
  ISSN = {1526-632X},
  url = {http://dx.doi.org/10.1212/WNL.0000000000006875},
  DOI = {10.1212/wnl.0000000000006875},
  number = {6},
  journal = {Neurology},
  publisher = {Ovid Technologies (Wolters Kluwer Health)},
  author = {Ossenkoppele,  Rik and Smith,  Ruben and Ohlsson,  Tomas and Strandberg,  Olof and Mattsson,  Niklas and Insel,  Philip S. and Palmqvist,  Sebastian and Hansson,  Oskar},
  year = {2019},
  month = feb 
}

@article{Cuingnet2011,
  title = {Automatic classification of patients with Alzheimer’s disease from structural MRI: A comparison of ten methods using the ADNI database},
  volume = {56},
  ISSN = {1053-8119},
  url = {http://dx.doi.org/10.1016/j.neuroimage.2010.06.013},
  DOI = {10.1016/j.neuroimage.2010.06.013},
  number = {2},
  journal = {NeuroImage},
  publisher = {Elsevier BV},
  author = {Cuingnet,  Rémi and Gerardin,  Emilie and Tessieras,  Jér\^ome and Auzias,  Guillaume and Lehéricy,  Stéphane and Habert,  Marie-Odile and Chupin,  Marie and Benali,  Habib and Colliot,  Olivier},
  year = {2011},
  month = may,
  pages = {766–781}
}

@article{Davatzikos2009,
  title = {Longitudinal progression of Alzheimer’s-like patterns of atrophy in normal older adults: the SPARE-AD index},
  volume = {132},
  ISSN = {1460-2156},
  url = {http://dx.doi.org/10.1093/brain/awp091},
  DOI = {10.1093/brain/awp091},
  number = {8},
  journal = {Brain},
  publisher = {Oxford University Press (OUP)},
  author = {Davatzikos,  C. and Xu,  F. and An,  Y. and Fan,  Y. and Resnick,  S. M.},
  year = {2009},
  month = may,
  pages = {2026–2035}
}

@article{Vogel2018,
  title = {Brain properties predict proximity to symptom onset in sporadic Alzheimer’s disease},
  volume = {141},
  ISSN = {1460-2156},
  url = {http://dx.doi.org/10.1093/brain/awy093},
  DOI = {10.1093/brain/awy093},
  number = {6},
  journal = {Brain},
  publisher = {Oxford University Press (OUP)},
  author = {Vogel,  Jacob W and Vachon-Presseau,  Etienne and Pichet Binette,  Alexa and Tam,  Angela and Orban,  Pierre and La Joie,  Renaud and Savard,  Mélissa and Picard,  Cynthia and Poirier,  Judes and Bellec,  Pierre and Breitner,  John C S and Villeneuve,  Sylvia},
  year = {2018},
  month = apr,
  pages = {1871–1883}
}

@article{Tam2019,
  title = {A highly predictive signature of cognition and brain atrophy for progression to Alzheimer’s dementia},
  volume = {8},
  ISSN = {2047-217X},
  url = {http://dx.doi.org/10.1093/gigascience/giz055},
  DOI = {10.1093/gigascience/giz055},
  number = {5},
  journal = {GigaScience},
  publisher = {Oxford University Press (OUP)},
  author = {Tam,  Angela and Dansereau,  Christian and Iturria-Medina,  Yasser and Urchs,  Sebastian and Orban,  Pierre and Sharmarke,  Hanad and Breitner,  John and Bellec,  Pierre},
  year = {2019},
  month = may 
}

@misc{wgangp_github,
  title={{Keras-GAN}},
  author={Linder-Norén, Erik},
  year={2021},
  howpublished="\url{https://github.com/eriklindernoren/Keras-GAN}",
}

@article{vanVeluw2015,
  title = {The Spectrum of MR Detectable Cortical Microinfarcts: A Classification Study with 7-Tesla Postmortem MRI and Histopathology},
  volume = {35},
  ISSN = {1559-7016},
  url = {http://dx.doi.org/10.1038/jcbfm.2014.258},
  DOI = {10.1038/jcbfm.2014.258},
  number = {4},
  journal = {Journal of Cerebral Blood Flow \& Metabolism},
  publisher = {SAGE Publications},
  author = {van Veluw,  Susanne J and Zwanenburg,  Jaco JM and Rozemuller,  Annemieke JM and Luijten,  Peter R and Spliet,  Wim GM and Biessels,  Geert Jan},
  year = {2015},
  month = jan,
  pages = {676–683}
}

@article{vanVeluw2012,
  title = {In Vivo Detection of Cerebral Cortical Microinfarcts with High-Resolution 7T MRI},
  volume = {33},
  ISSN = {1559-7016},
  url = {http://dx.doi.org/10.1038/jcbfm.2012.196},
  DOI = {10.1038/jcbfm.2012.196},
  number = {3},
  journal = {Journal of Cerebral Blood Flow \& Metabolism},
  publisher = {SAGE Publications},
  author = {van Veluw,  Susanne J and Zwanenburg,  Jaco JM and Engelen-Lee,  JooYeon and Spliet,  Wim GM and Hendrikse,  Jeroen and Luijten,  Peter R and Biessels,  Geert Jan},
  year = {2012},
  month = dec,
  pages = {322–329}
}

@article{Li2024,
  title = {Automatic segmentation of medial temporal lobe subregions in multi-scanner,  multi-modality MRI of variable quality},
  url = {http://dx.doi.org/10.1101/2024.05.21.595190},
  DOI = {10.1101/2024.05.21.595190},
  publisher = {Cold Spring Harbor Laboratory},
  author = {Li,  Yue and Xie,  Long and Khandelwal,  Pulkit and Wisse,  Laura E. M. and Brown,  Christopher A. and Prabhakaran,  Karthik and Tisdall,  M. Dylan and Mechanic-Hamilton,  Dawn and Detre,  John A. and Das,  Sandhitsu R. and Wolk,  David A. and Yushkevich,  Paul A.},
  year = {2024},
  month = may 
}

@article{lewis2019cortical,
  title={Cortical and subcortical T1 white/gray contrast, chronological age, and cognitive performance},
  author={Lewis, John D and Fonov, Vladimir S and Collins, D Louis and Evans, Alan C and Tohka, Jussi and Brain Development Cooperative Group and others},
  journal={NeuroImage},
  volume={196},
  pages={276--288},
  year={2019},
  publisher={Elsevier},
  doi = {https://doi.org/10.1016/j.neuroimage.2019.04.022}
}

@article{corpataux2025effect,
  title={Effect of age and sex on cardiac magnetic resonance native T1 mapping and synthetic extracellular volume},
  author={Corpataux, No{\'e} and Haider, Achi and Fuentes, Ruben and Wahl, Andreas and Gebert, Pimrapat and Mikail, Nidaa and Rossi, Alexia and Vinzens, Adriana and Bengs, Susan and Buechel, Ronny R and others},
  journal={International Journal of Cardiology},
  pages={133818},
  year={2025},
  doi = {https://doi.org/10.1016/j.ijcard.2025.133818},
  publisher={Elsevier}
}

@article{casamitjana2025probabilistic,
  title={A probabilistic histological atlas of the human brain for MRI segmentation},
  author={Casamitjana, Adri{\`a} and Mancini, Matteo and Robinson, Eleanor and Peter, Lo{\"\i}c and Annunziata, Roberto and Althonayan, Juri and Crampsie, Shauna and Blackburn, Emily and Billot, Benjamin and Atzeni, Alessia and others},
  journal={Nature},
  pages={1--8},
  year={2025},
  publisher={Nature Publishing Group UK London},
  doi={https://doi.org/10.1038/s41586-025-09708-2}
}

@article{zampeli2025does,
  title={Does 7 T MRI offer an added value in drug resistant temporal lobe epilepsy?},
  author={Zampeli, Ariadne and Malac, Miroslav and Bj{\"o}rkman-Burtscher, Isabella M and Hansson, Boel and Wennberg, Linda and Bloch, Karin Markenroth and K{\"a}ll{\'e}n, Kristina and Strandberg, Maria Compagno},
  journal={Seizure: European Journal of Epilepsy},
  year={2025},
  publisher={Elsevier},
  doi={https://doi.org/10.1016/j.seizure.2025.11.004}
}

@article{buugday2026triggering,
  title={Triggering hallucinations in model-based MRI reconstruction via adversarial perturbations},
  author={Bu{\u{g}}day, Suna and Saeys, Yvan and Peck, Jonathan},
  journal={arXiv e-prints},
  pages={arXiv--2602},
  year={2026},
  doi={https://doi.org/10.48550/arXiv.2602.18536}
}

@misc{kim2026hallugen,
  doi = {10.48550/ARXIV.2512.03345},
  url = {https://arxiv.org/abs/2512.03345},
  author = {Kim,  Seunghoi and Tregidgo,  Henry F. J. and Jin,  Chen and Figini,  Matteo and Alexander,  Daniel C.},
  title = {HalluGen: Synthesizing Realistic and Controllable Hallucinations for Evaluating Image Restoration},
  publisher = {arXiv},
  year = {2025},
  copyright = {arXiv.org perpetual,  non-exclusive license}
}

@inproceedings{kim2025tackling,
  title={Tackling hallucination from conditional models for medical image reconstruction with dynamicdps},
  author={Kim, Seunghoi and Tregidgo, Henry FJ and Figini, Matteo and Jin, Chen and Joshi, Sarang and Alexander, Daniel C},
  booktitle={International Conference on Medical Image Computing and Computer-Assisted Intervention},
  pages={593--603},
  year={2025},
  organization={Springer Nature Switzerland},
  isbn="978-3-032-04965-0",
  doi={https://doi.org/10.1007/978-3-032-04965-0_56}
}

@inproceedings{tivnan2024hallucination,
  title={Hallucination index: An image quality metric for generative reconstruction models},
  author={Tivnan, Matthew and Yoon, Siyeop and Chen, Zhennong and Li, Xiang and Wu, Dufan and Li, Quanzheng},
  booktitle={International Conference on Medical Image Computing and Computer-Assisted Intervention},
  pages={449--458},
  year={2024},
  organization={Springer},
  doi={https://doi.org/10.1007/978-3-031-72117-5_42}
}

\newpage 
\appendix

\section{Appendix}

\counterwithin{figure}{section}
\counterwithin{table}{section}
\setcounter{figure}{0}
\setcounter{table}{0}

\subsection{Participant characteristics}
\label{Participants}

\begin{figure}[!ht]
    \centering
    \includegraphics[width=0.6\textwidth]{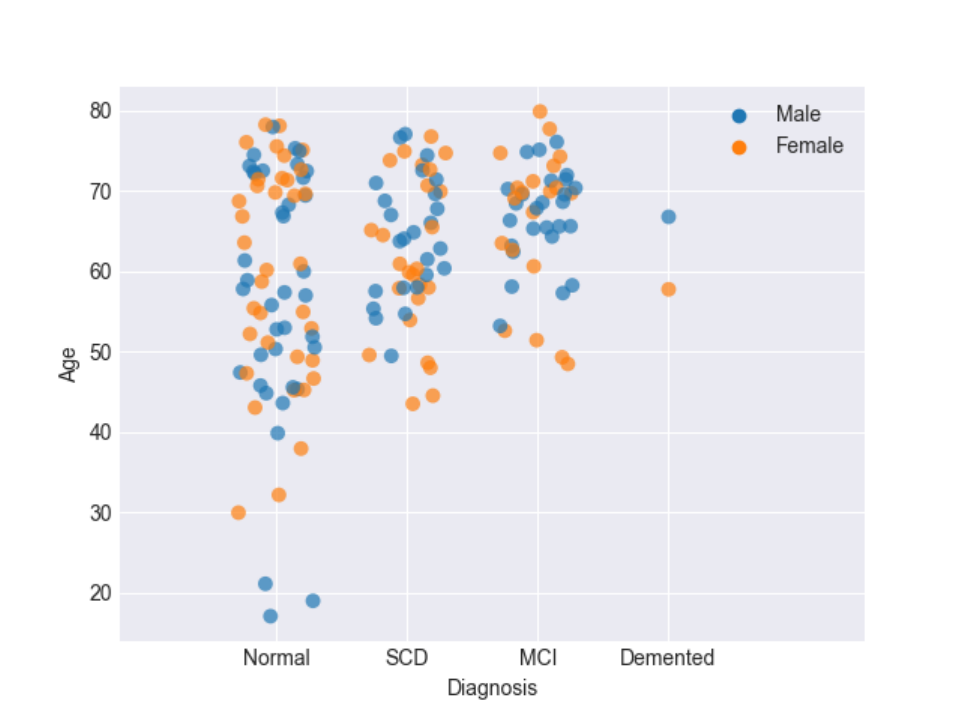}
    \small{\caption{Age and gender of the participants of the BioFINDER 2 dataset, according to their diagnosis. Each dot represents a participant and has a small random offset along the x-axis for more visual clarity. Mean age : $61.9\pm 11.8$, F/M : 82/90.}
    \label{fig: BF2 charac}}
\end{figure}

We display the distribution of the ages of the participants of BioFINDER 2 who underwent a 7T scan, according to their gender and diagnosis, in Figure~\ref{fig: BF2 charac}. 

We also display the characteristics of the patients on which the inference was performed for the downstream predictions in Table~\ref{tab: inf participants}. 

\begin{table}[!h]
    \centering
    \begin{tabular}{cccccccc}
        &  Scans & Age (mean $\pm$ SD)  &  Gender(F/M)  & Control & MCI & AD & Other Dementia\\ \hline
       participants  & 3168   & 69.20 $\pm$ 12.48    & 1559/1609   & 1641  & 687   & 421 & 419  \\ 
    \end{tabular}
    \caption{Summary of patient characteristics in the diagnostic prediction dataset. \enquote{MCI} mild cognitive impairment and \enquote{AD} stands for Alzheimer's disease dementia. }
    \label{tab: inf participants}
\end{table}

\subsection{Model drawings}
\label{Drawings}

The blocks used to build the U-Net are a residual block with positional encoding, a self- and a cross-attention block and a feed-forward block. 

\begin{figure}[!h]
    \centering
    \includegraphics[width=0.5\linewidth]{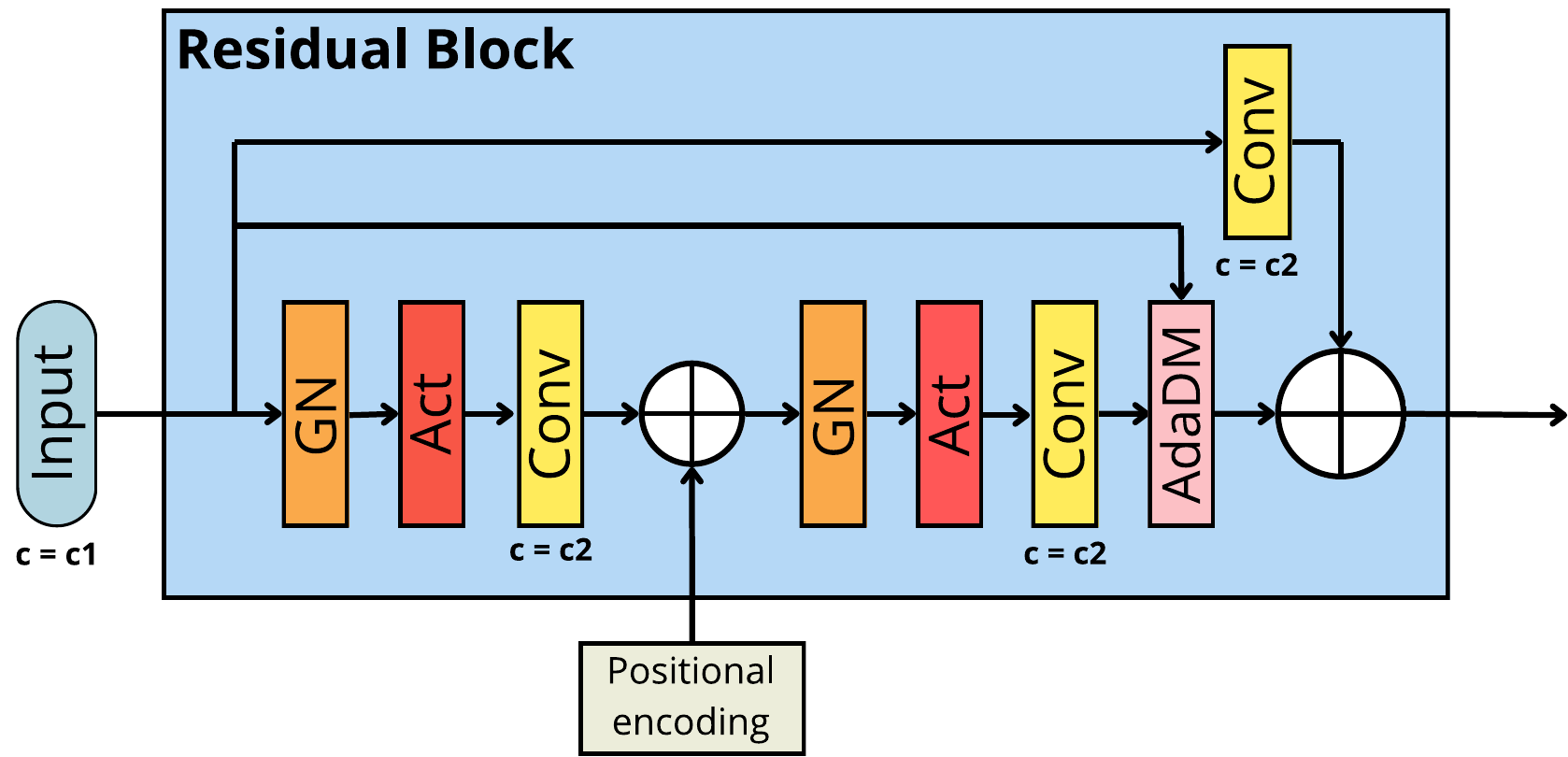}
    \caption{Drawing of the residual blocks from Ho et al.'s code \citep{ho2020denoising}, with an AdaDM layer \citep{liu2021adadm} used to sharpen the image edges. GN stands for group normalization, Act for activation function (here we used the Swish function), conv is a 2D convolutional layer. The notation $c_i$ is used for the number of channels outputted by the layer. Note that the convolution in the skip connection is used if $c_1=c_2$.}
    \label{fig: res block}
\end{figure}

The residual block is based on four layer types: group normalization (GN), activation function Swish (Act), a $3\times 3$ 2D convolution (Conv) and AdaDM \citep{liu2021adadm}. It also uses skip connections and the positional encoding mentioned previously. The architecture of the residual block is shown in Figure~\ref{fig: res block}.

The attention blocks rely on an attention mechanism, where $Q,K,V\in\mathbb{R}^{n\times d_k}$ are, respectively, the query, the key and the value matrices. The attention is then computed as
\begin{equation}
\label{eq: att}
    \text{Attention}(Q,K,V) = \text{softmax}\left(\frac{QK^T}{\sqrt{d_k}}\right)V.
\end{equation}

In the case of cross attention, $V$ contains information about the input, while $K \text{ and } Q$ contain the contextual information. In our application, the context is the age, diagnosis, gender and slice location.

To provide information about the location and order of pixels in an image to the attention mechanisms, we use a positional encoding matrix $PE\in\mathbb{R}^{c\times d}$, with $c$ being the number of channels and $d$ the pixel flattened position, is calculated as follows \citep{vaswani2017AttentionIsAllYouNeed}:
\begin{align*}
 \forall pos\in\{0, \dots, d-1\}, i\in\{0,\dots, \frac{c}{2}-1\} :  \\ 
     PE(2i,pos) = & \sin\left(pos/10000^{2i/c}\right),\\
     PE(2i+1,pos)& = \cos\left(pos/10000^{2i/c}\right).
\end{align*}

\begin{figure}[!h]
    \centering
    \includegraphics[width=1\linewidth]{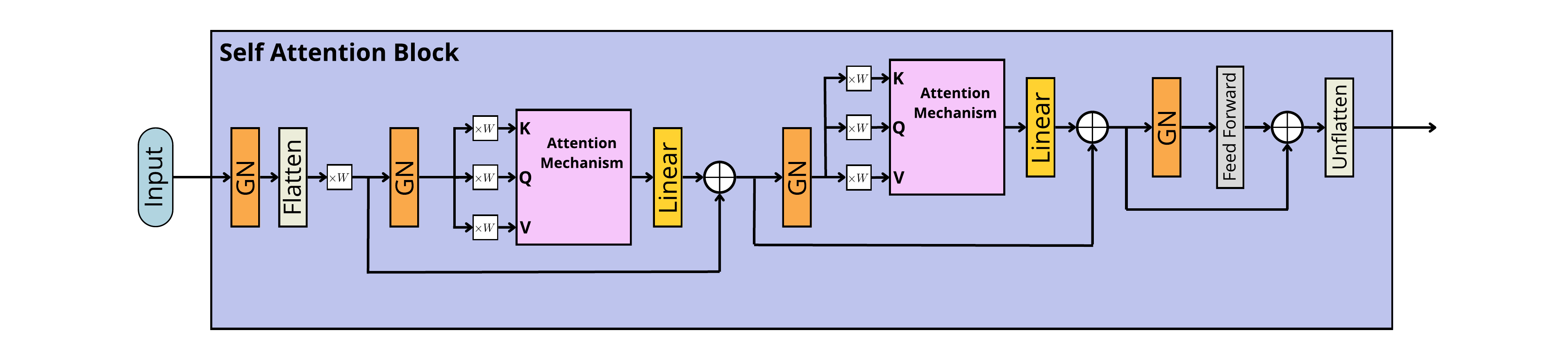}
    \caption{Self-attention block from Ho et al.'s code \citep{ho2020denoising}. $\times W$ is a trained linear projection. Linear performs a trained affine transform. See Figure~\ref{fig: feed forward} for the feed-forward layer.}
    \label{fig: self block}
\end{figure}

\begin{figure}[!h]
    \centering
    \includegraphics[width=0.8\linewidth]{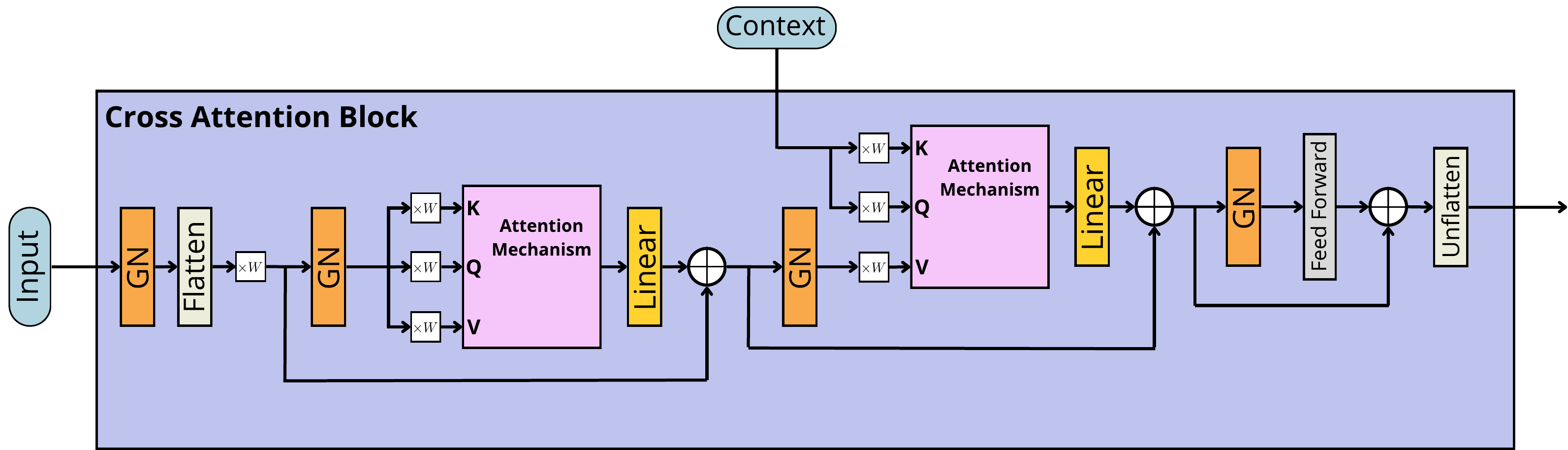}
    \caption{Cross-attention block from Ho et al.'s code \citep{ho2020denoising}, that calculates an attention map between an input and a context. For us the context is the gender, age, slice location and the diagnosis.}
    \label{fig: cross block}
\end{figure}

\begin{figure}[!h]
    \centering
    \includegraphics[width=0.3\linewidth]{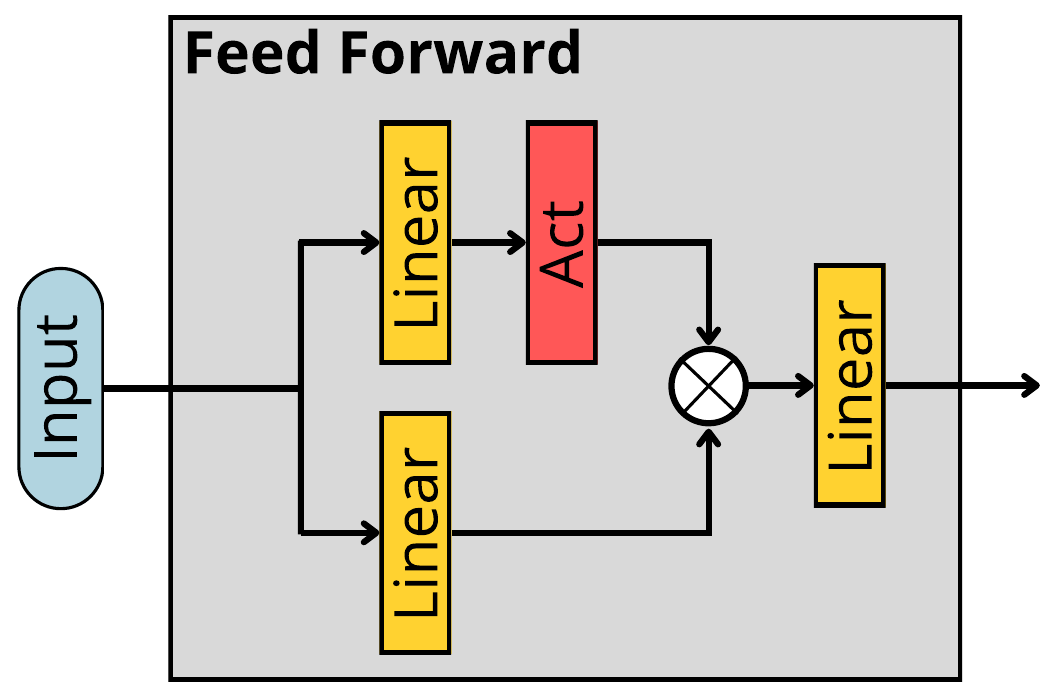}
    \caption{Feed-forward block used in self- and cross-attention mechanisms from Ho et al.'s code \citep{ho2020denoising}, shown in Figure~\ref{fig: self block} and \ref{fig: cross block}.}
    \label{fig: feed forward}
\end{figure}

\begin{figure}[!h]
    \centering
    \includegraphics[width=\linewidth]{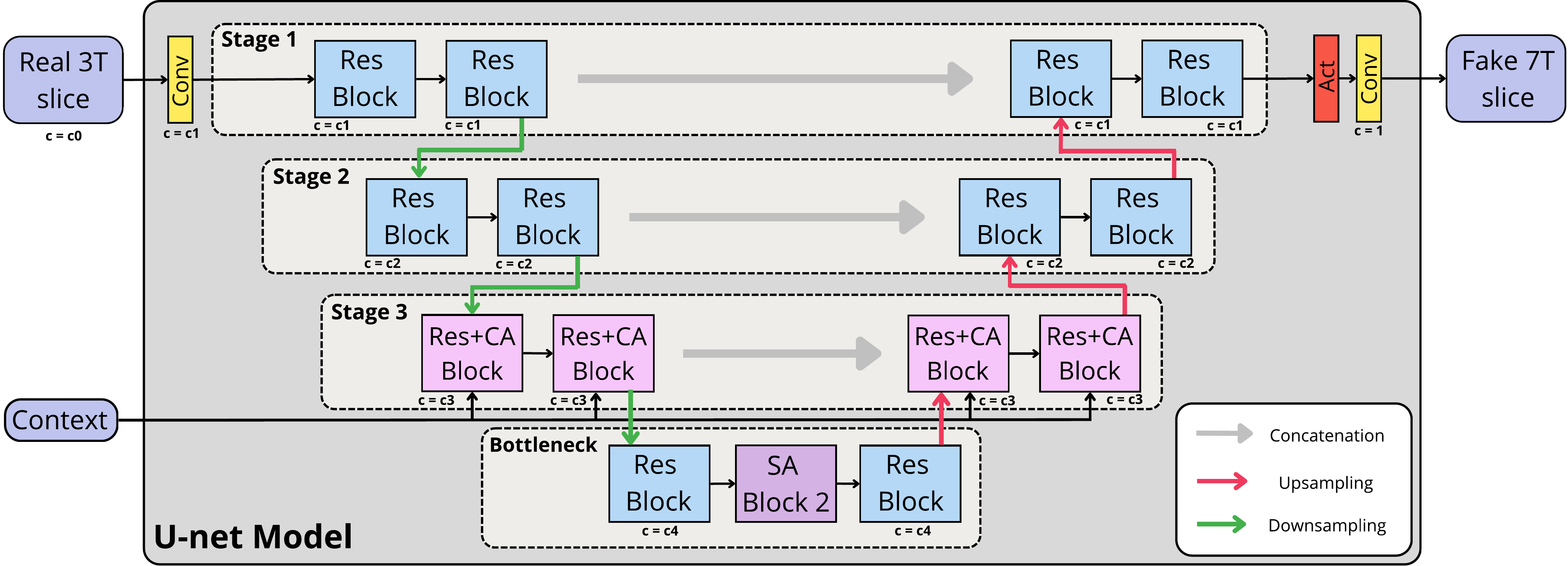}
    \caption{U-Net with cross-attention layers at the third stage. Res+CA block is a residual block drawn in Figure~\ref{fig: res block} followed by a cross attention block drawn in Figure~\ref{fig: cross block}.  $c_i$ is the number of channels at the stage, it is equal to $k_i\times c_0$, where $c_0$ is the initial number of channels and $k_i\in\mathbb{N}^*$.}
    \label{fig: attention Unet}
\end{figure}

The self- and cross-attention blocks are similar to each other and based on six layer types: group normalization (GN), flattening/unflattening layer, linear transforms (Linear), unbiased linear transforms ($\times $W), attention mechanism layer (based on equation \ref{eq: att}) and a feed-forward block (represented in Figure~\ref{fig: feed forward}). These self- and cross-attention blocks are presented in, respectively, Figures \ref{fig: self block} and \ref{fig: cross block}. 
Using these layers and blocks, we can build an attention U-Net. It is drawn in Figure~\ref{fig: attention Unet}, where Res+CA is a residual block followed by a cross attention mechanism, while SA block is a self attention block and context contains the external variables.

\begin{figure}[!h]
    \centering
    \includegraphics[width=0.8\linewidth]{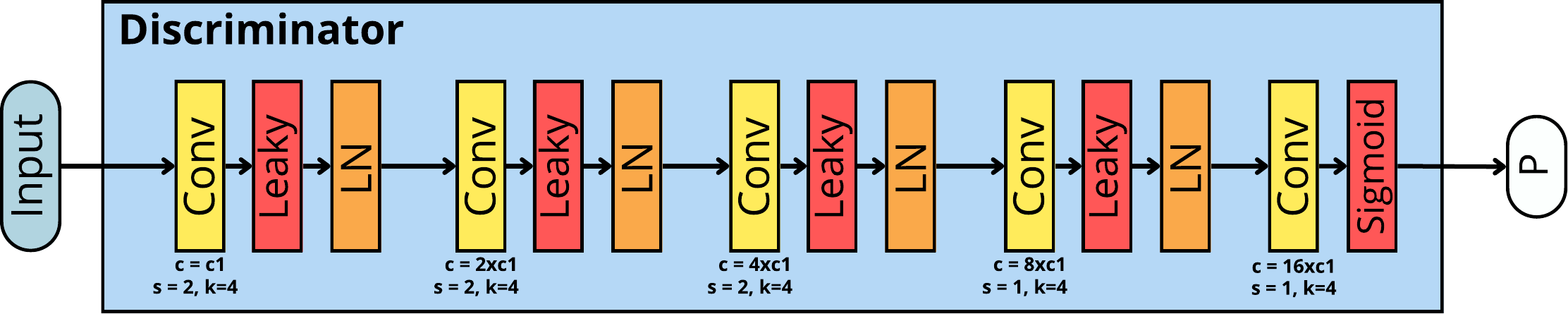}
    \caption{Drawing of our discriminator, k is the kernel size and s is the stride, s=2 implies that the image is downsampled by a factor two. The leakyReLU has a parameter of 0.2 ; LN stands for layer normalization. c1 is the initial number of channels. P is the matrix that contains the predicted probability for each patch to be fake. The number of patches is $ H/2^n\times W/2^n$, where $n$ is the number of convolutions minus one, $H$ is the height of the input and $W$ its width.}
    \label{fig: GAN discr}
\end{figure}

We also show our discriminator architecture in Figure~\ref{fig: GAN discr}, where LN is layer normalization, Leaky is Leaky Relu $(alpha=0.2)$. 

\newpage 
\newpage 
\newpage 
\subsection{Model hyperparameters}
\label{Hyperpara}

We describe the hyperparameters of the U-Net and show which values we used in Table~\ref{tab: params}.

\begin{itemize}
    \item $n_{epochs}\in\mathbb{N}^*$: number of epochs.
    \item $c_{init}\in\mathbb{N}^*$: initial number of channels ($c$ in Figure~\ref{fig: attention Unet}).
    \item Channel multiplication \textit{integer n-Tuple. $(1,k_1,...,k_n)\in\mathbb{N}^{n+1}$}, where $1\leq k_1\leq k_2\leq...\leq k_n$  and $n$ is the number of stages. This parameter controls the number of channels in the $i$-th stage $c_i$ (see Figure~\ref{fig: attention Unet}), as $c_{i+1}=c_i\times k_{i+1}$ and $c_1 = c_{init}$.
    For low level tasks such as super resolution, the first stages are the most important, so we set it to $(1,2,2,...,2)$.
    \item  $n_{groups}\in c\mathbb{N}^*$: number of groups in the group normalization layers, a multiple of $c$.
    \item $n_{res}\in\mathbb{N}^*$: number of residual blocks between two consecutive downsamplings or two consecutive upsamplings (the bottleneck always has 2 residual blocks).
    \item CA stages (\textit{int n-Tuple}): indicates at which stages to do cross attention.
    \item  $n_{input slices}\in2\mathbb{N}+1$: indicates how many 2D slices to include in the input (equivalent to the number of channels of the input).
    \item $\lambda_{perc}\in\mathbb{R}^+$: weight of the perceptual loss.
    \item lr$\in\mathbb{R}^*_+$: initial learning rate.
    \item lr schedule: describes how the learning rate decays during training.
    \item $\beta=(\beta_1,\beta_2)\in [0,1]^2$: $\beta$ parameters of the adam optimizer \citep{kingma2017adammethodstochasticoptimization}. $(0.9,0.999)$ is typically used for our generation purposes with Lp and perceptual losses.
    \item dropout$\in [0,1]$: for low-level tasks such as super resolution, this is said to have a bad effect \citep{kong2022dropout}, so we set it to 0.
    \item batch size : number of training examples used in each optimization step.
\end{itemize}

Most hyperparameters were chosen after a few manual tests and had the goal of maximizing usage of the 80GB of RAM of our GPUs. The hyperparameters of the U-Net with and without GAN are basically the same, except we added cross-attention mechanisms at one stage because not using a discriminator takes less GPU memory, which enabled us to increase the model size. 

\begin{table}[!h]
    \centering
    \begin{tabular}{ccc} 
        Parameters &  U-Net & GAN U-Net  \\ \hline 
        $n_{epochs}$ & 4 & 22  \\ 
        $c$  & 256 & 256  \\ 
        channel multiplication & (1,2,2,2) & (1,2,2,2) \\ 
        $n_{groups}$ & 64 & 64 \\ 
        $n_{res}$ & 3 & 3  \\ 
        CA stages & (3,4) & (4) \\ 
        batch size & 56 & 56  \\ 
        $n_{input slices}$ &  3 & 3 \\ 
        $\lambda_{perc}$ & $5.10^{-2}$ & $10^{-2}$  \\ 
        lr & $10^{-4}$ & $10^{-4}$  \\ 
        lr schedule & $\times 0.5/epoch$  &  $\times 0.9/epoch$ \\ 
        dropout & 0 & 0 \\ 
        betas & $(0.9,0.999)$ & $(0.9,0.999)$ \\ 
    \end{tabular}
    \caption{Parameters used for every U-Net model}
    \label{tab: params}
\end{table}

We also describe the hyperparameters specific to the GAN and show which one we chose in Table~\ref{tab: params GAN}.

\begin{itemize}
    \item $\beta=(\beta_1,\beta_2)\in [0,1]^2$: same parameter as for the U-Net, except we chose (0,0.9) for the discriminator's optimizer, following \cite{basu2024Collapse}.
    \item $n_{critic}\in\mathbb{N}^*$: this is a positive integer that tells how many times the discriminator should be trained every time the generator is trained. It is common to set it to 5 to have an efficient discriminator. During the first epoch, it is equal to 1 for warm-up.
    \item  $\lambda_{GAN}\in\mathbb{R}^+$: weight of the GAN loss during training. During the first epoch, we divide it by 10 for warm-up.
    \item $\lambda_{GP}$: weight of the gradient penalty for the , set to 10  following \cite{basu2024Collapse}.
    \item  $n_{layers}\in\mathbb{N}^*$: number of (convolution+activation+normalization) blocks in the discriminator (see Figure~\ref{fig: GAN discr}). It is also the number of times the input is downsampled. The patch size is $H/2^{n_{layers}}\times W/2^{n_{layers}}$, where $H$ is the height of the input and $W$ its width.
\end{itemize}

\begin{table}[!h]
    \centering
    \begin{tabular}{cc}
        Parameters &  Value \\ \hline 
        $n_{critic}$ & 5 \\ 
        lr & $2.10^{-5}$ \\ 
        $\lambda_{GAN}$ & 0.1 \\ 
        $c$ & 256 \\ 
        $n_{layers}$ & 5 \\ 
    \end{tabular}
    \caption{Parameters used for the discriminator of the GAN.}
    \label{tab: params GAN}
\end{table}

\newpage

\subsection{Model comparison}
\label{sec:supp_additional_results}

\begin{table}[h] 
   \centering
   \begin{tabular}{cccc} 
       { Image Type} & { SynthSeg Dice} $\uparrow$  & { NextBrain Dice, agg.} $\uparrow$  & { NextBrain Dice, all} $\uparrow$  \\ \hline
           {3T } &  $ 0.879 \pm 0.0089 $ & $ 0.784 \pm 0.015 $ & $ 0.498 \pm 0.019 $ \\ 
           { V-Net } & $ 0.877 \pm 0.0091 $ & $ 0.784 \pm 0.015 $  & $ 0.497 \pm 0.021 $ \\
           { WATNet } & $ 0.877 \pm 0.0095 $  & $ 0.787 \pm 0.015 $ & $ 0.499 \pm 0.023 $  \\
           { U-Net (ours) } &  $ 0.904 \pm 0.011 $  & $ 0.828 \pm 0.017 $ & $ 0.540 \pm 0.032 $ \\
           { GAN (ours) } & $ \textbf{0.909} \pm 0.0099 $ & $ \textbf{0.831} \pm 0.017 $ & $ \textbf{0.557} \pm 0.029 $  \\
   \end{tabular}	
   \caption{Dice scores comparing the generated 7T images using different models to the real 7T using both SynthSeg and NextBrain, for all regions (all) and after aggregating to larger regions (agg.). A comparison between the 3T image and the real 7T is included as a baseline. Computations are done on the 3D images without background or cerebellum artifacts and we show the average result over all regions in 29 images together with the standard deviation. Note that the Dice metrics are computed over a different set of region for each column.}
   \label{tab:supp_synthseg_nextbrain_dice_main}
\end{table}

\newpage
\begin{figure}[!ht]
    \centering
    \includegraphics[width=0.95\textwidth]{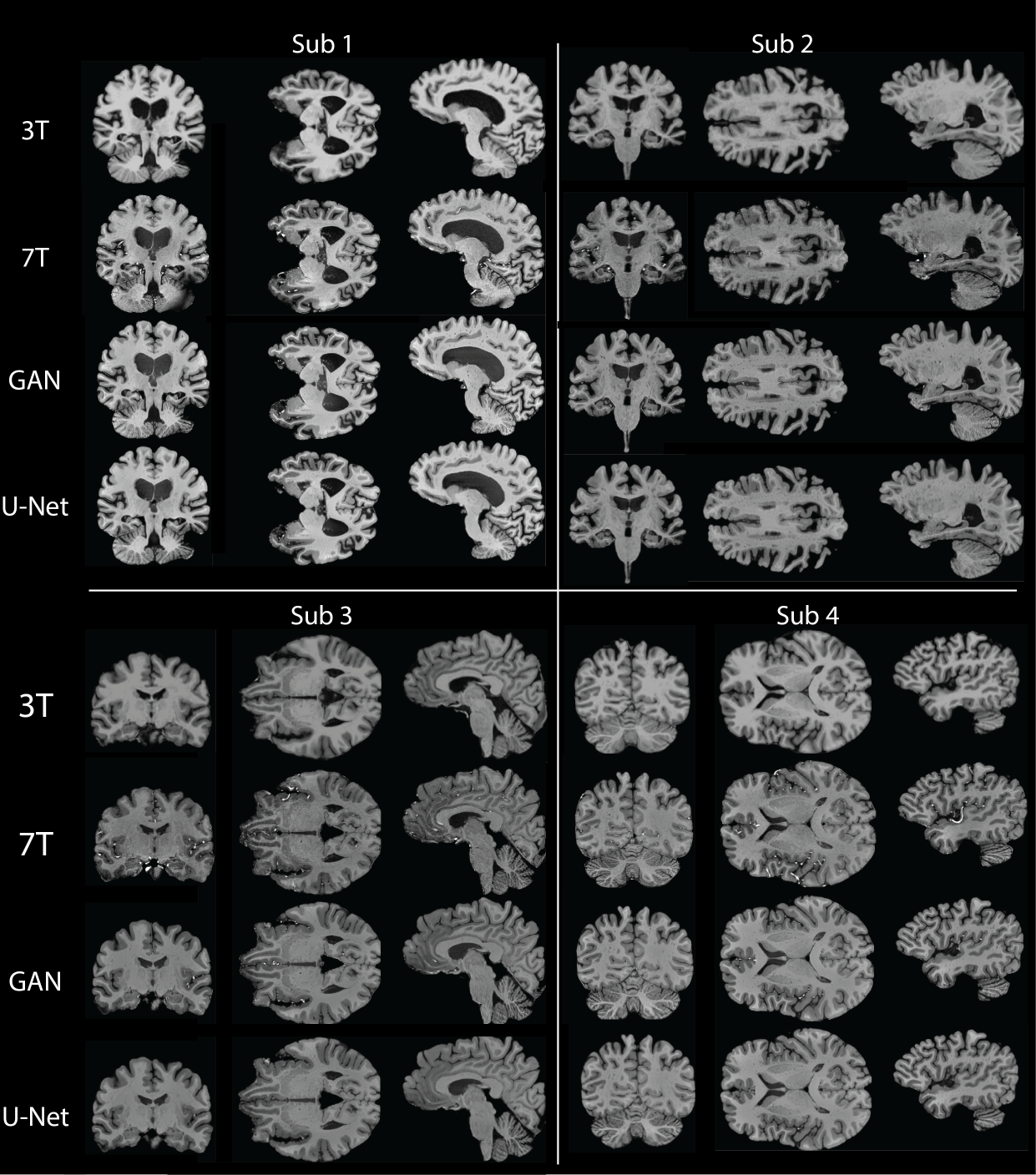}
    \caption{Side-by-side comparison of real 3T, real 7T, GAN U-Net synthetic 7T and U-Net synthetic 7T from the first four subjects in our dataset.}
    \label{fig:supp_qualcomp}
\end{figure}

\subsection{Survey results details and example}
\label{surveys}
\begin{figure}[!ht]
\centering
\begin{minipage}[!ht]{0.48\textwidth}
\includegraphics[width=1.15\textwidth]{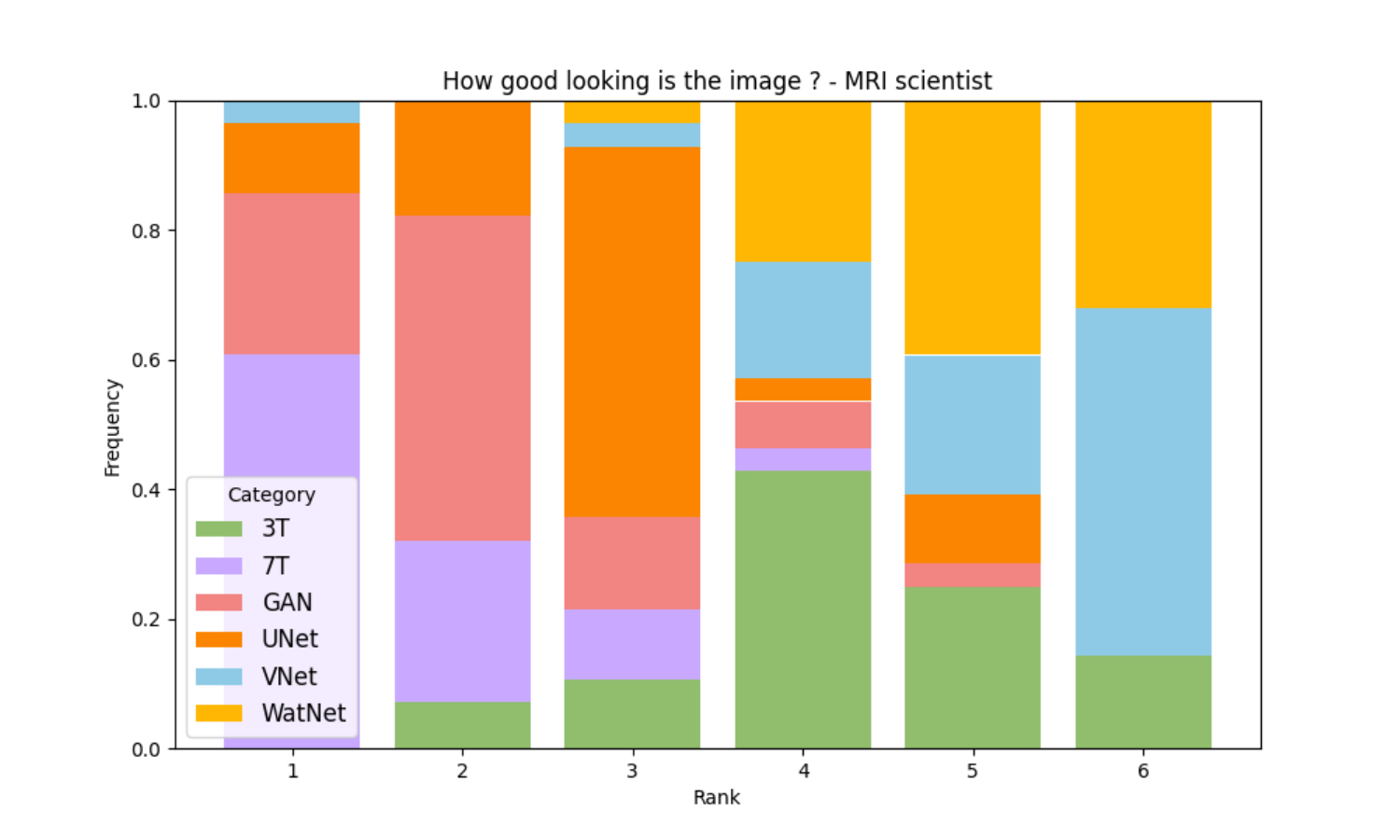}
\end{minipage}
\begin{minipage}[!ht]{0.48\textwidth}
\includegraphics[width=1.15\textwidth]{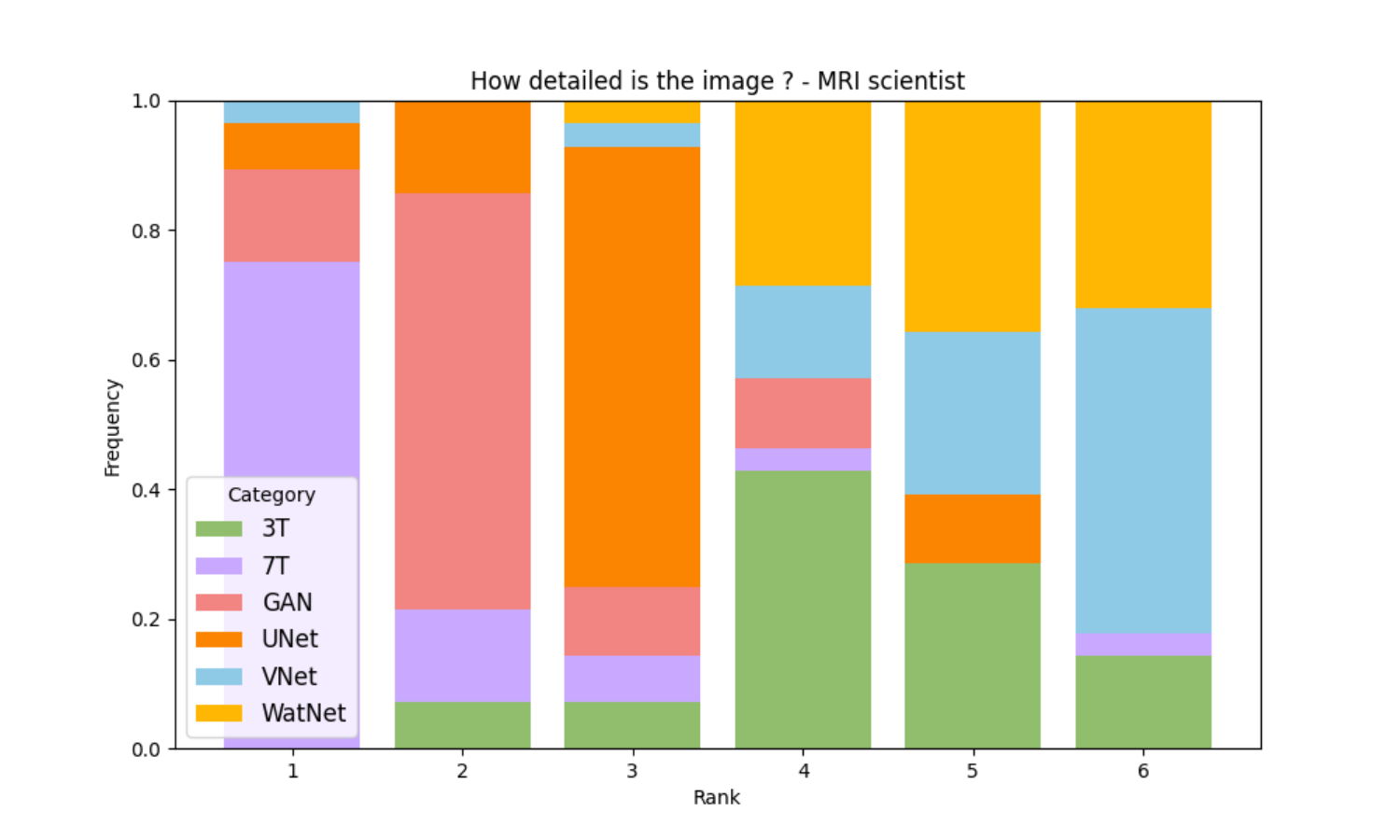}
\end{minipage}
\caption{Stacked bar graph of the results given by MRI scientist 1.}
\label{fig: survey 1}
\end{figure}

\begin{figure}[!ht]
\centering
\begin{minipage}[!ht]{0.48\textwidth}
\includegraphics[width=1.15\textwidth]{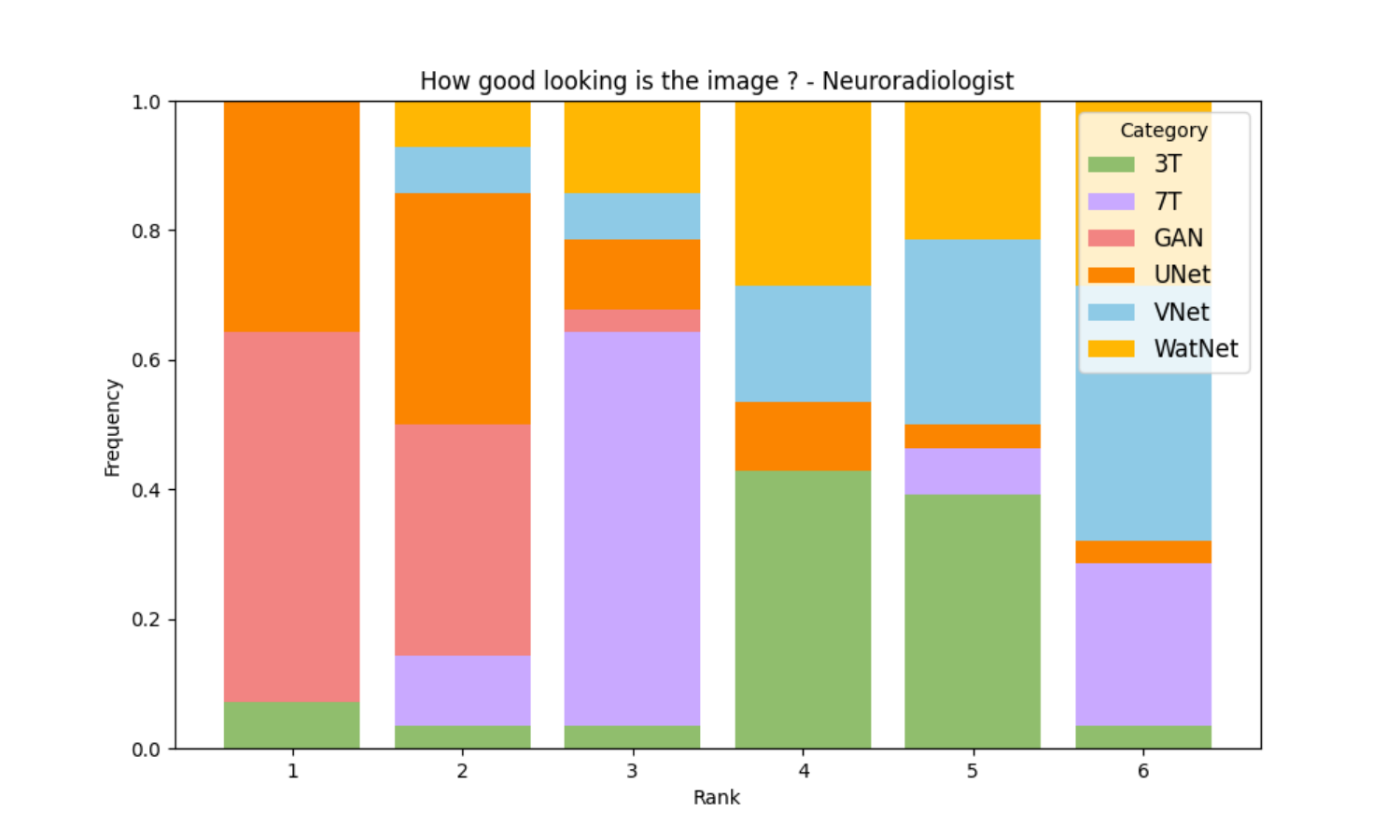}
\end{minipage}
\begin{minipage}[!ht]{0.48\textwidth}
\includegraphics[width=1.15\textwidth]{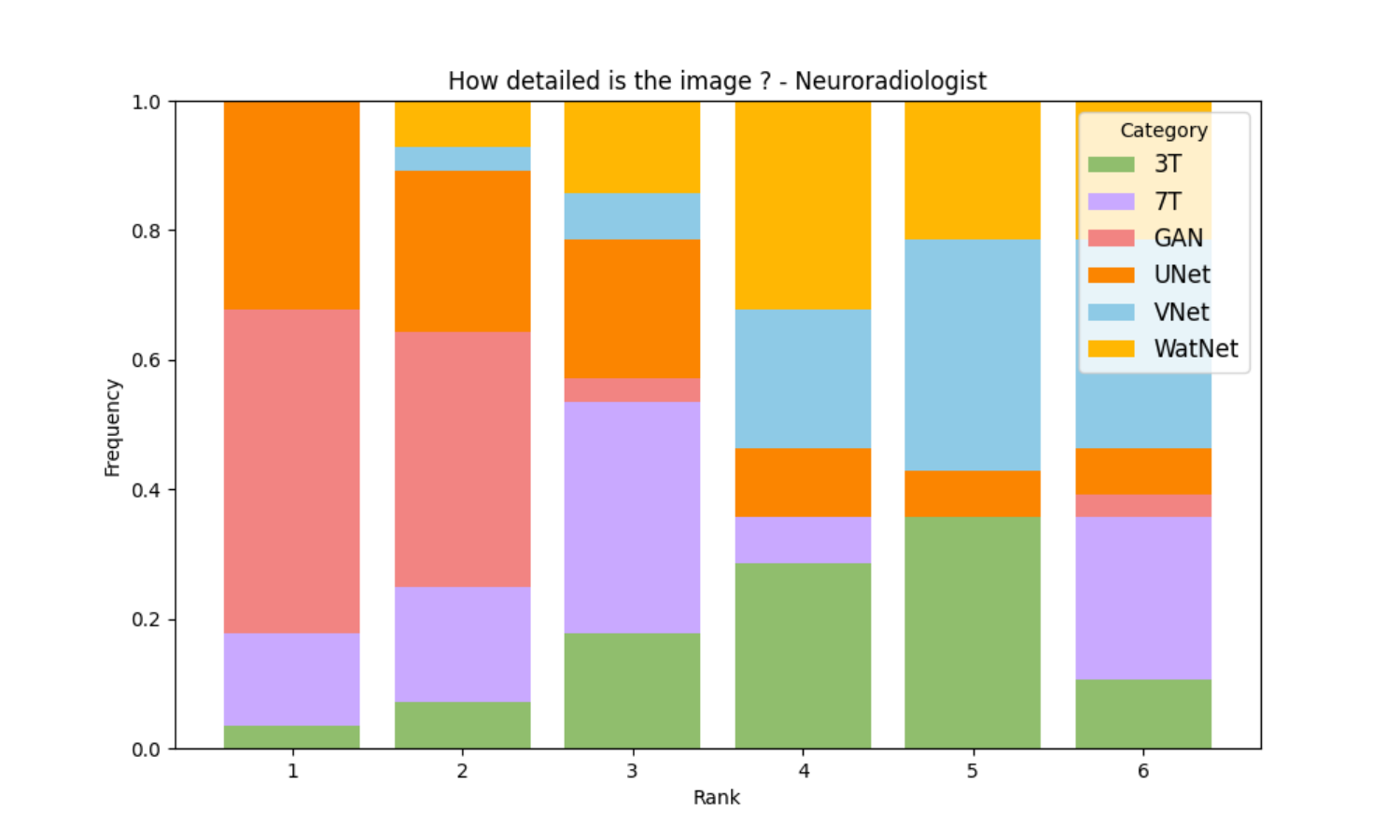}
\end{minipage}
\caption{Stacked bar graph of the results given by the neuroradiologist.}
\label{fig: survey 2}
\end{figure}

\begin{figure}[!ht]
\centering
\begin{minipage}[!ht]{0.48\textwidth}
\includegraphics[width=1.15\textwidth]{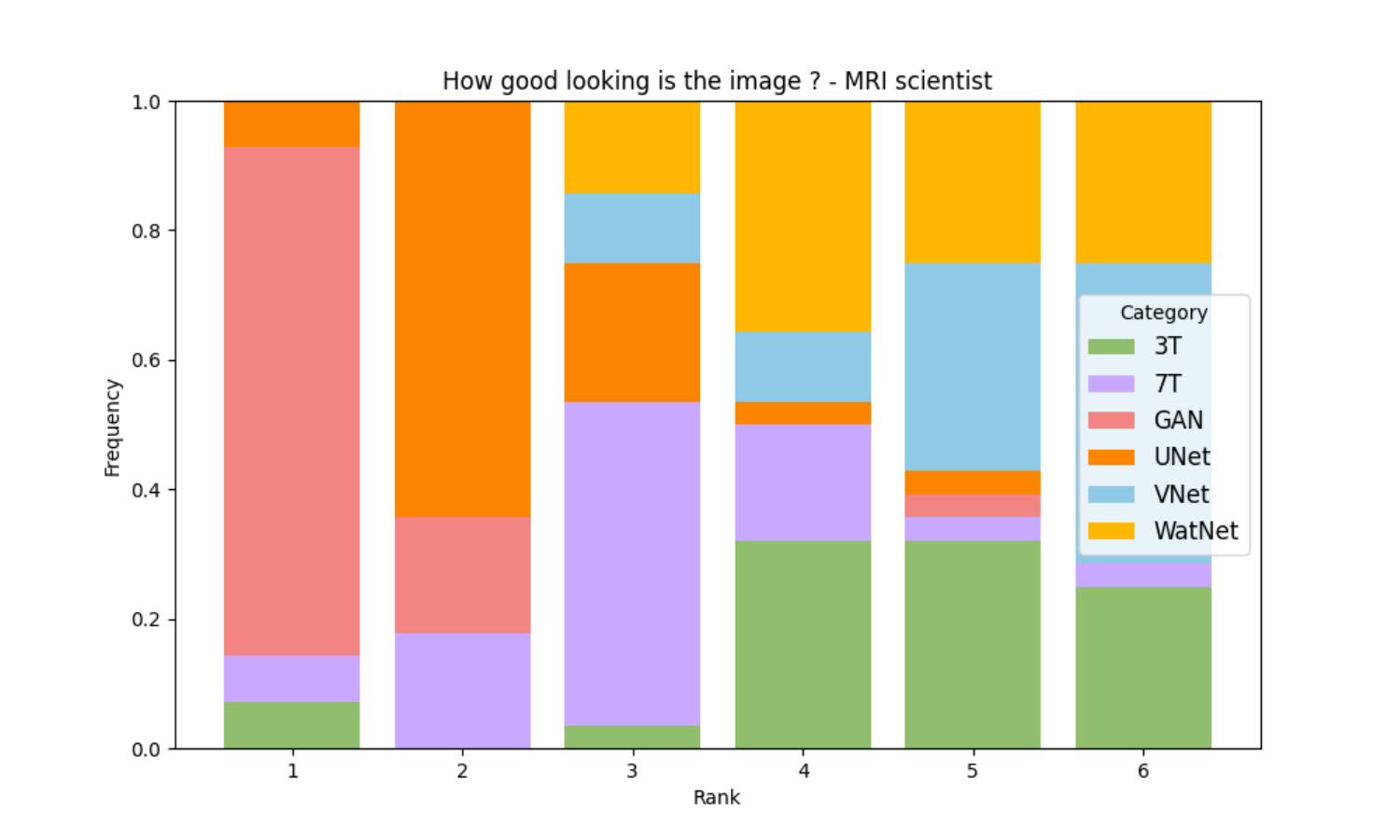}
\end{minipage}
\begin{minipage}[!ht]{0.48\textwidth}
\includegraphics[width=1.15\textwidth]{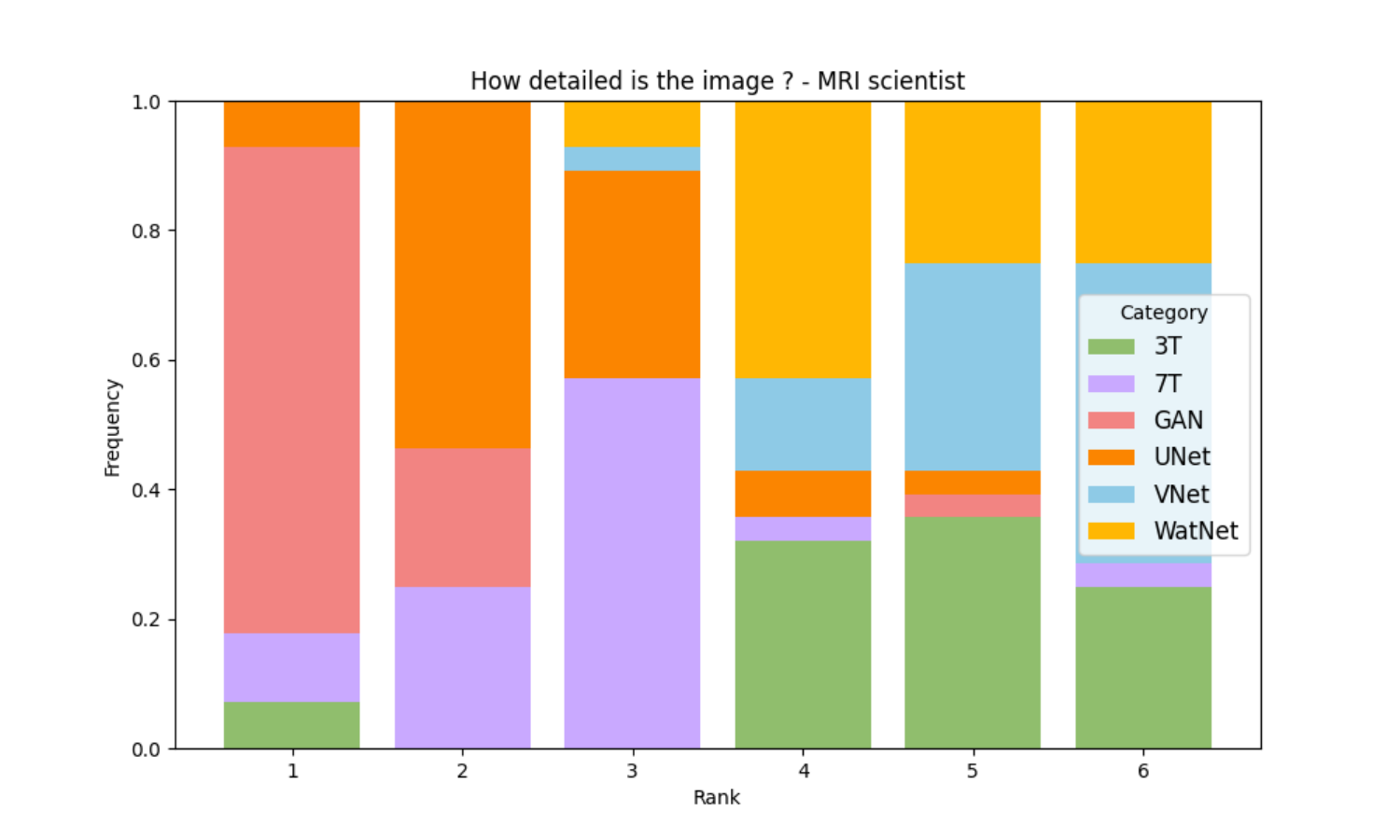}
\end{minipage}
\caption{Stacked bar graph of the results given by MRI scientist 2.}
\label{fig:survey 3 }
\end{figure}

\begin{figure}[!ht]
\centering
\begin{minipage}[!ht]{0.48\textwidth}
\includegraphics[width=1.15\textwidth]{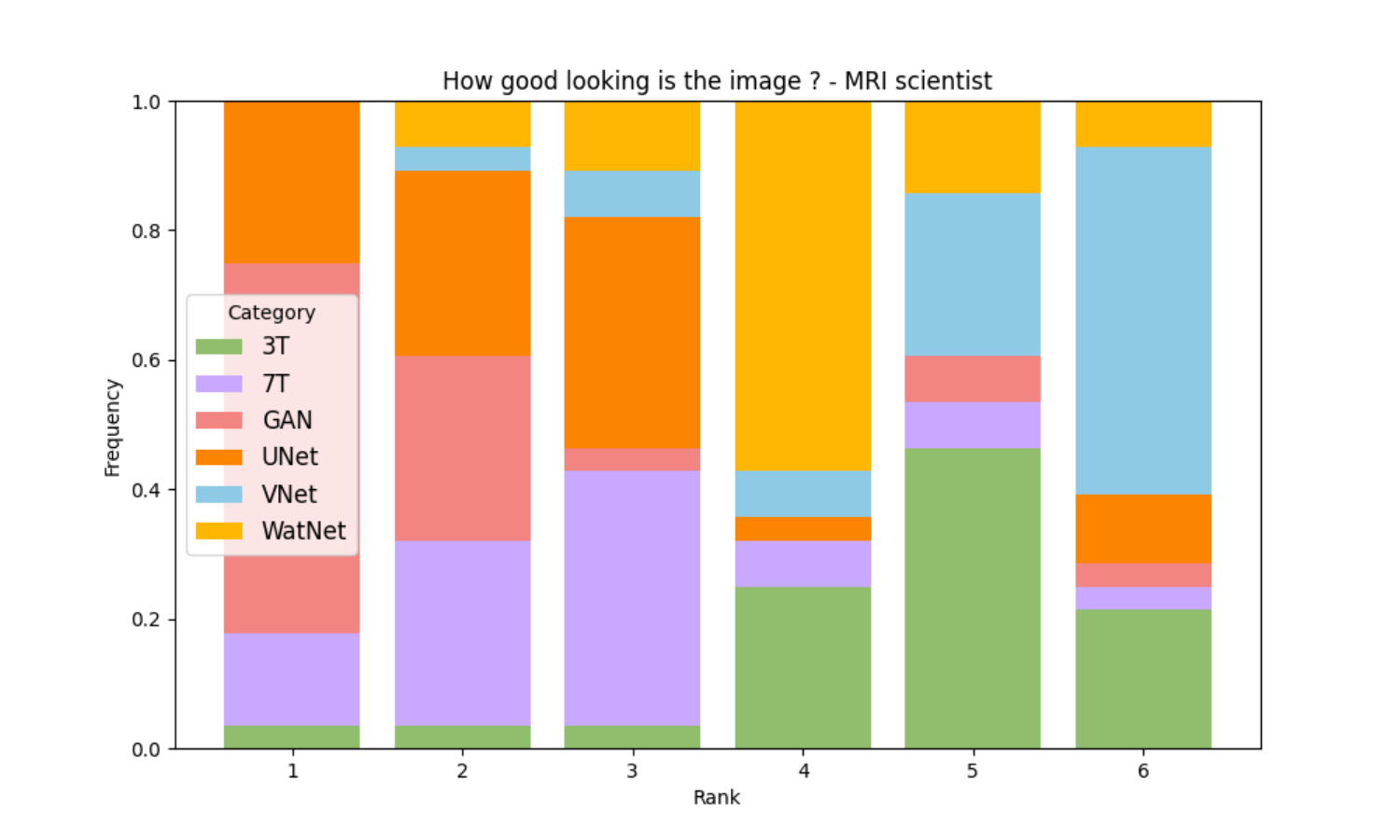}
\end{minipage}
\begin{minipage}[!ht]{0.48\textwidth}
\includegraphics[width=1.15\textwidth]{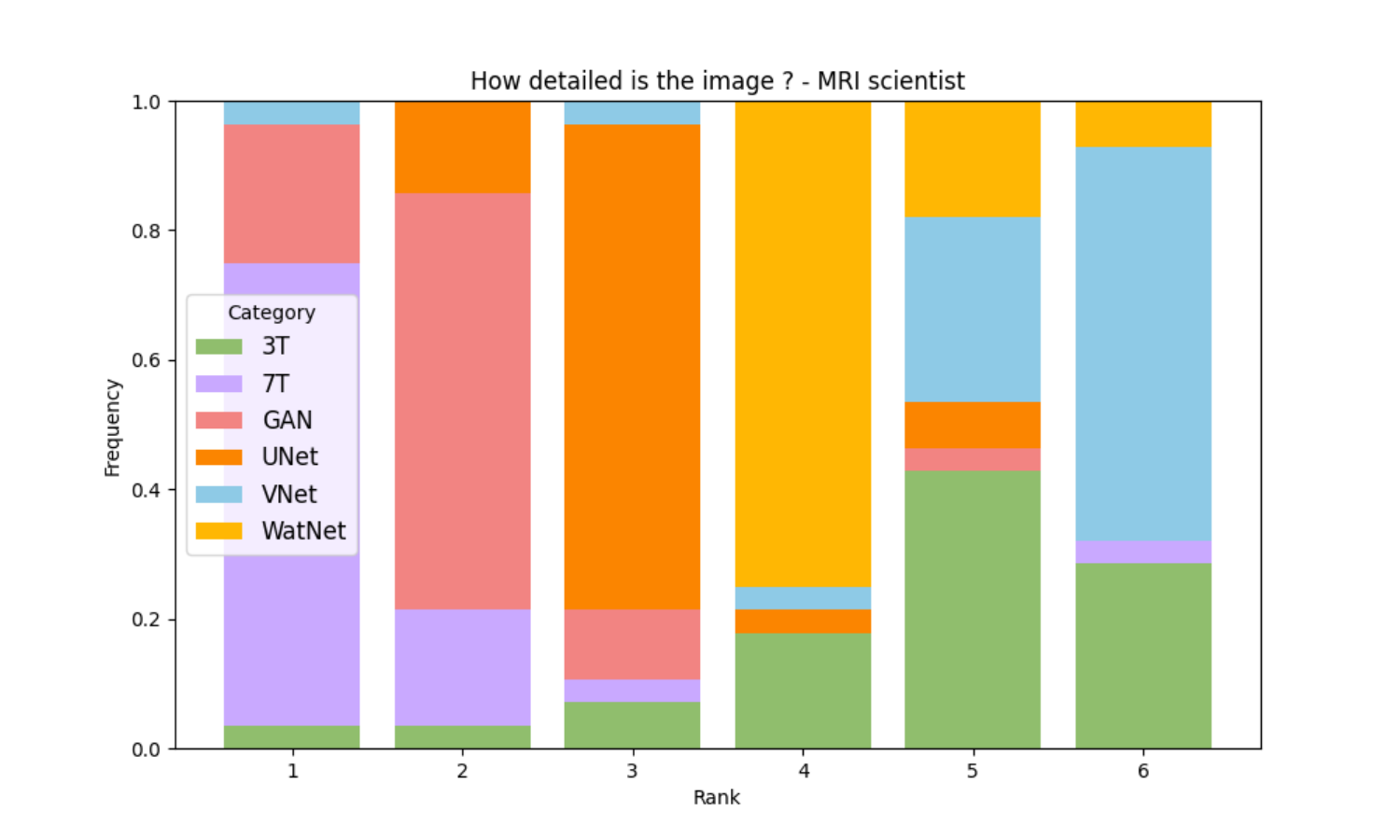}
\end{minipage}
\caption{Stacked bar graph of the results given by MRI scientist 3.}
\label{fig: survey 4}
\end{figure}

Here, we present more detailed results from the visual assessment survey. For each expert, we show the stacked bar graphs of the ranks given to every image and for each criteria in Figures \ref{fig: survey 1},\ref{fig: survey 2},\ref{fig:survey 3 }, \ref{fig: survey 4}. 

\newpage 

The following page is the first page of our survey.

\includepdf[pages={1}]{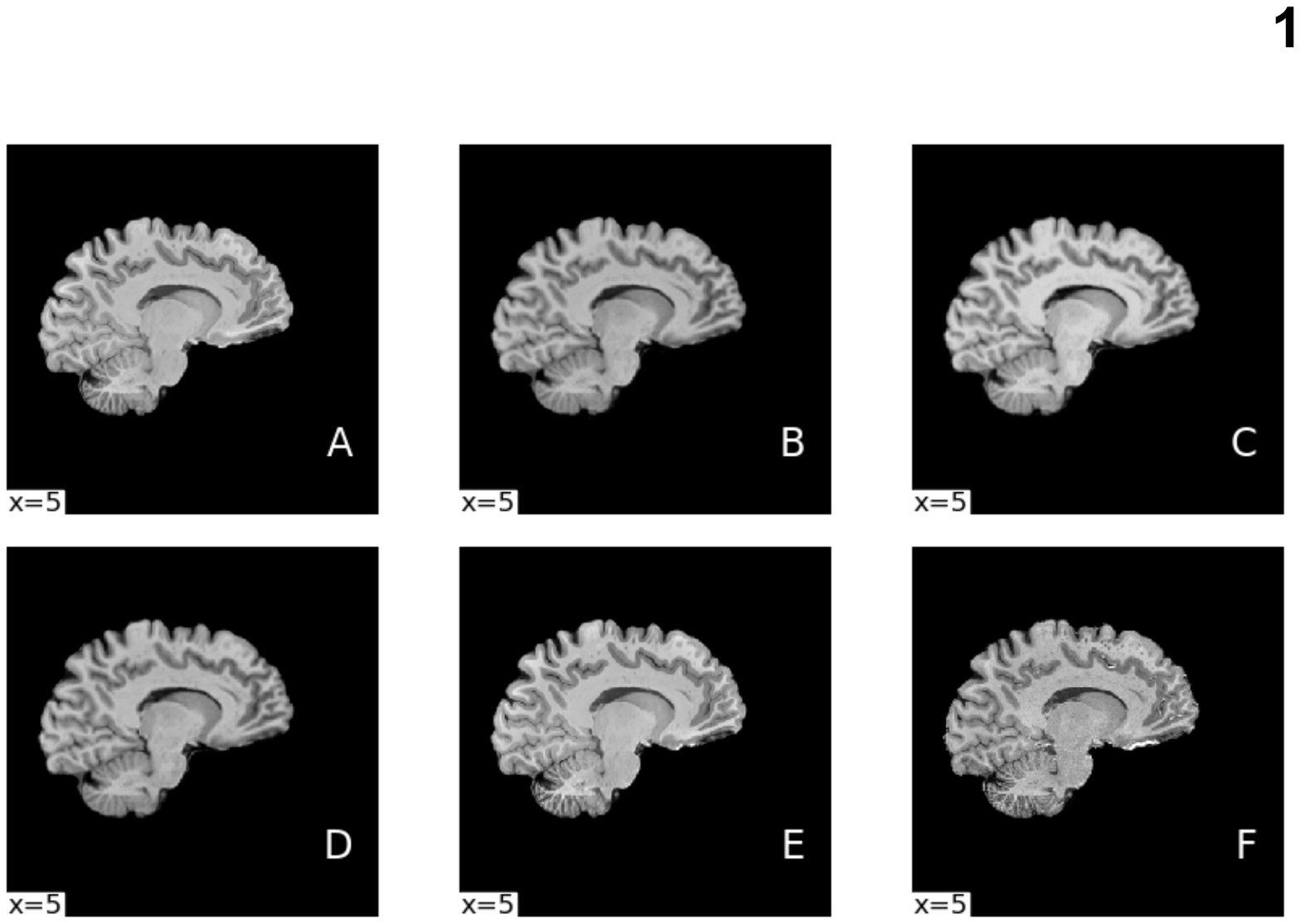}

\subsection{Ablation experiments}
\begin{figure}[!ht]
    \centering
    \includegraphics[width=0.7\textwidth]{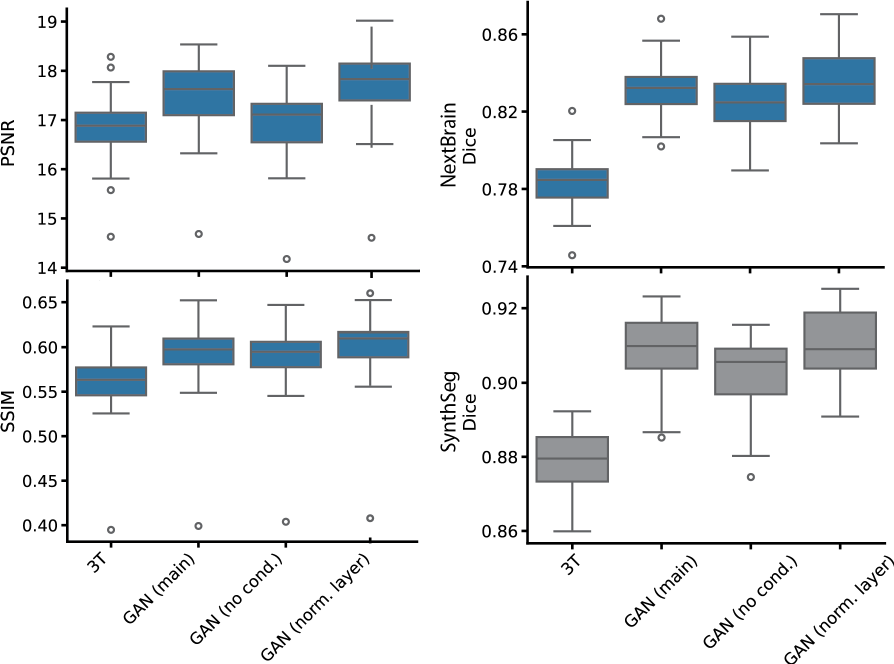}
    \caption{Results of ablation experiments. We investigated the effects of our modeling choices on modeling performance, namely conditioning on age and sex, and changes made to the final normalization layer of our models. We fit new GAN U-Net models without conditioning on age and sex (no cond.), as well as models with the AdaDM layer removed and the last normalization layer added back in (norm. layer). We compared these to our original GAN U-Net model (main) and the 3T scan for reference. PSNR, SSIM, and Dice from both NextBrain and SynthSeg segmentations are displayed, all with reference to the true 7T image. The PSNR and SSIM values are also shown in Table~\ref{tab:supp_psnr_ssim_ablation}.}
    \label{fig:supp_ablation}
\end{figure}

\begin{table}[h] 
    \centering

    \begin{tabular}{ccccc} 
        {Image Type} & {PSNR} $\uparrow$ & {SSIM} $\uparrow$  & SynthSeg Dice $\uparrow$  & NextBrain Dice $\uparrow$ \\ \hline
            { 3T } & $ 16.8 \pm 0.78 $ & $ 0.559 \pm 0.040 $ &$ 0.879 \pm 0.0089 $ & $ 0.784 \pm 0.0150 $ \\
            {GAN (main) } & $ 17.5 \pm 0.81 $ & $ 0.591 \pm 0.045 $ & $ 0.909 \pm 0.0099 $ & $ 0.831 \pm 0.0150 $  \\
            {GAN (no cond.) } & $ 16.9 \pm 0.84 $ & $ 0.588 \pm 0.044 $ & $ 0.902 \pm 0.011 $ & $ 0.823 \pm 0.0168 $ \\
            { GAN (norm. layer) } & $ \textbf{17.7} \pm 0.88 $ & $ \textbf{0.600} \pm 0.046 $ & $ \textbf{0.910} \pm 0.0096 $ & $ \textbf{0.836} \pm 0.0163 $ \\
    \end{tabular}	
    
    \caption{PSNR and SSIM scores comparing the generated 7T images using different versions of our GAN model to the real 7T, as well as Dice scores for SynthSeg and aggregated NextBrain segmentations. We show results from the GAN U-Net models without conditioning on age and sex (no cond.), a model with the AdaDM layer removed and the last normalization layer added back in (norm. layer), our original GAN U-Net model (main) and the 3T scan for reference. Computations are done on the 3D images without background or cerebellum artifacts and we show the average result over all regions in 28 images together with the standard deviation. The results are visulaized in the left part of Figure~\ref{fig:supp_ablation}.}
    \label{tab:supp_psnr_ssim_ablation}
\end{table}

\end{document}